\documentclass[referee]{raa}            % referee version: for submission

\UseRawInputEncoding
\usepackage{graphicx}             %for PS/EPS graphics inclusion, new
\usepackage{times}
\usepackage{natbib}
\usepackage{appendix}
\usepackage{amsmath}
\usepackage{amssymb}
\usepackage{epsfig}
\usepackage{rotating}
\usepackage{color}
\usepackage{longtable}
\usepackage{pdflscape,lscape}
\usepackage{float}
\usepackage{placeins}
\usepackage{booktabs}
\usepackage{upgreek}
\usepackage{url}
\usepackage{hyperref}
\hypersetup{colorlinks = true, linkcolor = green, anchorcolor = red, citecolor = blue, filecolor = red, pagecolor = red, urlcolor = red}
\newcommand{\teff}{$T_{\rm eff}$}
\newcommand{\logg}{$\log g$}
\newcommand{\vsini}{$v \sin i$}

\newcommand{\ds}{$\delta$\,Scuti}

\begin{document}

   \title{Candidate eclipsing binary systems with a $\delta$ Scuti star in Northern TESS field
%\footnotetext{$*$ Supported by the National Natural Science Foundation of China.}
}
%   \subtitle{I. Place Your Subtitle Here}

   \volnopage{Vol.0 (20xx) No.0, 000--000}      %%preserved for Editor. DOn't remove!
   \setcounter{page}{1}          %%starting page, preserved for Editor. DOn't remove!

   \author{F. Kahraman Ali\c{c}avu\c{s}
      \inst{1,2}
   \and D. G\"{u}m\"{u}\textcommabelow{s}
      \inst{3}
      \and \"{O}. K{\i}rm{\i}z{\i}ta\textcommabelow{s}
      \inst{4}
    \and \"{O}. Ekinci
      \inst{5}
       \and S. \c{C}avu\c{s}
      \inst{4}
   \and Y. T. Kaya
      \inst{6}
    \and F. Ali\c{c}avu\c{s}
      \inst{1,2}
   }

   \institute{\c{C}anakkale Onsekiz Mart University, 
Faculty of Sciences and Arts, Physics Department, 17100, \c{C}anakkale, Turkey; {\it filizkahraman01@gmail.com}\\
        \and
             \c{C}anakkale Onsekiz Mart University, Astrophysics Research Center and Ulup\u0131nar Observatory, TR-17100, 크nakkale, Turkey\\
             \and
             Istanbul  University,  Institute  of  Graduate  Studies  in  Science, Programme  of  Astronomy  and  Space  Sciences,34116, Beyaz{\i}t,Istanbul, Turkey\\
    \and
    \c{C}anakkale Onsekiz Mart University, School of Graduate Studies, Department of Physics, TR-17100, 크nakkale, Turkey\\
    \and
\c{C}anakkale Onsekiz Mart University, School of Graduate Studies, Department of Space Sciences and Technologies, TR-17100, 크nakkale, Turkey\\
\and
\c{C}anakkale Onsekiz Mart University, Faculty of Engineering
Computer Engineering, TR-17100, 크nakkale, Turkey\\
\vs\no
   {\small Received~~20xx month day; accepted~~20xx~~month day}}

\abstract{Existence of pulsating stars in eclipsing binaries have been known for decades. These types of objects are extremely valuable systems for astronomical studies as they exhibit both eclipsing and pulsation variations. The eclipsing binaries are the only way to directly measure the mass and radius of stars with a good accuracy ($\leq$\,1\%), while the pulsations are a unique way to probe the stellar interior via oscillation frequencies. There are different types of pulsating stars existing in eclipsing binaries. One of them is the \ds\, variables. Currently, the known number of \ds\, stars in eclipsing binaries is around 90 according to the latest catalog of these variables. An increasing number of these kinds of variables is important to understand the stellar structure, evolution and the effect of binarity on the pulsations. Therefore, in this study, we focus on discovering new eclipsing binaries with \ds\, component(s). We searched for the northern TESS field with a visual inspection by following some criteria such as light curve shape, the existence of pulsation like variations in the out-of-eclipse light curve and the \teff\, values of the targets. As a result of these criteria, we determined some targets. The TESS light curves of the selected targets first were removed from the binarity and frequency analysis was performed on the residuals. The luminosity, absolute and bolometric magnitudes of the targets were calculated as well. To find how much of these parameters represent the primary binary component (more luminous) we also computed the flux density ratio of the systems by utilizing the area of the eclipses. In addition, the positions of the systems in the H-R diagram was examined considering the flux density ratios. As a consequence of the investigation, we defined 38 candidates \ds\, and also one Maia variable in eclipsing binary systems. 
 }

   \authorrunning{Kahraman Ali\c{c}avu\c{s} et. al.}            %author_head in even pages
   \titlerunning{Eclipsing binary systems with a $\delta$ Scuti star}  % title_head in odd pages

   \maketitle
   
%%%%%%%%%%%%%%%%%%%%%%%%%%%%%%%%%%%%%%%%%%%%%%%%%%%%%%%%%%%%%%%%%%%%%%%%%%%%%%
\section{Introduction}           %% first-level sections will be auto-capitalized
\label{sect:intro}

Space telescopes have created a revolution in astronomical studies. The primary mission of some of these telescopes is mainly discovering new exoplanets, however, in addition to their success in finding new exoplanets, they have provided a huge amount of photometric data of stellar systems. Especially, the \textit{Kepler} \citep{2010Sci...327..977B} and the Transiting Exoplanet Survey Satellite (TESS, \citeauthor{2014SPIE.9143E..20R} \citeyear{2014SPIE.9143E..20R}) have a big impact on this. TESS has already finished its two-years of primary mission and is currently continuing its extended mission by observing the almost entire sky. The high-quality photometric data of these space telescopes allow us to deeply investigate some phenomena in stellar systems and understand their structures. 

To comprehend the evolution and the structure of stars, the eclipsing binaries and the pulsating stars are substantial systems. The eclipsing binaries are the only way to precisely determine the mass ($M$) and radius ($R$) parameters of the binary components with the help of modelling the photometric light curves and radial velocity measurements. The accuracy of the measured $M$ and $R$ values can be lower than 1\% \citep{2015ApJ...802..102L, 2013A&A...557A.119S}. On the other hand, pulsating stars are unique systems that allow us to probe the stellar interior via oscillation frequencies \citep{2010aste.book.....A}. Thanks to the analysis of high-quality space-based photometric data of pulsating stars we had information on some important phenomena such as internal rotation, core overshooting and angular momentum \citep[e.g.][]{2017ApJ...849...38L, 2015MNRAS.447.3264S}. Therefore, the eclipsing binaries with a pulsating component(s) are such crucial systems for deeply exploring the stellar evolution and structure.

The presence of pulsating stars in eclipsing binaries has been known for decades \citep{1971IBVS..596....1T}. There are different types of oscillating variables present in eclipsing binaries for instance $\beta$\,Cephei, $\delta$\,Scuti and $\gamma$\,Doradus stars \citep{2021Galax...9...28L, 2021Obs...141..282S}. Currently, the known number of $\delta$\,Scuti stars in eclipsing binaries is higher than the other type of pulsating variables because of their relatively shorter pulsation periods \citep{2021Galax...9...28L, 2017MNRAS.470..915K, 2017MNRAS.465.1181L}. The \ds\, stars are early A to F type variable and their luminosity class changes from dwarf to giant \citep{2010aste.book.....A}. These variables have their own instability strip where theoretically \ds\,-type variations are expected \citep{2005A&A...435..927D}. The \ds\, stars generally exhibit pressure mode oscillation with an oscillation period range of 18-min. to 8h \citep{2010aste.book.....A}. According to the recent catalog of $\delta$\,Scuti stars in eclipsing binaries (DSEB), there are around 90 DSEB \citep{2017MNRAS.470..915K}. Additionally, there are also $\delta$\,Scuti stars present in other types of binary systems \citep{2017MNRAS.465.1181L}. 

The effect of binarity on pulsations has been shown in some studies \citep{2020NatAs...4..684H, 2017MNRAS.470..915K, 2006MNRAS.366.1289S, 2004A&A...419.1015M}. Because of the gravitational effects of the components on each other, it was expected that the pulsations in oscillating components differ from single pulsating stars. The effect of binarity on the pulsating $\delta$\,Scuti components was shown by a relationship between the pulsation and orbital period \citep{2006MNRAS.366.1289S}. It was obtained that the pulsation period ($P_{puls}$) decreases when the orbital period ($P_{orb}$) declines. The smaller $P_{orb}$ means that the binary components are closer to each other, so the semi-major axis ($a$) is shorter. According to the gravity law, applied gravitational force on the pulsating components is increased by the declined $P_{orb}$. A comparison of the properties of single and eclipsing binary member $\delta$\,Scuti stars was given by \cite{2017MNRAS.470..915K}. They showed that $P_{puls}$ and pulsation amplitude ($A_{puls}$) of single classical $\delta$\,Scuti stars are longer and higher, respectively, comparing to eclipsing binary member $\delta$\,Scuti variables. In this study, it was also presented that single $\delta$\,Scuti stars have a higher rotation velocity (\vsini) on average compared to DSEB. These are the results of the effects of gravitational forces between the binary components. In the same study, lots of relationships were examined between the pulsation properties ($P_{puls}$, $A_{puls}$) and other stellar parameters such as $M$, $R$, effective temperature (\teff) and surface gravity (\logg).

In a recent study, by the help of TESS data, \cite{2020NatAs...4..684H} first time showed that in a close binary system the pulsation axis of the oscillating component can be aligned by tidal forces. There are now a few samples of this kind of objects including an eclipsing binary system and they are called to be "Tidally tilted pulsators" \citep{2020MNRAS.494.5118K,2020MNRAS.498.5730F, 2020NatAs...4..684H}. The high-quality TESS data allowed us to find such variables which have been searched for years and has provided a good opportunity to deeply probe DSEB. A comprehensive research of DSEB is quite important for comprehending stellar structure, evolution and testing evolutionary models. For this reason, an increased number of such systems will offer us more opportunities for understanding stellar objects. Hence, in this study, we focus on DSEB. 

In this study, we present our TESS northern field research for discovering new \ds\, stars in eclipsing binaries. The paper is organised as follows. In Sect.\,2, information about the observational data and target selection are introduced. In Sect.\,3, the frequency analysis of the selected targets is given. In Sect.\,4, calculations of some physical parameters and the position of the systems are presented. In Sect.\,5 and Sect.\,6, discussion and conclusions are given, respectively.
%%%%%%%%%%%%%%%%%%%%%%%%%%%%%%%%%%%%%%%%%%%%%%%%%%%%%%%%%%%%%%%%%%%%%%%%%%%%%%%%%%

\section{OBSERVATIONAL DATA AND TARGET SELECTION} \label{sec:obs}
To discover new DSEB we searched for the northern TESS field. TESS was launched in April 2018 and it is the main goal is detecting exoplanets. TESS has four identical CCD cameras that have $24^o$\,$\times$\,$24^o$field of view (FOV) \citep{2014SPIE.9143E..20R}. TESS monitors the sky with a wide red-bandpass filter by dividing it into sectors. Each sector has around 27 days of photometric observations. In the first two years of the mission, TESS observed many targets with 2-min short cadence (SC) and 30-min long cadence (LC) and now in its extended mission, LC was reduced to 10-min. According to the position of the target in the sky, some objects were observed by TESS in more than one sector while some have only one sector data. The TESS data are public in the Barbara A. Mikulski Archive for Telescopes (MAST)\footnote{https://mast.stsci.edu} archive where the data is present in two kinds of flux; the simple aperture photometry (SAP) and the pre-search data conditioning SAP flux (PDCSAP). 

In the current study, to find new eclipsing binaries with a \ds\, component, we had a visual inspection for all TESS sectors in the northern hemisphere. Basically, in the first step, we searched for eclipsing binary like variations in the TESS data. In the next step, the out-of-eclipse variations of the determined eclipsing binaries were examined and the systems that exhibit both eclipsing binary variations and oscillation like changes in the out-of-eclipse light curves were chosen as targets. In the final step, the atmospheric parameters (\teff\,, \logg) of the targets were checked. As \ds\, components are searched for, we considered that targets which have \teff\, and  \logg\, parameters in the range given for \ds\, stars. According to the study of \cite{2001A&A...366..178R}, typical \teff\, and \logg\, ranges for \ds\, stars are $6300-8500$\,K and $3.2-4.3$, respectively. Therefore, for the final list the targets, which have \teff\, and \logg\, in the given ranges within errors, were chosen. The atmospheric parameters of the targets were taken from the TESS input catalog \citep[TIC,][]{2019AJ....158..138S}. The final list of the targets is given in Table\,\ref{tab1}. There is one target (TIC\,13037534) which has \teff\, lower than the given \teff\, for \ds\, stars even considering a possible error. However, the target illustrates significant \ds\,-type variations and its spectral type is given as F1V in the catalog of \cite{2013AN....334..860A}. Considering a possible error in \teff\, we included this target in our list as well. As a result, we have 39 candidates of DSEB listed in Table\,\ref{tab1}.

\begin{table*}
\centering
\caption{The List of Candidate DSEB and Their Properties. The Atmospheric Parameters Were Taken From the TIC \citep{2019AJ....158..138S} and an Average Error is Given For These Parameters.} \label{tab1}
\begin{tabular}{llllllll} \toprule
 TIC Number& RA(J2000)                     & DEC(J2000)                    &  P$_{orb}$   &    V     &  \teff              & \logg\,        & Sectors \\
           &                               &                               & (day)        & (mag)    &  (K)\,$\pm$\,181    &  \,$\pm$\,0.09 &  \\ 
\hline
 8669966   & 16$^{h}$ 33$^{m}$ 29$^{s}$.1  & +30$^{o}$ 29$^{'}$ 56$^{"}$.6 & 3.400272 (6) & 6.89     &  7651               & 3.54           & 25 \\
 10057647  & 08$^{h}$ 53$^{m}$ 19$^{s}$.3  & +53$^{o}$ 44$^{'}$ 08$^{"}$.9 & 16.259725 (5) & 8.52     & 7249                & 4.06           & 20, 47 \\
 13037534$^{a}$ & 22$^{h}$ 55$^{m}$ 30$^{s}$.2  & +64$^{o}$ 00$^{'}$ 31$^{"}$.0 & 30.220888 (9)& 11.29& 5694            & $-$      & 17, 18, 24 \\
 14948284  & 12$^{h}$ 41$^{m}$ 07$^{s}$.8  & +30$^{o}$ 26$^{'}$ 13$^{"}$.6 & 2.699312 (8) & 6.95     & 7178                & 3.86           & 22 \\ 
 48084398 & 18$^{h}$ 47$^{m}$ 29$^{s}$.6  & +49$^{o}$ 25$^{'}$ 55$^{"}$.3 & 4.243430 (7) & 7.20     & 6814                & 3.29           & 14, 15, 26, 40\\ 
 71613490  & 13$^{h}$ 29$^{m}$ 56$^{s}$.1  & +34$^{o}$ 31$^{'}$ 27$^{"}$.4 & 1.313542 (7) & 7.69     & 7737                & 3.84           & 23\\ 
 72839144$^{b}$  & 16$^{h}$ 56$^{m}$ 28$^{s}$.7  & +37$^{o}$ 39$^{'}$ 18$^{"}$.9 & 1.755764 (2)  & 10.03   & 7189                & 3.68           & 25\\
 75593781$^{c}$  & 05$^{h}$ 41$^{m}$ 34$^{s}$.9 & +25$^{o}$ 59$^{'}$ 52$^{"}$.9 & 1.532149 (6) & 11.39    & 8300                & 4.33           & 43, 44, 45\\
 78148497   & 05$^{h}$ 54$^{m}$ 24$^{s}$.5  & +26$^{o}$ 18$^{'}$ 31$^{"}$.7 & 2.708848 (5) & 10.96    & 7287                & 3.93           & 43, 45\\
 85600400   & 17$^{h}$ 16$^{m}$ 49$^{s}$.9  & +38$^{o}$ 21$^{'}$ 58$^{"}$.7 & 1.471461 (5) & 12.19    & 7235                & 3.85           & 25, 26\\
 116334565 & 05$^{h}$ 40$^{m}$ 28$^{s}$.4  & +30$^{o}$ 58$^{'}$ 27$^{"}$.3 & 3.475191 (7) & 11.32    & 7546                & 3.60           & 43, 45\\
 165618747  & 17$^{h}$ 20$^{m}$ 07$^{s}$.8  & +13$^{o}$ 39$^{'}$ 57$^{"}$.6 & 0.648264 (3) & 11.67    & 7603                & 4.11           & 25, 26\\ 
 172431974  & 19$^{h}$ 58$^{m}$ 50$^{s}$.1  & +39$^{o}$ 19$^{'}$ 52$^{"}$.1 & 2.387408 (3) & 10.96        & 7236                & 3.71           & 14, 15, 41\\
 193774939$^{d}$ & 17$^{h}$ 53$^{m}$ 12$^{s}$.7 & +43$^{o}$ 46$^{'}$ 23$^{"}$.2 & 1.305766 (7) & 10.16& 7358               & 4.05           & 25, 26\\ 
 197755658$^{e}$ & 22$^{h}$ 28$^{m}$ 01$^{s}$.7 & +53$^{o}$ 41$^{'}$ 00$^{"}$.1  & 3.158785 (4) & 11.46& 7645           & 3.86           & 16, 17\\ 
 197757000$^{f}$  & 22$^{h}$ 28$^{m}$ 49$^{s}$.9  & +53$^{o}$ 46$^{'}$ 15$^{"}$.9 & 2.185923 (7) & 11.10& 6985           & 3.92           & 16, 17\\
 240962482 & 01$^{h}$ 15$^{m}$ 58$^{s}$.9  & +52$^{o}$ 46$^{'}$ 40$^{"}$.0 & 4.475000 (1) & 10.12     & 7748               & 3.81           & 18\\
 241013310 & 01$^{h}$ 20$^{m}$ 12$^{s}$.8  & +48$^{o}$ 36$^{'}$ 41$^{"}$.4 & 2.144879 (6) & 10.11     & 7093               & 4.01           & 17, 18\\ 
 256640561 & 21$^{h}$ 51$^{m}$ 35$^{s}$.4  & +71$^{o}$ 53$^{'}$ 08$^{"}$.7 & 1.712080 (2)   & 8.41        & 7891               & 3.48           & 17, 18, 24, 25\\
 272822330 & 23$^{h}$ 46$^{m}$ 17$^{s}$.9  & +62$^{o}$ 01$^{'}$ 33$^{"}$.4 & 4.501230 (4) & 10.06     & 7340               & 3.56           & 17, 18, 24\\ 
 289947843 & 06$^{h}$ 40$^{m}$ 16$^{s}$.6  & +79$^{o}$ 35$^{'}$ 58$^{"}$.3 &   $-$   & 6.75      & 9175               & 4.12           & 19, 20, 26, 40, 47\\
 301909087$^{g}$ & 02$^{h}$ 41$^{m}$ 16$^{s}$.5  & +48$^{o}$ 56$^{'}$ 18$^{"}$.8 & 4.448759 (6) & 10.81& 6847              & 3.54           & 18\\ 
 305633328 & 21$^{h}$ 02$^{m}$ 23$^{s}$.9  & +60$^{o}$ 04$^{'}$ 41$^{"}$.9 & 2.506158 (3) & 11.76     & 6899               & $-$              & 16, 17\\
 322428763 & 18$^{h}$ 35$^{m}$ 03$^{s}$.3  & +28$^{o}$ 49$^{'}$ 40$^{"}$.1 & 6.714251 (8) & 11.18     & 6151               & 3.27           & 40\\ 
 327121759$^{h}$ & 00$^{h}$ 13$^{m}$ 30$^{s}$.1  & +58$^{o}$ 17$^{'}$ 00$^{"}$.8 & 2.990910 (1) & 11.67     & 7159         & 3.69           & 17, 18, 24 \\ 
 337094559  & 22$^{h}$ 22$^{m}$ 32$^{s}$.6 & +63$^{o}$ 35$^{'}$ 10$^{"}$.6 & 1.777515 (8) & 9.98 & 7424 & 3.72 & 24 \\
 338159479$^{i}$ & 22$^{h}$ 32$^{m}$ 15$^{s}$.5  & +64$^{o}$ 58$^{'}$ 40$^{"}$.3 & 1.414305 (1) & 12.07& $-$           & $-$              & 17, 18, 24 \\ 
 354926863 & 02$^{h}$ 37$^{m}$ 46$^{s}$.2  & +67$^{o}$ 51$^{'}$ 19$^{"}$.9  & 1.436540 (3) & 10.98     & 7388              & 4.07           & 18, 19, 25 \\
 358613523$^{j}$ & 11$^{h}$ 40$^{m}$ 24$^{s}$.7  & +80$^{o}$ 14$^{'}$ 09$^{"}$.5  & 1.327830 (5) & 10.31& 7271              & 4.28           & 14, 20, 21$^{n}$\\
 393894013$^{k}$ & 13$^{h}$ 13$^{m}$ 33$^{s}$.4 & +47$^{o}$ 47$^{'}$ 51$^{"}$.9 & 0.816127 (4) & 9.34  & 7866   & 4.19           & 22 \\ 
 396134795 & 23$^{h}$ 48$^{m}$ 23$^{s}$.6 & +36$^{o}$ 18$^{'}$ 40$^{"}$.3 & 2.586765 (4) & 10.11 & 6946                    & 3.80           & 17 \\
 396201681 & 23$^{h}$ 58$^{m}$ 06$^{s}$.1 & +67$^{o}$ 36$^{'}$ 11$^{"}$.4 & 7.039790 (1) & 10.16 & 7767                    & $-$              & 17, 18, 24, 25 \\
 420114772 & 19$^{h}$ 21$^{m}$ 43$^{s}$.2 & +74$^{o}$ 45$^{'}$ 45$^{"}$.5 & 6.470060 (5) & 10.47 & 7015                    & $-$              & 14$-$26, 41, 47 \\
 421714420 & 21$^{h}$ 38$^{m}$ 45$^{s}$.8 & +55$^{o}$ 47$^{'}$ 29$^{"}$.9 &    $-$       & 7.90  & 8820                    & 3.81           & 16,17  \\
 428003183 & 23$^{h}$ 16$^{m}$ 53$^{s}$.1 & +44$^{o}$ 29$^{'}$ 18$^{"}$.4 & 0.742981 (7) & 11.11 & 7345                    & 4.05           & 16, 17  \\
 430808126 & 22$^{h}$ 15$^{m}$ 58$^{s}$.4 & +53$^{o}$ 18$^{'}$ 43$^{"}$.1 & 6.481440 (7) & 11.56 & 8460                    & $-$              & 16, 17 \\
 440003271$^{l}$ & 00$^{h}$ 10$^{m}$ 03$^{s}$.2 & +46$^{o}$ 23$^{'}$ 25$^{"}$.1 & 2.639210 (1) & 7.51& 8334                & 4.16           & 17 \\
 456905229$^{m}$ & 01$^{h}$ 44$^{m}$ 53$^{s}$.5 & +19$^{o}$ 51$^{'}$ 24$^{"}$.5 & 1.692620 (8) & 8.44& 7041                 & $-$             & 17, 42, 43 \\
 467354611 & 21$^{h}$ 46$^{m}$ 23$^{s}$.5 & +77$^{o}$ 22$^{'}$ 19$^{"}$.7 &      $-$ & 9.76 & 6235                    & 3.35           & 17, 19, 25, 26 \\
     \bottomrule
     \hline
\end{tabular}
\begin{description}
     \item[] $^{a}$F1V\,\citep{2013AN....334..860A}, $^{b}$F0\,$+$\,G9 IV\,\citep{2004yCat.5124....0S}, $^{c}$B9\,\citep{2013AN....334..860A}, $^{d}$F1V\,$+$\,K6\,\citep{2013AN....334..860A}, $^{e}$A9.5V\,$+$\,F3.5V\,\citep{2013AN....334..860A}, $^{f}$A5\,\citep{2013AN....334..860A}, $^{g}$A2\,$+$\,G6IV\,\citep{2004A&A...417..263B}, $^{h}$A0\citep{2013AN....334..860A}, $^{i}$A8\,$+$\,G8IV\,\citep{2004A&A...417..263B}, $^{j}$A5\,$+$\,G6IV\,\citep{2004A&A...417..263B}, $^{k}$A3\,\citep{2013AN....334..860A}, $^{l}$A3\,\citep{2013AN....334..860A}, $^{m}$F0\,\citep{2013AN....334..860A}, $^{n}$14, 20, 21, 26, 40, 41 
   \end{description}
\end{table*}

For the analysis of the candidate DSEB, we preferred to use only SC data because the Nyquist frequency for the SC data reaches $\sim$360\,$d^{-1}$. Taking into account the typical frequency range of \ds\, variables ($\sim$4$-$80\,$d^{-1}$ \citeauthor{2010aste.book.....A} \citeyear{2010aste.book.....A}), the SC data are the most suitable data for an examination of \ds\,-type variations. In Table\,\ref{tab1}, the available TESS sectors for each target is given. The SAP and PDCSAP fluxes of our targets were controlled and they found similar and as the PDCSAP fluxes are the long term trends removed fluxes and mostly the cleaner data, we preferred to use the PDCSAP fluxes. Each of fluxes were converted into magnitude\footnote{Flux [e{-}/s]=10$^{(20.44-TESS_{mag}/2.5)}$} to use in the analysis.

\begin{figure}[hbt!]
\centering
\includegraphics[width=\textwidth]{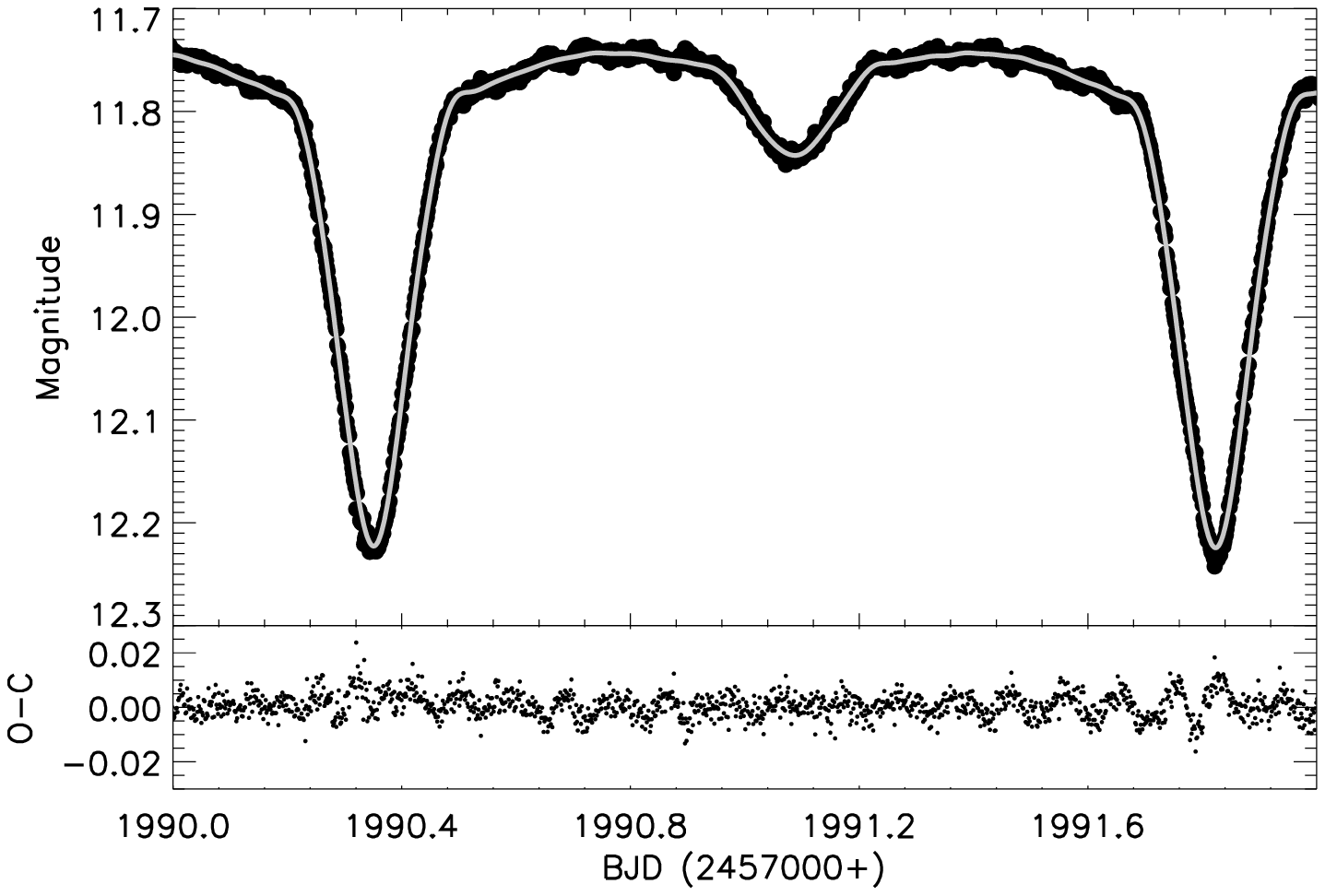}
\caption{Upper panel: the fit (grey line) of the orbital frequency and its harmonics to the TESS data of TIC\,85600400. Lower panel: residuals.}\label{fig1}
\end{figure}

%%%%%%%%%%%%%%%%%%%%%%%%%%%%%%%%%%%%%%%%%%%%%%%%%%%%%%%%%%%%%%%%%%%%%%%%%%%%%%%%%

\section{FREQUENCY ANALYSIS}\label{sec:fre}

In the current study, our main goal is revealing \ds\,-type variations in our candidate systems. Therefore, we carried out a frequency analysis for each system. We used the {\sc Period04} program which derives individual frequencies from astronomical data including gaps and also allows us to find the combination and harmonic frequencies \citep{2005CoAst.146...53L}. Our analysis consists of two steps. In the first step, the binary variations were removed from all available data of each target to obtain only the variation of oscillations. The binary variations were removed from the data with a phenomenological fit including the frequency of orbital periods and their harmonics \citep{2022MNRAS.510.1413K}. Before starting this analysis, the orbital periods of each target were calculated by performing a frequency analysis and these orbital periods were used in the current research. We could not determine the orbital periods for some systems because there is only one primary eclipse in their data. The derived orbital parameters are given in Table\,\ref{tab1}. In Fig.\,\ref{fig1}, we show one example of orbital period frequency fits to the TESS data and the residuals. As clearly seen from the figure, the binary variations were extracted and only the light curve of the pulsations was obtained. In the light curves of some systems, there are only one or two eclipses in available data. For these systems, no orbital frequency fit was applied, only the eclipse(s) was removed from the light curves and the rest was analysed.

\begin{figure*}
 \centering
  \begin{minipage}[b]{0.45\textwidth}
  \includegraphics[height=4.6cm, width=1\textwidth]{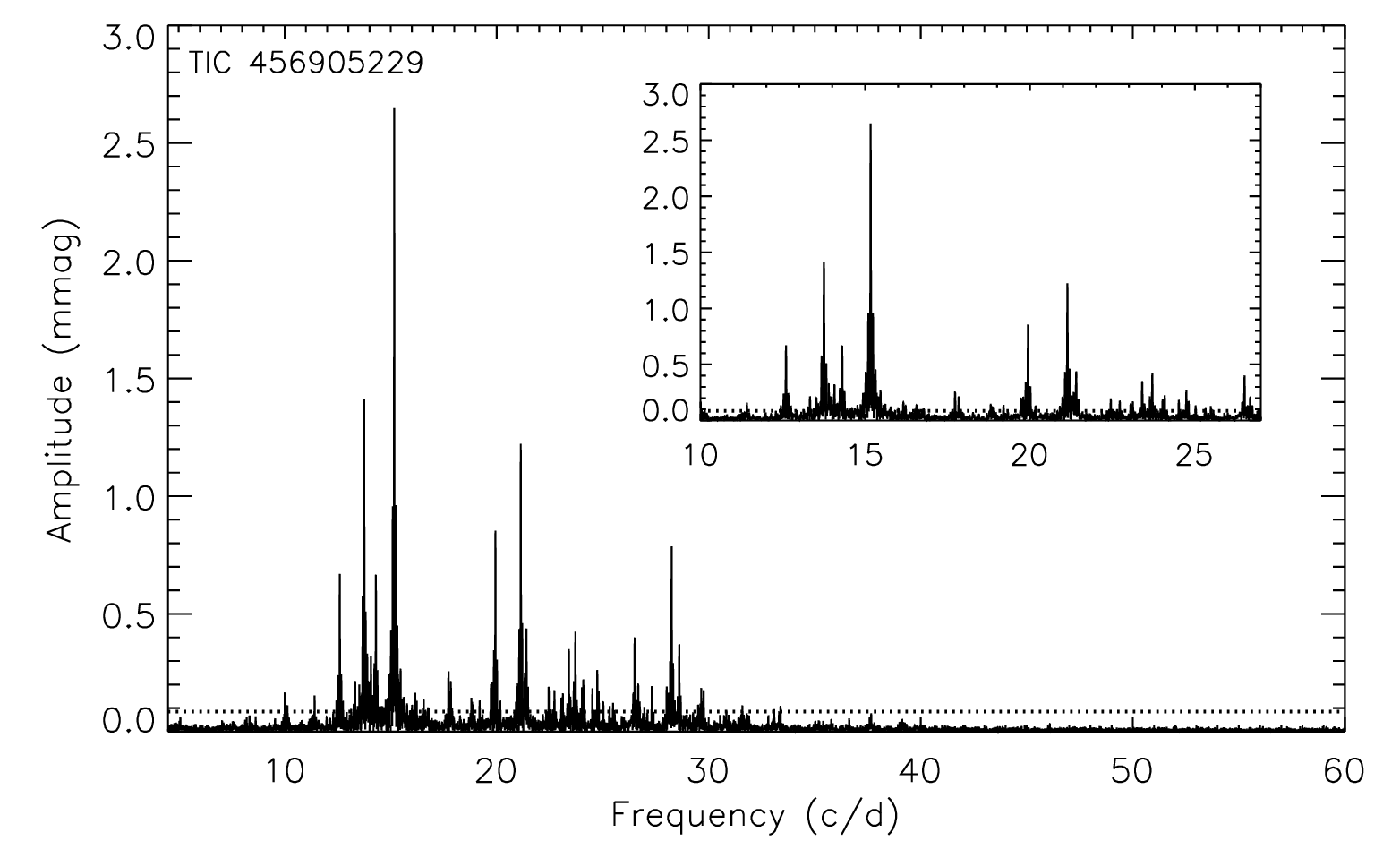}
%   \caption{1}
 \end{minipage}
 \begin{minipage}[b]{0.45\textwidth}
  \includegraphics[height=4.6cm, width=1\textwidth]{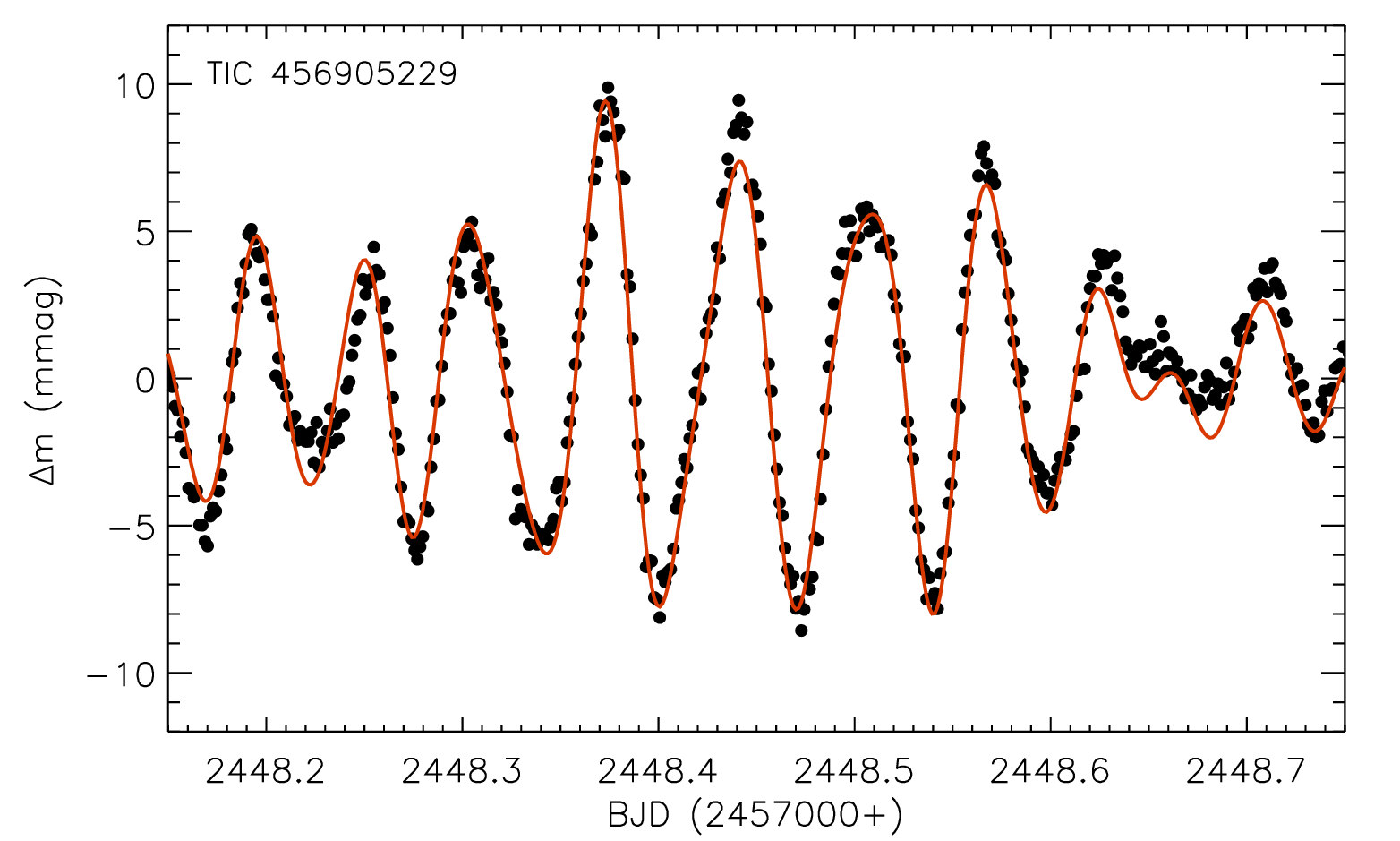}
  \end{minipage}
  \caption{Left panel: the frequency spectrum of TIC\,456905229. The dashed line represents the 4.5-$\sigma$ level. Right panel: theoretical fit (red solid line) to observed data (dots) for TIC\,456905229.}\label{spec1}
\end{figure*}

After the binary variations were removed from the TESS data of all targets, we carried out a frequency analysis of the residuals in the second step of the analysis. The independent, harmonic and combination frequencies were searched for.
%Even if the binary variations were cleaned from the light curves, we still controlled whether there are any harmonic orbital frequencies.
During the analysis, the frequencies having a signal-to-noise (S/N) ratio over 4.5 were expected as significant. A typical significance limit for the detected frequencies is given as 4.0 by \cite{1993A&A...271..482B}. However, \cite{2021AcA....71..113B} showed that this limit should be higher for TESS data and by taking into account the results obtained in their study, we took the significance limit as 4.5. The analysis was carried out for a range of $\sim$\,$4-80$\,$d^{-1}$ considering the typical pulsation period of \ds\, stars (see, Sect.\,\ref{sec:int}). One object in our targets (TIC\,396201681) clearly show long-term $\gamma$\,Doradus-type variation in its light curve. Therefore for this system the frequency analysis was performed for $\sim$\,$0-80$\,$d^{-1}$ range of frequency, as $\gamma$\,Doradus stars typically exhibit pulsations with a frequency changing from $\sim$\,0.3 to 3 $d^{-1}$ \citep{2010aste.book.....A}. Consequently, the range and the number of the detected frequencies are listed in Table\,\ref{table2}. The first five highest amplitude frequencies are also given in Table\,\ref{tableap} for each target. The full table is given in electronic form. The frequency spectrum and the fits of the calculated frequencies to the observations are shown in Fig.\,\ref{spec1} for one sample and in Fig.\,\ref{spec2} for the others.

%%%%%%%%%%%%%%%%%%%%%%%%%%%%%%%%%%%%%%%%%%%%%%%%%%%%%%%%%%%%%%%%%%%%%%%%%%%%%%%%%%
\section{CALCULATING PHYSICAL PARAMETERS} \label{sec:fun}

For all systems, we calculated some physical parameters such as luminosity ($L$), absolute ($M_{V}$) and bolometric ($M_{bol}$) magnitudes. However, one should keep in mind that these parameters represent the binary system and do not belong to only one binary component. Before starting to calculate these parameters, we estimated flux densities of binary components to estimate the flux contributions of each component to the total. It is known that the area ratio of eclipses approximately gives the flux density ratio of the binary components \citep{1960pdss.book.....B}. In the primary eclipse (deeper one) the loss of light is more than the secondary eclipse, so in the primary eclipse, the star with the higher surface luminosity hence hotter is obscured by other. We call this star as primary (p) and the other cool and less luminous one as secondary (s) component. The area of primary ($A_{p}$) and secondary ($A_{s}$) eclipses were measured by using IRAF\footnote{http://iraf.noao.edu/}\citep{1986SPIE..627..733T} splot task. The ratio of these areas is equal to flux density ($I$) ratio as shown in the following equation \citep{1960pdss.book.....B}:

\begin{equation}
\frac{A_{p}}{A_{s}}\,=\,\frac{I_{p}}{I_{s}}
\end{equation}
\noindent
The calculated $I$ ratios are listed in Table\,\ref{table2}. For some systems, there is only one eclipse available in the light curves, therefore flux ratio value could not be determined for these systems. As can be seen from the $I$ ratios, in most systems $I$ of the primary and secondary components close each other. We assumed that if the $I$ ratio is over $\sim$4 the calculated physical parameters mostly belong to the primary component. 

The physical parameters of all systems were calculated using the distance module and the Pogson equation. To compute these parameters, the distances of the systems were taken from the Gaia EDR3 \citep{2021A&A...649A...1G} and also extinction coefficient (A$_{\upsilon}$) was calculated utilizing the interstellar extinction maps of \cite{2005AJ....130..659A}. First the $M_{V}$ parameters were calculated by using the following equation:

\begin{equation}
V\,-\,M_{V} = 5logd\,-\,5\,+\,A_{\upsilon}
\end{equation}
\noindent
where $V$ is the visual magnitude, d is the distance of the systems. After the $M_{V}$ values were derived the $M_{bol}$ parameters were computed taking into account the TIC $T_{\rm eff}$ values and bolometric corrections from \cite{2020MNRAS.496.3887E}. If there is no \teff\, value for a system we estimated this value from the target's spectral type by using the calibration between the spectral type and \teff\, given by \cite{1992oasp.book.....G}. Additionally, for TIC\,13037534, \teff\, determined from the spectral type-\teff\, calibration was used. Then, the $L$ parameters were computed with the below equation: 
\begin{equation}
\frac{L}{L_{\odot}}\,=\,10^{(M_{bol\odot}\,-\,M_{bol})/2.5}
\end{equation}
\noindent
$M_{bol{\odot}}$ value is taken as 4$^{m}$.74 according to IAU 2015 General Assembly resolution B2\footnote{$https://www.iau.org/static/resolutions/IAU2015_English.pdf$}. The calculated parameters are given in Table\,\ref{table2}. Uncertainties of the computed parameters were estimated considering the errors in the input parameters. According to the $I$ ratio of the primary and secondary components, these calculated parameters mostly represents the binary systems. However, there are some systems in which the given $I$ ratio is over $\sim$4. In these systems the flux coming from the primary is significantly higher than the secondary and we assumed the calculated physical parameters mostly represents the hotter primary components and probably the pulsating one. 

\begin{table*}
\centering
\footnotesize
\caption{Calculated Physical Parameters, $I$ Ratios of the Binary Components and the Range and the Number of the Detected Frequency. For TIC\,430808126 Parameters could not be Calculated as It has no Parallax.}\label{table2}
\begin{tabular*}{1\linewidth}{@{\extracolsep{\fill}}cccccclc@{}}
\hline
 TIC     &   A$_{\upsilon}$                 &   $M_{V}$      & $M_{bol}$    & log (L/L$_{\odot}$)  & $I_{p}$/$I_{s}$ & Frequency        &     Number of \\  
number   &  (mag)\,$\pm$\,0.002             &  (mag)         &  (mag)       &                      &                 & range ($d^{-1}$) & frequency \\ 
 \hline
  8669966 & 0.031 & 0.716\,$\pm$\,0.025 & 0.767\,$\pm$\,0.025 & 1.610\,$\pm$\,0.045 & 1.23 &13.103\,$-$\,14.21 & 6\\
 10057647 & 0.192 & 2.164\,$\pm$\,0.032 & 2.237\,$\pm$\,0.032 & 1.030\,$\pm$\,0.052 & 1.40 &13.77\,$-$\,51.41 & 23\\
13037534 & 0.720 & 0.571\,$\pm$\,0.025 & 0.588\,$\pm$\,0.024 & 1.668\,$\pm$\,0.045 & 6.64 & 8.84\,$-$\,12.11 & 8\\
14948284 & 0.069 & 1.807\,$\pm$\,0.024 & 1.883\,$\pm$\,0.029 & 1.173\,$\pm$\,0.044 & 1.20 & 8.09\,$-$\,24.92 & 21   \\
48084398 & 0.048 & 0.555\,$\pm$\,0.029 & 0.635\,$\pm$\,0.032 & 1.674\,$\pm$\,0.049 & 1.04 &8.98\,$-$\,13.22 & 21\\
71613490 & 0.029 & 1.395\,$\pm$\,0.032 & 1.441\,$\pm$\,0.007 & 1.338\,$\pm$\,0.052 & 1.04 &24.47\,$-$\,32.34 & 19\\
72839144 & 0.077 & 1.416\,$\pm$\,0.007 & 1.492\,$\pm$\,0.028 & 1.329\,$\pm$\,0.027 & 3.55 &4.11 & 1\\
75593781 & 0.708 & 1.364\,$\pm$\,0.029 & 1.355\,$\pm$\,0.016 & 1.350\,$\pm$\,0.049 & 4.47 &46.11\,$-$\,57.94 & 15\\
 78148497 & 0.663 & 1.324\,$\pm$\,0.016 & 1.395\,$\pm$\,0.032 & 1.367\,$\pm$\,0.036 & 2.80 &9.77\,$-$\,32.97 & 22\\
 85600400 & 0.087 & 1.651\,$\pm$\,0.032 & 1.724\,$\pm$\,0.027 & 1.236\,$\pm$\,0.052 & 3.85 &9.70\,$-$\,19.97 & 17\\
 116334565 & 0.809 & 0.849\,$\pm$\,0.027 & 0.908\,$\pm$\,0.031 & 1.557\,$\pm$\,0.047 & 1.27 &6.55\,$-$\,22.85 & 12\\ 
 165618747 & 0.298 & 2.107\,$\pm$\,0.031 & 2.162\,$\pm$\,0.011 & 1.053\,$\pm$\,0.051 & 1.97 &15.13\,$-$\,40.18 & 28 \\ 
 172431974 & 0.047 & 1.557\,$\pm$\,0.011 & 1.631\,$\pm$\,0.032 & 1.273\,$\pm$\,0.031 & 1.01 &4.24\,$-$\,6.40 & 10\\
193774939 & 0.080 & 2.220\,$\pm$\,0.033 & 4.974\,$\pm$\,0.019 & 1.008\,$\pm$\,0.053 & 5.16 &16.59\,$-$\,31.21 & 21\\
197755658 & 0.029 & 2.074\,$\pm$\,0.019 & 2.126\,$\pm$\,0.013 & 1.066\,$\pm$\,0.039 & 1.94 &46.63\,$-$\,51.26 & 3\\
197757000 & 0.123 & 2.576\,$\pm$\,0.013 & 2.657\,$\pm$\,0.009 & 0.865\,$\pm$\,0.033 & 4.46 &21.17\,$-$\,41.98 & 21\\
 240962482 & 0.190 & 1.689\,$\pm$\,0.010 & 1.733\,$\pm$\,0.007 & 1.221\,$\pm$\,0.030 & 11.84 &24.96\,$-$\,29.43 & 5\\
 241013310 & 0.110 & 2.335\,$\pm$\,0.007 & 2.414\,$\pm$\,0.033 & 0.962\,$\pm$\,0.027 & 4.02 &4.82\,$-$\,14.77 & 4    \\
256640561 & 0.079 & 0.823\,$\pm$\,0.033 & 0.855\,$\pm$\,0.007 & 1.567\,$\pm$\,0.053 & 1.86 &10.12\,$-$\,24.42 & 12\\
 272822330 & 0.165 & 0.997\,$\pm$\,0.007 & 1.066\,$\pm$\,0.030 & 1.497\,$\pm$\,0.027 & 1.91 &14.03\,$-$\,20.34 & 11\\
289947843 & 0.083 & 0.912\,$\pm$\,0.030 & 0.791\,$\pm$\,0.012 & 1.531\,$\pm$\,0.050 & $-$ &5.97\,$-$\,14.44 & 11\\
 301909087 & 0.624 & 1.137\,$\pm$\,0.012 & 1.218\,$\pm$\,0.026 & 1.441\,$\pm$\,0.032 & 1.04 &5.58\,$-$\,17.44 & 10\\
 305633328 & 0.245 & 1.680\,$\pm$\,0.026 & 1.760\,$\pm$\,0.014 & 1.224\,$\pm$\,0.046 & 1.08 &9.38\,$-$\,15.37 & 8\\
 322428763 & 0.174 & 1.441\,$\pm$\,0.014 & 1.517\,$\pm$\,0.022 & 1.483\,$\pm$\,0.014 & 1.53 &11.99\,$-$\,49.78 & 11\\ 
 327121759 & 0.726 & 1.374\,$\pm$\,0.022 & 1.451\,$\pm$\,0.011 & 1.346\,$\pm$\,0.042 & 3.43 &22.90\,$-$\,30.97 & 6\\ 
 337094559 & 0.097 & 2.049\,$\pm$\,0.011 & 2.114\,$\pm$\,0.061 & 1.077\,$\pm$\,0.031 & 1.78 &7.86\,$-$\,19.47 & 8\\
 338159479 & 0.292 & 2.332\,$\pm$\,0.062 & 2.383\,$\pm$\,0.011 & 0.963\,$\pm$\,0.081 & 2.01 &21.75\,$-$\,28.51 & 17\\ 
 354926863 & 0.298 & 2.755\,$\pm$\,0.011 & 2.822\,$\pm$\,0.007 & 0.794\,$\pm$\,0.031 &  1.06 &4.66\,$-$\,39.11 & 18\\ 
 358613523 & 0.097 & 2.813\,$\pm$\,0.007 & 2.885\,$\pm$\,0.034 & 0.771\,$\pm$\,0.027 & 5.16 &24.59\,$-$\,46.44 & 15\\
393894013 & 0.044 & 2.216\,$\pm$\,0.035 & 2.250\,$\pm$\,0.023 & 1.010\,$\pm$\,0.055 & 1.49 &8.55\,$-$\,69.65 & 19\\
396134795 & 0.628 & 1.290\,$\pm$\,0.023 & 1.371\,$\pm$\,0.038 & 1.380\,$\pm$\,0.043 & 1.63 &14.20\,$-$\,28.47 & 7  \\
396201681 & 0.184 & 1.595\,$\pm$\,0.038 & 1.638\,$\pm$\,0.071 & 1.258\,$\pm$\,0.058 & 1.14 &1.21\,$-$\,23.52 & 11\\
420114772 & 0.127 & 0.656\,$\pm$\,0.071 & 0.736\,$\pm$\,0.016 & 1.634\,$\pm$\,0.091 & 1.05 &5.02\,$-$\,41.95 & 30\\
421714420 & 0.097 & -0.083\,$\pm$\,0.016 & -0.154\,$\pm$\,0.014 & 1.929\,$\pm$\,0.036 & 10.00& 4.01\,$-$\,53.19 & 11\\
 428003183 & 0.106 & 2.438\,$\pm$\,0.014  & 2.506\,$\pm$\,0.015 &  0.921\,$\pm$\,0.034 & 2.56 &12.93\,$-$\,32.77 & 22\\ 
430808126 & $-$ & $-$ & $-$ & $-$ & 1.07 &5.26\,$-$\,27.63 & 22\\ 
440003271 & 0.030 & 1.524\,$\pm$\,0.026 & 1.510\,$\pm$\,0.026 & 1.287\,$\pm$\,0.046 & 1.10&12.19\,$-$\,37.33 & 19\\
456905229 & 0.048 & 2.060\,$\pm$\,0.074 & 2.139\,$\pm$\,0.074 & 1.072\,$\pm$\,0.094 & 1.12 &10.13\,$-$\,33.39 & 44\\
467354611 & 0.094 & 1.825\,$\pm$\,0.023 & 1.166\,$\pm$\,0.043 & 1.876\,$\pm$\,0.027 & $-$ &11.58\,$-$\,31.04 & 41\\
 \hline
 \bottomrule
\end{tabular*}
\end{table*}

%%%%%%%%%%%%%%%%%%%%%%%%%%%%%%%%%%%%%%%%%%%%%%%%%%%%%%%%%%%%%%%%%%%%%%%%%%%%%%
\section{DISCUSSION} \label{sec:dis}
In this section, we examined some properties of our candidate DSEB. 

\subsection{Pulsation type}
In this study, we present the analysis of some targets showing \ds\, like variations. While determining the candidate DSEB, one of the important criteria was the \teff\, of the eclipsing binary systems. However, we know that these \teff\, values are an average of both binary components. So real \teff\, of pulsating components, could be higher or cooler than the TIC \teff. In some systems, we found that the primary components have significantly more flux density compared to secondary (see, Table\,\ref{table2}). In these systems\footnote{TIC\,13037534, TIC\,75593781, TIC\,85600400, TIC\,193774939, TIC\,197757000, TIC\,240962482, TIC\,241013310, TIC\,358613523, TIC\,421714420}, the \teff\, mostly represents the hotter primary and probably the pulsating components. When we examined the \teff\, values \footnote{for TIC\,13037534, the spectral type was taken into account} of these systems, we found that their \teff\, are in the range of the \teff\, given for \ds\, stars. Their pulsation amplitudes and frequencies are also consistent with the values given for \ds\, stars \citep{2010aste.book.....A}.
%Therefore, these systems could consider as real DSEBs. 

For the other systems which have the $I$ ratio lover than 4, probably the TIC \teff\, values are substantially different than the real \teff\, values of the pulsating components. Inside the pulsating stars, there are two different types that exhibit frequencies like \ds\, variables. One of them is $\beta$\,Cephei stars. The $\beta$\,Cephei stars mostly show frequencies between 3 and 12\,$d^{-1}$ and these pulsators have $B0-B3$ spectral type \citep{2010aste.book.....A}. Although these pulsators are quite hotter than the \ds\, variables, if a $\beta$\,Cephei star have a very cool binary component, the total \teff\, of the system will be cooler than the value expected for the $\beta$\,Cephei stars. However, our targets have a \teff\, value in a range of $\sim$\,$6300-9100$\,K and even if a $\beta$\,Cephei star has a cool binary component, the average \teff\, of the binary system could not as low as our targets' \teff\, range. 

Another pulsating star group is Maia variables. The existence of Maia variables has not been exactly confirmed, however, for decades Maia variables are considered as a new group of pulsating stars \citep{2005A&A...431..615A, 2016MNRAS.460.1318B, 2020MNRAS.493.5871B}. The Maia variables are located between the $\beta$\,Cephei and \ds\, stars, so they are cooler than $\beta$\,Cephei variables and hotter than the \ds\, stars. Additionally, Maia variables demonstrate oscillations approximately in a similar frequency range with the \ds\,stars \citep{2016MNRAS.460.1318B, 2020MNRAS.493.5871B}. Even the existence of Maia variables has not been approved, if they are a new type of variables, they could be a member of binary systems and seen cooler than expected if a Maia variable has a cooler binary component. In this case, they could be considered as a \ds\, variable. To have an idea about the variability type of our targets, their positions in the Hertzsprung-Russell (H-R) diagram should be examined by considering the $I$ ratios.  Therefore, we showed the positions of our systems in the H-R diagram by using the parameters given in Table\,\ref{tab1} and Table\,\ref{table2}.

\begin{figure}
\centering
\includegraphics[width=\textwidth]{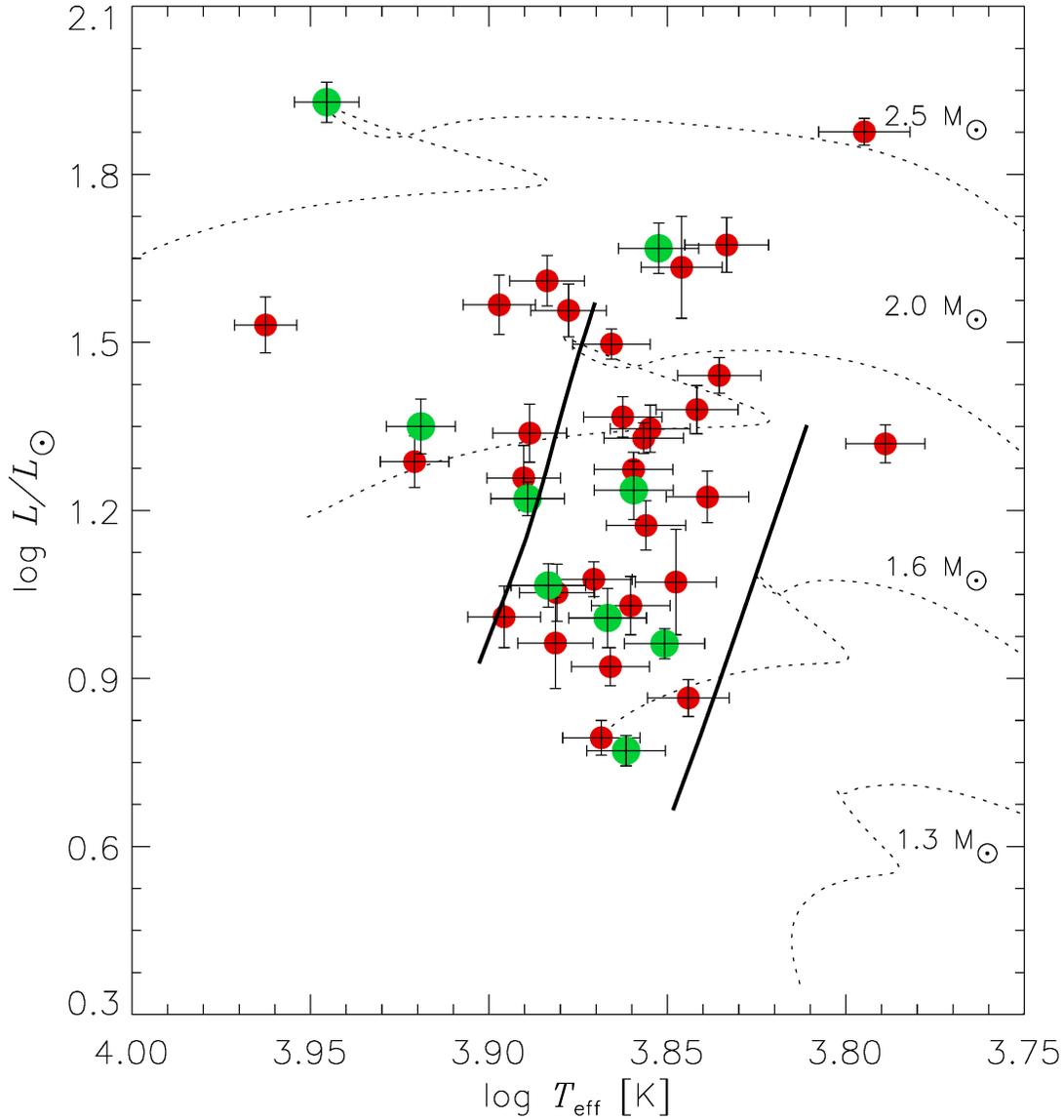}
\caption{Positions of the targets in the H-R diagram. Green dots represent the systems having $I$ ratio higher than $\sim$\,4, while red smaller dots shows the other systems. The solid and dashed lines are the borders of the \ds\, instability strip \citep{2005A&A...435..927D} and the evolutionary tracks taken from the MESA Isochrones and Stellar Tracks (MIST) \citep{2011ApJS..192....3P, 2016ApJS..222....8D, 2016ApJ...823..102C}, respectively.}\label{hr}
\end{figure}

As can be seen from Fig.\,\ref{hr}, most of the systems, which have an $I$ ratio over around 4, are located in the \ds\, instability strip and for these systems, we assumed the $L$ and \teff\, mostly represent the primary, probably the pulsating, binary component. Inside these systems, there is one object (TIC\,421714420) that is placed beyond the hot border of \ds\, instability strip. In a detailed study of \textit{Kepler} field \ds\, and related stars, \cite{2011A&A...534A.125U} showed that there are some real \ds\, variables located beyond the borders of the \ds\, instability strip. However, this system is noticeable far from the \ds\, instability strip, and very close to the place where Maia variables are located \citep{2020MNRAS.493.5871B}. Therefore we classified this system as a candidate of Maia variable in an eclipsing binary. There is another system located beyond the hotter border of \ds\, stars, TIC\,289947843, unfortunately, we could not measure the $I$ ratio for this system because there is not enough data. For the other systems having $I$ ratio $<$4 the $L$ value of the pulsating component should be lower than the calculated one and depending on \teff\, of the other binary component the \teff\, of the pulsating star could be lower or higher than the used in the H-R diagram. Taking into account these conditions, we could say that most pulsating components could be located inside the \ds\, instability strip however it is difficult to have an idea about the variability of the systems considering this. Therefore, as a result of this examination, we classified 38 systems as candidate DSEB and one of them, TIC\,421714420, as a Maia candidate in an eclipsing binary system.

\subsection{Consistency with the known relationship for DSEBs}

\begin{figure}[hbt!]
\centering
\includegraphics[width=\textwidth]{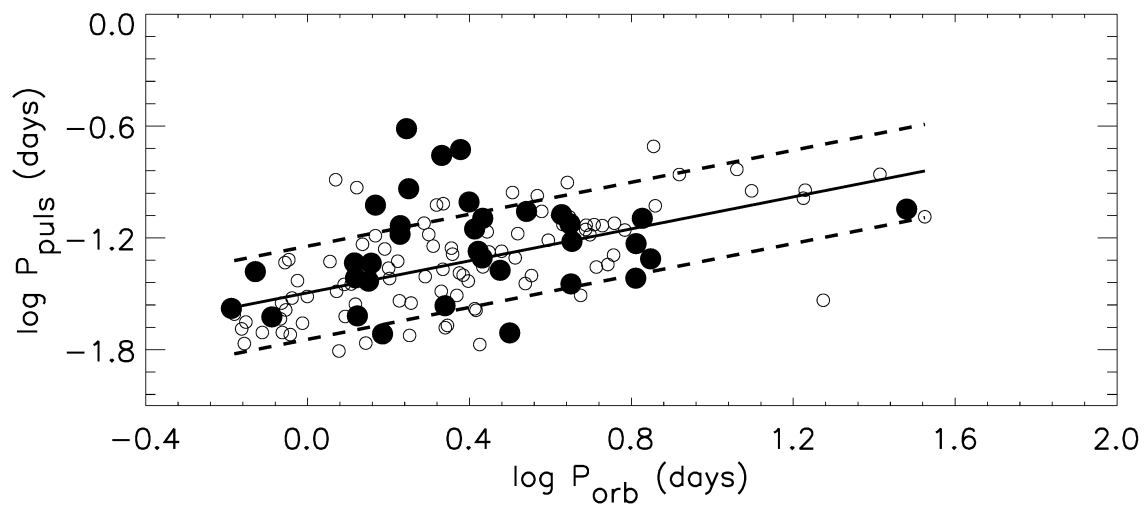}
\caption{Consistency of the candidate DSEB with P$_{orb}$\,$-$\,P$_{puls}$ relationship. The filled and empty circles represent our candidate and the known DSEB \citep{2017MNRAS.470..915K}, respectively. The solid and dashed lines show the correlation and 1-$\sigma$ level, respectively.}\label{porbrel}
\end{figure}

DSEB have been investigated for a long time and it was shown that there are some relationships between the pulsation period, amplitude and the other parameters such as $P_{orb}$, $R$ and $\log g$ \citep{2017MNRAS.470..915K, 2017MNRAS.465.1181L}. For DSEB, the well-known relationship is between the P$_{orb}$ and P$_{puls}$. In the latest study of \cite{2017MNRAS.470..915K}, it was shown that known DSEB obey this relationship within error bars. We examined whether our candidate systems suit this relationship. In this section, we only investigate the candidate DSEB. As can be seen from Fig.\ref{porbrel}, most of our candidates are consistent with the relationship within errors. There are a few objects placed outside of the 1-$\sigma$ level, TIC\,72839144, TIC\,172431974, TIC\,197755658 and TIC\,241013310. The reason for these could be the additional effect in the binary system such as mass transfer between binary components if these systems are DSEB.  

More relationships between the P$_{puls}$ and other parameters are given by \cite{2017MNRAS.470..915K} for detached and semi-detached systems. These parameters help us approximately estimate some physical parameters of the \ds\, pulsating components in eclipsing binary systems. As we do not know the Roche geometry of our systems we could not classify the binary configurations of our targets. However, for both detached and semi-detached binary configurations, some relationships give a good correlation between the P$_{puls}$ such as \logg\, and radius ($R$) of the pulsating component. Therefore, we calculated \logg\, and $R$ parameters of our systems using the equations given in the study of \cite{2017MNRAS.470..915K}. In this calculation, we excluded the systems found outside of the  P$_{orb}$\,$-$\,P$_{puls}$ relationship. The computed \logg\, and $R$ parameters are listed in Table\,\ref{tablecalculated}. By utilizing the calculated \logg\, and $R$, we also estimated the mass ($M$) values of the pulsating component. These values are also given in Table\,\ref{tablecalculated}. It should keep in mind that these parameters are just an estimation and do not give exact values and the real errors should be higher.  

\begin{table}
\centering
\footnotesize
\caption{Calculated Parameters from the Given Relationships by \cite{2017MNRAS.470..915K}.}\label{tablecalculated}
\begin{tabular}{cccc}
\hline
 TIC     &   \logg\,           &   $R$                  & $M$     \\  
number   &  (cgs)              &  ($R_{\odot}$)         &  ($M_{\odot}$)       \\ 
 \hline
  8669966  & 3.65\,$\pm$\,0.20  & 3.53\,$\pm$\,0.39 & 2.07\,$\pm$\,0.30 \\
 10057647  & 4.41\,$\pm$\,0.24  & 1.39\,$\pm$\,0.24 & 1.85\,$\pm$\,0.38  \\
 13037534  & 3.51\,$\pm$\,0.28  & 3.93\,$\pm$\,0.51 & 1.85\,$\pm$\,0.40 \\
 14948284  & 3.93\,$\pm$\,0.10  & 2.74\,$\pm$\,0.32 & 2.38\,$\pm$\,0.10 \\
 48084398  & 3.56\,$\pm$\,0.26  & 3.80\,$\pm$\,0.47 & 1.93\,$\pm$\,0.36 \\
 71613490  & 4.10\,$\pm$\,0.11  & 2.25\,$\pm$\,0.05 & 2.39\,$\pm$\,0.06 \\
 75593781  & 4.59\,$\pm$\,0.34  & 0.88\,$\pm$\,0.39 & 1.13\,$\pm$\,0.36 \\
 78148497  & 3.59\,$\pm$\,0.24  & 3.70\,$\pm$\,0.44 & 1.98\,$\pm$\,0.34 \\
 116334565 & 3.53\,$\pm$\,0.28  & 3.87\,$\pm$\,0.49 & 1.88\,$\pm$\,0.38 \\ 
 165618747 & 4.37\,$\pm$\,0.22  & 1.52\,$\pm$\,0.20 & 1.97\,$\pm$\,0.34 \\ 
 193774939 & 3.97\,$\pm$\,0.12  & 2.62\,$\pm$\,0.12 & 2.40\,$\pm$\,0.19 \\
 197757000 & 4.35\,$\pm$\,0.20  & 1.57\,$\pm$\,0.18 & 2.03\,$\pm$\,0.30 \\ 
 240962482 & 4.16\,$\pm$\,0.14  & 2.10\,$\pm$\,0.03 & 2.36\,$\pm$\,0.09 \\
 256640561 & 3.44\,$\pm$\,0.32  & 4.11\,$\pm$\,0.56 & 1.75\,$\pm$\,0.44 \\
 272822330 & 3.79\,$\pm$\,0.12  & 3.14\,$\pm$\,0.28 & 2.25\,$\pm$\,0.20 \\
 289947843 & 3.69\,$\pm$\,0.18  & 3.41\,$\pm$\,0.36 & 2.13\,$\pm$\,0.28 \\
 301909087 & 3.64\,$\pm$\,0.22  & 3.57\,$\pm$\,0.40 & 2.05\,$\pm$\,0.32 \\
 305633328 & 3.45\,$\pm$\,0.32  & 4.10\,$\pm$\,0.56 & 1.75\,$\pm$\,0.44\\ 
 322428763 & 3.59\,$\pm$\,0.24  & 3.70\,$\pm$\,0.44 & 1.98\,$\pm$\,0.34 \\ 
 327121759 & 4.04\,$\pm$\,0.10  & 2.44\,$\pm$\,0.17 & 2.41\,$\pm$\,0.22 \\ 
 338159479 & 4.13\,$\pm$\,0.10  & 2.17\,$\pm$\,0.01 & 2.38\,$\pm$\,0.09 \\ 
 354926863 & 3.98\,$\pm$\,0.10  & 2.61\,$\pm$\,0.12 & 2.40\,$\pm$\,0.17 \\ 
 358613523 & 4.43\,$\pm$\,0.26  & 1.32\,$\pm$\,0.26 & 1.77\,$\pm$\,0.42 \\
 393894013 & 4.44\,$\pm$\,0.26  & 1.30\,$\pm$\,0.26 & 1.74\,$\pm$\,0.44 \\
 396134795 & 3.68\,$\pm$\,0.10  & 3.44\,$\pm$\,0.37 & 2.11\,$\pm$\,0.28 \\
 396201681 & 3.94\,$\pm$\,0.10  & 2.72\,$\pm$\,0.15 & 2.38\,$\pm$\,0.17  \\
 420114772 & 4.11\,$\pm$\,0.10  & 2.24\,$\pm$\,0.05 & 2.39\,$\pm$\,0.07  \\
 421714420 & 4.27\,$\pm$\,0.16  & 1.78\,$\pm$\,0.24 & 2.20\,$\pm$\,0.22   \\
 428003183 & 4.05\,$\pm$\,0.10  & 2.40\,$\pm$\,0.06 & 2.41\,$\pm$\,0.11 \\ 
 430808126 & 3.81\,$\pm$\,0.06  & 3.09\,$\pm$\,0.26 & 2.27\,$\pm$\,0.18\\ 
 440003271 & 3.87\,$\pm$\,0.14  & 2.90\,$\pm$\,0.21 & 2.34\,$\pm$\,0.14 \\
 456905229 & 3.73\,$\pm$\,0.16  & 3.31\,$\pm$\,0.33 & 2.18\,$\pm$\,0.24 \\
 467354611 & 3.71\,$\pm$\,0.16  & 3.35\,$\pm$\,0.34 & 2.16\,$\pm$\,0.26 \\
 \hline
 \bottomrule
\end{tabular}
\end{table}

%%%%%%%%%%%%%%%%%%%%%%%%%%%%%%%%%%%%%%%%%%%%%%%%%%%%%%%%%%%%%%%%%%%%%%%%%%
\section{CONCLUSIONS}

In this study, we present the results of our northern TESS field search to discover new eclipsing binaries with \ds\, components. We first determined 39 targets and examined the pulsational properties (pulsation amplitude and frequencies) of these systems after removing the eclipsing variations. In addition to determining pulsation amplitude and frequencies, we also estimated the $I$ ratios of binary components to find how much binary components contribute to total flux relative to each other. To estimate whether our systems could be DSEB or not, we also controlled the positions of the targets in the H-R diagram. For this, we calculated $L$ parameters of the systems. By considering the positions of the systems in the H-R diagram and the $I$ ratios, we showed that one of our targets (TIC\,421714420) could be a candidate of Maia variable in an eclipsing binary system. The other targets in the study are classified as candidate DSEB. However, to be sure about the positions of the systems in the H-R diagram and the real \teff\, values of them, detailed spectroscopic analysis and binary modelling are necessary. With the spectroscopy the \teff\, value of each binary component can be derived and with the binary modelling, the real $L$ parameters could be reached. 

We know that the pulsating eclipsing binary systems are quite important to deeply understand stellar systems. An increasing number of these kinds of systems would contribute to improving our knowledge about the stellar evolution and structure. Therefore, this study would be useful for both probing stellar structure, evolution and understanding the pulsation behaviour of oscillating stars in eclipsing binary systems. 

%%%%%%%%%%%%%%%%%%%%%%%%%%%%%%%%%%%%%%%%%%%%%%%%%%%%%%%%%%%%%%%%%%%%%%%%%%

\begin{acknowledgements}
This work has been supported in part by the Scientific and Technological Research Council (TUBITAK) under the grant number 120F330. The TESS data presented in this paper were obtained from the Mikulski Archive for Space Telescopes (MAST). FKA thanks Prof. Gerald HANDLER for showing how to clean binarity with a phenomenological fit. Funding for the TESS mission is provided by the NASA Explorer Program. This work has made use of data from the European Space Agency (ESA) mission Gaia (http://www.cosmos.esa.int/gaia), processed by the Gaia Data Processing and Analysis Consortium (DPAC, http://www.cosmos.esa.int/web/gaia/dpac/consortium). Funding for the DPAC has been provided by national institutions, in particular the institutions participating in the Gaia Multilateral Agreement. This research has made use of the SIMBAD data base, operated at CDS, Strasbourq, France.
\end{acknowledgements}

%%%%%%%%%%%%%%%%%%%%%%%%%%%%%%%%%%%%%%%%%%%%%%%%%%%%%%%%%%%%%%%%%%%%%%%%%%%%%%

%%%%%%%%%%%%%%%%%%%%%%%%%%%%%%%%%%%%%%%%%%%%%%%%%%%%%%%%%%%%%%%%%%%%%%%%%

\appendix                  %%appendicial material is supported

\setcounter{table}{3}
\begin{landscape}

\begin{table}
 \begin{scriptsize}
\label{tab:apt}
\caption{The List of the First Five Highest Amplitude Frequency Detected in This Study. The Full List of the Frequencies is Given in Electronic Form.}\label{tableap}
\begin{tabular}{llccc|llccc|llccc} \toprule
\hline\noalign{\smallskip}
   & Frequency  &  Amplitude  & Phase  &S/N     &    &Frequency  &  Amplitude   & Phase  &S/N & &Frequency  &  Amplitude   & Phase  &S/N\\
   & (d$^{-1}$) & (mmag)      & (rad)  &        &    &(d$^{-1}$) & (mmag)       & (rad)  &    & &(d$^{-1}$) & (mmag)       & (rad)  &    \\
   \hline\noalign{\smallskip}
\multicolumn{5}{c}{\hrulefill TIC\,8669966\hrulefill} & \multicolumn{5}{c}{\hrulefill TIC\,10057647\hrulefill} & \multicolumn{5}{c}{\hrulefill TIC\,13037534\hrulefill} \\  
   \hline\noalign{\smallskip}
$\nu_1$  &13.62189\,$\pm$\,0.00015 &0.717\,$\pm$\,0.005 &0.5562\,$\pm$\,0.0011  & 62 & $\nu_1$ &40.11281\,$\pm$\,0.00019 &0.727\,$\pm$\,0.007  &0.2288\,$\pm$\,0.0015  & 54 & $\nu_1$  &11.10889\,$\pm$\,0.00004 &0.526\,$\pm$\,0.012 &0.8332\,$\pm$\,0.0036  & 20\\
$\nu_2$  &13.14363\,$\pm$\,0.00017 &0.618\,$\pm$\,0.005 &0.0358\,$\pm$\,0.0013  & 61 & $\nu_2$ &30.95924\,$\pm$\,0.00020 &0.686\,$\pm$\,0.007  &0.1324\,$\pm$\,0.0015  & 69 & $\nu_2$  &8.84272\,$\pm$\,0.00006  &0.459\,$\pm$\,0.012 &0.3634\,$\pm$\,0.0048  & 24 \\
$\nu_3$  &13.32788\,$\pm$\,0.00038 &0.278\,$\pm$\,0.005 &0.0784\,$\pm$\,0.0028  & 26 & $\nu_3$ &38.78309\,$\pm$\,0.00028 &0.481\,$\pm$\,0.007  &0.8545\,$\pm$\,0.0022  & 31 & $\nu_3$  &9.74327\,$\pm$\,0.00006  &0.323\,$\pm$\,0.012 &0.4381\,$\pm$\,0.0055  & 13\\
$\nu_4$  &13.03191\,$\pm$\,0.00042 &0.251\,$\pm$\,0.005 &0.0809\,$\pm$\,0.0031  & 25 & $\nu_4$ &43.60348\,$\pm$\,0.00034 &0.399\,$\pm$\,0.007  &0.4591\,$\pm$\,0.0027  & 28 & $\nu_4$  &10.63240\,$\pm$\,0.00015 &0.210\,$\pm$\,0.012 &0.2293\,$\pm$\,0.0045  & 8 \\
$\nu_5$  &13.73165\,$\pm$\,0.00049 &0.217\,$\pm$\,0.005 &0.5787\,$\pm$\,0.0036  & 19 & $\nu_5$ &36.01630\,$\pm$\,0.00048 &0.283\,$\pm$\,0.007  &0.6612\,$\pm$\,0.0037  & 24 & $\nu_5$  &10.41897\,$\pm$\,0.00026 &0.207\,$\pm$\,0.019 &0.8755\,$\pm$\,0.0495  & 8 \\
\hline
\multicolumn{4}{c}{\hrulefill TIC\,14948284\hrulefill} & \multicolumn{5}{c}{\hrulefill TIC\,48084398  \hrulefill} & \multicolumn{5}{c}{\hrulefill TIC\,71613490\hrulefill} &  \\  
   \hline\noalign{\smallskip}
$\nu_1$ &20.31697\,$\pm$\,0.00019 &3.130\,$\pm$\,0.028  &0.7751\,$\pm$\,0.0014  & 22 & $\nu_1$  &11.88739\,$\pm$\,0.00006 &1.099\,$\pm$\,0.006 &0.5054\,$\pm$\,0.0008  & 30 & $\nu_1$ &25.98838\,$\pm$\,0.00049 &0.323\,$\pm$\,0.007  &0.9887\,$\pm$\,0.0035  & 24 \\
$\nu_2$ &11.20278\,$\pm$\,0.00021 &2.776\,$\pm$\,0.028  &0.1298\,$\pm$\,0.0016  & 20 & $\nu_2$  &9.49467\,$\pm$\,0.00006  &1.009\,$\pm$\,0.006 &0.9585\,$\pm$\,0.0009  & 24 & $\nu_2$ &31.57531\,$\pm$\,0.00054 &0.289\,$\pm$\,0.007  &0.8029\,$\pm$\,0.0039  & 26 \\
$\nu_3$ &20.11472\,$\pm$\,0.00038 &1.567\,$\pm$\,0.028  &0.5509\,$\pm$\,0.0029  & 11 & $\nu_3$  &11.41491\,$\pm$\,0.00007 &0.861\,$\pm$\,0.006 &0.9264\,$\pm$\,0.0010  & 25  & $\nu_3$ &30.81604\,$\pm$\,0.00057 &0.274\,$\pm$\,0.007  &0.2540\,$\pm$\,0.0041  & 23 \\
$\nu_4$  &18.05449\,$\pm$\,0.00038 &1.541\,$\pm$\,0.028  &0.0802\,$\pm$\,0.0029  & 16 & $\nu_2$+$\nu_3$-$\nu_1$&9.02219\,$\pm$\,0.00008 &0.784\,$\pm$\,0.006 &0.3972\,$\pm$\,0.0011&19&$\nu_4$&31.33778\,$\pm$\,0.00061&0.257\,$\pm$\,0.007&0.7478\,$\pm$\,0.0043&22 \\
$\nu_5$  &17.67646\,$\pm$\,0.00048 &1.247\,$\pm$\,0.028  &0.1414\,$\pm$\,0.0036  & 14 & $\nu_5$  &11.02949\,$\pm$\,0.00010 &0.632\,$\pm$\,0.006 &0.1951\,$\pm$\,0.0014  & 18 & $\nu_5$ &26.25839\,$\pm$\,0.00065 &0.242\,$\pm$\,0.007  &0.9569\,$\pm$\,0.0046  & 15 \\
\hline
\multicolumn{4}{c}{\hrulefill TIC\,72839144 \hrulefill} & \multicolumn{5}{c}{\hrulefill TIC\,75593781 \hrulefill} & \multicolumn{5}{c}{\hrulefill TIC\,78148497 \hrulefill} \\  
   \hline\noalign{\smallskip}
$\nu_1$  &4.11102\,$\pm$\,0.00133  &0.328\,$\pm$\,0.020 &0.8042\,$\pm$\,0.0098  & 6.0 & $\nu_1$ &51.91704\,$\pm$\,0.00011 &1.544\,$\pm$\,0.023  &0.1718\,$\pm$\,0.0024  & 46 & $\nu_1$  &12.44590\,$\pm$\,0.00009 &1.383\,$\pm$\,0.017 &0.3311\,$\pm$\,0.0020  & 39 \\
         &                         &                    &                       &     & $\nu_2$ &48.85966\,$\pm$\,0.00018 &0.914\,$\pm$\,0.023  &0.8996\,$\pm$\,0.0040  & 30 & $\nu_2$  &16.39778\,$\pm$\,0.00014 &0.900\,$\pm$\,0.017 &0.5090\,$\pm$\,0.0031  & 19\\
         &                         &                    &                       &     & $\nu_3$ &52.40034\,$\pm$\,0.00022 &0.754\,$\pm$\,0.023  &0.5367\,$\pm$\,0.0048  & 22 & $\nu_3$  &24.48400\,$\pm$\,0.00019 &0.662\,$\pm$\,0.017 &0.2477\,$\pm$\,0.0042  & 16 \\
         &                         &                    &                       &     & $\nu_4$ &47.75704\,$\pm$\,0.00028 &0.588\,$\pm$\,0.023  &0.1146\,$\pm$\,0.0062  & 20 & $\nu_4$  &23.67636\,$\pm$\,0.00018 &0.677\,$\pm$\,0.017 &0.9442\,$\pm$\,0.0041  & 16 \\
         &                         &                    &                       &     & $\nu_5$ &50.61103\,$\pm$\,0.00026 &0.622\,$\pm$\,0.023  &0.7971\,$\pm$\,0.0058  & 19 & $\nu_5$  &26.07247\,$\pm$\,0.00020 &0.616\,$\pm$\,0.017 &0.5684\,$\pm$\,0.0045  & 17 \\
\hline
\multicolumn{5}{c}{\hrulefill TIC\,85600400\hrulefill} & \multicolumn{5}{c}{\hrulefill TIC\,116334565\hrulefill} & \multicolumn{5}{c}{\hrulefill TIC\,165618747\hrulefill}\\  
   \hline\noalign{\smallskip}
$\nu_1$  &10.57549\,$\pm$\,0.00018 &1.666\,$\pm$\,0.027  &0.6049\,$\pm$\,0.0026  & 29 & $\nu_1$  &11.42838\,$\pm$\,0.00007 &2.212\,$\pm$\,0.117 &0.4961\,$\pm$\,0.0005  & 36 & $\nu_1$ &37.87282\,$\pm$\,0.00039 &0.765\,$\pm$\,0.025  &0.5868\,$\pm$\,0.0060  & 23 \\
$\nu_2$  &11.01141\,$\pm$\,0.00021 &1.386\,$\pm$\,0.027  &0.2475\,$\pm$\,0.0031  & 24 & $\nu_2$  &14.76193\,$\pm$\,0.00007 &1.360\,$\pm$\,0.117 &0.8032\,$\pm$\,0.0005  & 20 & $\nu_2$ &37.54370\,$\pm$\,0.00039 &0.675\,$\pm$\,0.025  &0.4108\,$\pm$\,0.0059  & 20 \\
$\nu_3$  &16.63944\,$\pm$\,0.00025 &1.171\,$\pm$\,0.027  &0.3368\,$\pm$\,0.0037  & 22 & $\nu_3$  &15.15892\,$\pm$\,0.00008 &0.1207\,$\pm$\,0.117 &0.7195\,$\pm$\,0.0006  & 17 & $\nu_3$ &35.11748\,$\pm$\,0.00060 &0.474\,$\pm$\,0.025  &0.1010\,$\pm$\,0.0090  & 14 \\
$\nu_4$  &11.39975\,$\pm$\,0.00025 &1.183\,$\pm$\,0.027  &0.8613\,$\pm$\,0.0037  & 20 & $\nu_4$  &7.97122\,$\pm$\,0.00010  &1.212\,$\pm$\,0.117 &0.7470\,$\pm$\,0.0007  & 17 & $\nu_4$ &23.54175\,$\pm$\,0.00056 &0.430\,$\pm$\,0.025  &0.9865\,$\pm$\,0.0084  & 14 \\
$\nu_5$  &17.99572\,$\pm$\,0.00030 &0.971\,$\pm$\,0.027  &0.3860\,$\pm$\,0.0045  & 22 & $\nu_5$  &6.54611\,$\pm$\,0.00012  &1.073\,$\pm$\,0.117 &0.7175\,$\pm$\,0.0008  & 13 & $\nu_5$ &30.49806\,$\pm$\,0.00069 &0.414\,$\pm$\,0.025  &0.6068\,$\pm$\,0.0103  & 12 \\
 \hline
\end{tabular}
 \end{scriptsize}
\end{table}
 \end{landscape}

\setcounter{table}{3}
\begin{landscape}

\begin{table}
 \begin{scriptsize}
\label{tab:apt}
\caption{Continuation.}\label{tableap}
\begin{tabular}{llccc|llccc|llccc} \toprule
\hline\noalign{\smallskip}
   & Frequency  &  Amplitude  & Phase  &S/N     &    &Frequency  &  Amplitude   & Phase  &S/N & &Frequency  &  Amplitude   & Phase  &S/N\\
   & (d$^{-1}$) & (mmag)      & (rad)  &        &    &(d$^{-1}$) & (mmag)       & (rad)  &    & &(d$^{-1}$) & (mmag)       & (rad)  &    \\
   \hline\noalign{\smallskip}
\multicolumn{5}{c}{\hrulefill TIC\,172431974\hrulefill} & \multicolumn{5}{c}{\hrulefill TIC\,193774939\hrulefill} & \multicolumn{5}{c}{\hrulefill TIC\,197755658\hrulefill} \\  
   \hline\noalign{\smallskip}
$\nu_1$  &5.32323\,$\pm$\,0.00015  &1.854\,$\pm$\,0.026 &0.3324\,$\pm$\,0.0022  & 42 & $\nu_1$ &21.59652\,$\pm$\,0.00014 &0.825\,$\pm$\,0.010  &0.6836\,$\pm$\,0.0020  & 38 & $\nu_1$  &51.25620\,$\pm$\,0.00044 &0.562\,$\pm$\,0.024 &0.9177\,$\pm$\,0.0067  & 22 \\
$\nu_2$  &4.27899\,$\pm$\,0.00053  &0.528\,$\pm$\,0.026 &0.5921\,$\pm$\,0.0078  & 7  & $\nu_2$ &26.67014\,$\pm$\,0.00014 &0.783\,$\pm$\,0.010  &0.7211\,$\pm$\,0.0021  & 45 & $\nu_2$  &46.63643\,$\pm$\,0.00080 &0.310\,$\pm$\,0.024 &0.7649\,$\pm$\,0.0121  & 11 \\
$\nu_3$  &5.12970\,$\pm$\,0.00054  &0.443\,$\pm$\,0.026 &0.6626\,$\pm$\,0.0093  & 10 & $\nu_3$ &16.69181\,$\pm$\,0.00017 &0.642\,$\pm$\,0.010  &0.9187\,$\pm$\,0.0026  & 42 & $\nu_3$  &49.19415\,$\pm$\,0.00097 &0.256\,$\pm$\,0.024 &0.7373\,$\pm$\,0.0147  & 8  \\
$\nu_4$  &4.24755\,$\pm$\,0.00058  &0.489\,$\pm$\,0.026 &0.7757\,$\pm$\,0.0085  & 7  & $\nu_4$ &23.58869\,$\pm$\,0.00039 &0.290\,$\pm$\,0.010  &0.2894\,$\pm$\,0.0058  & 17 &          &                         &                    &                       &   \\
$\nu_5$  &4.69256\,$\pm$\,0.00079  &0.356\,$\pm$\,0.026 &0.4970\,$\pm$\,0.0116  & 7  & $\nu_5$ &18.01112\,$\pm$\,0.00049 &0.228\,$\pm$\,0.010  &0.4957\,$\pm$\,0.0073  & 14 &          &                         &                    &                       &   \\
\hline
\multicolumn{4}{c}{\hrulefill TIC\,197757000\hrulefill} & \multicolumn{5}{c}{\hrulefill TIC\,240962482 \hrulefill} & \multicolumn{5}{c}{\hrulefill TIC\,241013310\hrulefill} &  \\  
   \hline\noalign{\smallskip}
$\nu_1$ &36.65383\,$\pm$\,0.00021 &0.859\,$\pm$\,0.016  &0.3463\,$\pm$\,0.0030  & 35 & $\nu_1$  &27.98413\,$\pm$\,0.00035 &1.147\,$\pm$\,0.016 &0.2308\,$\pm$\,0.0022  & 23 & $\nu_1$ &5.72370\,$\pm$\,0.00025  &0.615\,$\pm$\,0.014  &0.9089\,$\pm$\,0.0035  & 13 \\
$\nu_2$ &29.57429\,$\pm$\,0.00029 &0.606\,$\pm$\,0.016  &0.9506\,$\pm$\,0.0023  & 23 & $\nu_2$  &27.65528\,$\pm$\,0.00064 &0.633\,$\pm$\,0.016 &0.5763\,$\pm$\,0.0040  & 12 & $\nu_2$ &6.25826\,$\pm$\,0.00041  &0.370\,$\pm$\,0.014  &0.3633\,$\pm$\,0.0058  & 8 \\
$\nu_3$ &26.64726\,$\pm$\,0.00033 &0.541\,$\pm$\,0.016  &0.4796\,$\pm$\,0.0048  & 19 & $\nu_3$  &27.29194\,$\pm$\,0.00071 &0.567\,$\pm$\,0.016 &0.0469\,$\pm$\,0.0045  & 10 & $\nu_3$ &14.76784\,$\pm$\,0.00042 &0.361\,$\pm$\,0.014  &0.4021\,$\pm$\,0.0060  & 17 \\
$\nu_4$ &21.64643\,$\pm$\,0.00038 &0.466\,$\pm$\,0.016  &0.2506\,$\pm$\,0.0056  & 18 & $\nu_4$  &29.43289\,$\pm$\,0.00082 &0.490\,$\pm$\,0.016 &0.9475\,$\pm$\,0.0052  & 11 & $\nu_4$ &4.81999\,$\pm$\,0.00057  &0.264\,$\pm$\,0.014  &0.4349\,$\pm$\,0.0082  & 5 \\
$\nu_5$ &38.32894\,$\pm$\,0.00043 &0.411\,$\pm$\,0.016  &0.3283\,$\pm$\,0.0063  & 17 & $\nu_5$  &24.96472\,$\pm$\,0.00078 &0.517\,$\pm$\,0.016 &0.7146\,$\pm$\,0.0049  & 17 &         &                         &                     &                       &  \\
\hline
\multicolumn{4}{c}{\hrulefill TIC\,256640561\hrulefill} & \multicolumn{5}{c}{\hrulefill TIC\,272822330 \hrulefill} & \multicolumn{5}{c}{\hrulefill TIC\,289947843\hrulefill} \\  
   \hline\noalign{\smallskip}
$\nu_1$  &10.11866\,$\pm$\,0.00018 &0.217\,$\pm$\,0.004 &0.8695\,$\pm$\,0.0028  & 36 & $\nu_1$ &16.55851\,$\pm$\,0.00001 &0.506\,$\pm$\,0.013  &0.7901\,$\pm$\,0.0042  & 11 & $\nu_1$  &14.42398\,$\pm$\,0.00003 &0.357\,$\pm$\,0.003 &0.1029\,$\pm$\,0.0015  & 41 \\
$\nu_2$  &16.27639\,$\pm$\,0.00037 &0.105\,$\pm$\,0.004 &0.2941\,$\pm$\,0.0057  & 19 & $\nu_2$ &17.89061\,$\pm$\,0.00001 &0.471\,$\pm$\,0.013  &0.7731\,$\pm$\,0.0045  & 10 & $\nu_2$  &11.61625\,$\pm$\,0.00005 &0.224\,$\pm$\,0.003 &0.9873\,$\pm$\,0.0023  & 22  \\
$\nu_3$  &15.11399\,$\pm$\,0.00039 &0.100\,$\pm$\,0.004 &0.6635\,$\pm$\,0.0060  & 19 & $\nu_3$ &15.37946\,$\pm$\,0.00001 &0.332\,$\pm$\,0.013  &0.9197\,$\pm$\,0.0064  & 8 & $\nu_3$  &11.62266\,$\pm$\,0.00005 &0.217\,$\pm$\,0.003 &0.1661\,$\pm$\,0.0024  & 21  \\
$\nu_4$  &17.43972\,$\pm$\,0.00043 &0.090\,$\pm$\,0.004 &0.0847\,$\pm$\,0.0067  & 16 & $\nu_4$ &20.33903\,$\pm$\,0.00001 &0.326\,$\pm$\,0.013  &0.7286\,$\pm$\,0.0065  & 7 & $\nu_4$  &14.42912\,$\pm$\,0.00008 &0.150\,$\pm$\,0.003 &0.5904\,$\pm$\,0.0035  & 17  \\
$\nu_5$  &11.00822\,$\pm$\,0.00063 &0.062\,$\pm$\,0.004 &0.8177\,$\pm$\,0.0097  & 10 & $\nu_5$ &17.61348\,$\pm$\,0.00001 &0.306\,$\pm$\,0.013  &0.7254\,$\pm$\,0.0070  & 6 & $\nu_5$  &12.40031\,$\pm$\,0.00008 &0.146\,$\pm$\,0.003 &0.1397\,$\pm$\,0.0035  & 15 \\
\hline
\multicolumn{5}{c}{\hrulefill TIC\,301909087\hrulefill} & \multicolumn{5}{c}{\hrulefill TIC\,305633328\hrulefill} & \multicolumn{5}{c}{\hrulefill TIC\,322428763\hrulefill}\\  
   \hline\noalign{\smallskip}
$\nu_1$ &13.31209\,$\pm$\,0.00019 &2.861\,$\pm$\,0.023  &0.1732\,$\pm$\,0.0013  & 39 & $\nu_1$  &10.17360\,$\pm$\,0.00037 &0.806\,$\pm$\,0.026 &0.2727\,$\pm$\,0.0052  & 19 & $\nu_1$ &12.44685\,$\pm$\,0.00002 &27.487\,$\pm$\,0.024 &0.2984\,$\pm$\,0.0001  & 380 \\
$\nu_2$ &11.41486\,$\pm$\,0.00041 &1.341\,$\pm$\,0.023  &0.0983\,$\pm$\,0.0028  & 25 & $\nu_2$  &14.84142\,$\pm$\,0.00043 &0.699\,$\pm$\,0.026 &0.2635\,$\pm$\,0.0060  & 18 & 2$\nu_1$ &24.89370\,$\pm$\,0.00013 &3.479\,$\pm$\,0.024  &0.0780\,$\pm$\,0.0011  & 81 \\
$\nu_3$ &10.42865\,$\pm$\,0.00057 &0.964\,$\pm$\,0.023  &0.9806\,$\pm$\,0.0038  & 20 & $\nu_3$  &15.31862\,$\pm$\,0.00055 &0.541\,$\pm$\,0.026 &0.1905\,$\pm$\,0.0078  & 15 & $\nu_3$  &15.92382\,$\pm$\,0.00037 &1.259\,$\pm$\,0.024  &0.9422\,$\pm$\,0.0030  & 26 \\
$\nu_4$ &13.09508\,$\pm$\,0.00061 &0.896\,$\pm$\,0.023  &0.5211\,$\pm$\,0.0041  & 13 & $\nu_4$  &15.36912\,$\pm$\,0.00094 &0.320\,$\pm$\,0.026 &0.2315\,$\pm$\,0.0132  & 9  & 3$\nu_1$ &37.34056\,$\pm$\,0.00038 &1.223\,$\pm$\,0.024  &0.4383\,$\pm$\,0.0031  & 54 \\
$\nu_5$ &11.63832\,$\pm$\,0.00068 &0.803\,$\pm$\,0.023  &0.9980\,$\pm$\,0.0046  & 14 & $\nu_5$  &11.05172\,$\pm$\,0.00119 &0.253\,$\pm$\,0.026 &0.1454\,$\pm$\,0.0166  & 5  & $\nu_5$  &12.14271\,$\pm$\,0.00062 &0.749\,$\pm$\,0.024  &0.9277\,$\pm$\,0.0050  & 11 \\
 \hline
\end{tabular}
 \end{scriptsize}
\end{table}
 \end{landscape}

\setcounter{table}{3}
\begin{landscape}

\begin{table}
 \begin{scriptsize}
\label{tab:apt}
\caption{Continuation.}\label{tableap}
\begin{tabular}{llccc|llccc|llccc} \toprule
\hline\noalign{\smallskip}
   & Frequency  &  Amplitude  & Phase  &S/N     &    &Frequency  &  Amplitude   & Phase  &S/N & &Frequency  &  Amplitude   & Phase  &S/N\\
   & (d$^{-1}$) & (mmag)      & (rad)  &        &    &(d$^{-1}$) & (mmag)       & (rad)  &    & &(d$^{-1}$) & (mmag)       & (rad)  &    \\
   \hline\noalign{\smallskip}
\multicolumn{5}{c}{\hrulefill TIC\,327121759\hrulefill} & \multicolumn{5}{c}{\hrulefill TIC\,337094559\hrulefill} & \multicolumn{5}{c}{\hrulefill TIC\,338159479\hrulefill} \\  
   \hline\noalign{\smallskip}
$\nu_1$  &23.65750\,$\pm$\,0.00008 &1.016\,$\pm$\,0.030 &0.3121\,$\pm$\,0.0048  & 18  & $\nu_1$ &8.63508\,$\pm$\,0.00035  &1.561\,$\pm$\,0.026  &0.6307\,$\pm$\,0.0027  & 16 & $\nu_1$  &27.06439\,$\pm$\,0.00006 &1.347\,$\pm$\,0.024 &0.0505\,$\pm$\,0.0029  & 54 \\
$\nu_2$  &24.50198\,$\pm$\,0.00009 &0.872\,$\pm$\,0.030 &0.7576\,$\pm$\,0.0056  & 17  & $\nu_2$ &8.26695\,$\pm$\,0.00054  &1.020\,$\pm$\,0.026  &0.2042\,$\pm$\,0.0041  & 11 & $\nu_2$  &23.55252\,$\pm$\,0.00004 &0.642\,$\pm$\,0.015 &0.5789\,$\pm$\,0.0038  & 23\\
$\nu_3$  &22.90103\,$\pm$\,0.00011 &0.690\,$\pm$\,0.030 &0.4459\,$\pm$\,0.0070  & 13  & $\nu_3$ &7.86106\,$\pm$\,0.00056  &0.990\,$\pm$\,0.026  &0.8082\,$\pm$\,0.0043  & 12 & $\nu_3$  &24.96688\,$\pm$\,0.00004 &0.461\,$\pm$\,0.014 &0.8664\,$\pm$\,0.0049  & 18 \\
$\nu_4$  &23.56995\,$\pm$\,0.00013 &0.573\,$\pm$\,0.030 &0.6620\,$\pm$\,0.0085&10&16$\nu_orb$+2$\nu_1$-$\nu_3$&18.46141\,$\pm$\,0.00069&0.923\,$\pm$\,0.026 &0.3725\,$\pm$\,0.0053  & 8 & $\nu_4$  &28.51068\,$\pm$\,0.00005 &0.390\,$\pm$\,0.014 &0.3448\,$\pm$\,0.0058  & 15 \\
$\nu_5$  &30.96969\,$\pm$\,0.00016 &0.482\,$\pm$\,0.030 &0.2912\,$\pm$\,0.0101  & 10 & $\nu_5$ &10.20012\,$\pm$\,0.00069 &0.802\,$\pm$\,0.026  &0.5700\,$\pm$\,0.0062  & 6 & $\nu_5$  &23.06376\,$\pm$\,0.00008 &0.280\,$\pm$\,0.014 &0.7468\,$\pm$\,0.0084  & 10 \\
\hline
\multicolumn{4}{c}{\hrulefill TIC\,354926863\hrulefill} & \multicolumn{5}{c}{\hrulefill TIC\,358613523 \hrulefill} & \multicolumn{5}{c}{\hrulefill TIC\,393894013\hrulefill} &  \\  
   \hline\noalign{\smallskip}
$\nu_1$ &21.66089\,$\pm$\,0.00002 &1.813\,$\pm$\,0.013  &0.3126\,$\pm$\,0.0011  & 89 & $\nu_1$  &41.55243\,$\pm$\,0.00024 &0.465\,$\pm$\,0.011 &0.0551\,$\pm$\,0.0039  & 24 & $\nu_1$ &42.07023\,$\pm$\,0.00056 &0.468\,$\pm$\,0.013  &0.6493\,$\pm$\,0.0043  & 14 \\
$\nu_2$ &23.97610\,$\pm$\,0.00010 &0.311\,$\pm$\,0.013  &0.8778\,$\pm$\,0.0064  & 15 & $\nu_2$  &32.82261\,$\pm$\,0.00036 &0.309\,$\pm$\,0.011 &0.7414\,$\pm$\,0.0058  & 17 & $\nu_2$ &50.86278\,$\pm$\,0.00060 &0.444\,$\pm$\,0.013  &0.3516\,$\pm$\,0.0045  & 15 \\
$\nu_3$ &29.74307\,$\pm$\,0.00011 &0.291\,$\pm$\,0.013  &0.2144\,$\pm$\,0.0068  & 20 & $\nu_3$  &30.25640\,$\pm$\,0.00038 &0.298\,$\pm$\,0.011 &0.6099\,$\pm$\,0.0060  & 16 & $\nu_3$ &52.98499\,$\pm$\,0.00074 &0.359\,$\pm$\,0.013  &0.3011\,$\pm$\,0.0056  & 13 \\
$\nu_4$ &7.55377\,$\pm$\,0.00011  &0.278\,$\pm$\,0.013  &0.2867\,$\pm$\,0.0071  & 15 & $\nu_4$  &36.31599\,$\pm$\,0.00047 &0.240\,$\pm$\,0.011 &0.3481\,$\pm$\,0.0075  & 13 & $\nu_4$ &45.70912\,$\pm$\,0.00077 &0.345\,$\pm$\,0.013  &0.3654\,$\pm$\,0.0058  & 12 \\
$\nu_5$ &17.71561\,$\pm$\,0.00013 &0.238\,$\pm$\,0.013  &0.5979\,$\pm$\,0.0084  & 14 & $\nu_5$  &45.76655\,$\pm$\,0.00050 &0.224\,$\pm$\,0.011 &0.7536\,$\pm$\,0.0080  & 14 & $\nu_5$ &40.74380\,$\pm$\,0.00087 &0.305\,$\pm$\,0.013  &0.7169\,$\pm$\,0.0066  & 10 \\
\hline
\multicolumn{4}{c}{\hrulefill TIC\,396134795\hrulefill} & \multicolumn{5}{c}{\hrulefill TIC\,396201681 \hrulefill} & \multicolumn{5}{c}{\hrulefill TIC\,420114772\hrulefill} \\     \hline\noalign{\smallskip}
$\nu_1$  &14.19761\,$\pm$\,0.00025 &2.507\,$\pm$\,0.026 &0.6492\,$\pm$\,0.0016  & 31 & $\nu_1$ &2.09187\,$\pm$\,0.00003  &3.890\,$\pm$\,0.011  &0.7970\,$\pm$\,0.0004  & 90 & $\nu_1$  &26.07772\,$\pm$\,0.00012 &0.402\,$\pm$\,0.009 &0.9829\,$\pm$\,0.0035  & 35 \\
$\nu_2$  &23.06211\,$\pm$\,0.00052 &1.190\,$\pm$\,0.026 &0.1993\,$\pm$\,0.0035  & 20 & $\nu_2$ &1.96327\,$\pm$\,0.00004  &3.284\,$\pm$\,0.011  &0.4973\,$\pm$\,0.0005  & 72 & $\nu_2$  &19.65159\,$\pm$\,0.00012 &0.400\,$\pm$\,0.009 &0.4690\,$\pm$\,0.0035  & 34 \\
$\nu_3$  &27.50727\,$\pm$\,0.00087 &0.711\,$\pm$\,0.026 &0.3443\,$\pm$\,0.0058  & 9  & $\nu_3$ &20.53793\,$\pm$\,0.00006 &1.973\,$\pm$\,0.011  &0.3971\,$\pm$\,0.0009  & 93 & $\nu_3$  &21.29354\,$\pm$\,0.00012 &0.390\,$\pm$\,0.009 &0.8579\,$\pm$\,0.0036  & 36 \\
$\nu_4$  &28.46996\,$\pm$\,0.00097 &0.637\,$\pm$\,0.026 &0.1605\,$\pm$\,0.0065  & 8  & $\nu_4$ &25.52245\,$\pm$\,0.00008 &1.433\,$\pm$\,0.011  &0.0180\,$\pm$\,0.0012  & 68 & $\nu_4$  &22.36688\,$\pm$\,0.00014 &0.338\,$\pm$\,0.009 &0.0937\,$\pm$\,0.0041  & 32 \\
$\nu_5$  &25.46129\,$\pm$\,0.00162 &0.380\,$\pm$\,0.026 &0.8832\,$\pm$\,0.0109  & 8  & $\nu_5$ &20.07776\,$\pm$\,0.00014 &0.873\,$\pm$\,0.011  &0.8036\,$\pm$\,0.0020  & 48 & $\nu_5$  &22.30157\,$\pm$\,0.00015 &0.300\,$\pm$\,0.009 &0.8503\,$\pm$\,0.0047  & 28\\
\hline
\multicolumn{5}{c}{\hrulefill TIC\,421714420\hrulefill} & \multicolumn{5}{c}{\hrulefill TIC\,428003183\hrulefill} & \multicolumn{5}{c}{\hrulefill TIC\,430808126\hrulefill}\\  
   \hline\noalign{\smallskip}
$\nu_1$ &33.00852\,$\pm$\,0.00001  &0.566\,$\pm$\,0.005 &0.8806\,$\pm$\,0.0013  & 94 & $\nu_1$  &24.04691\,$\pm$\,0.00025 &0.921\,$\pm$\,0.021 &0.4706\,$\pm$\,0.0037  & 21 & $\nu_1$ &17.01396\,$\pm$\,0.00030 &0.780\,$\pm$\,0.022  &0.9924\,$\pm$\,0.0045  & 21 \\
$\nu_2$ &16.63040\,$\pm$\,0.00001  &0.223\,$\pm$\,0.005 &0.2679\,$\pm$\,0.0033  & 38 & $\nu_2$  &22.69924\,$\pm$\,0.00051 &0.592\,$\pm$\,0.021 &0.2310\,$\pm$\,0.0074  & 15 & $\nu_2$ &21.80517\,$\pm$\,0.00044 &0.540\,$\pm$\,0.022  &0.1305\,$\pm$\,0.0064  & 12 \\
$\nu_3$ &53.15438\,$\pm$\,0.00001  &0.172\,$\pm$\,0.005 &0.0092\,$\pm$\,0.0043  & 33 & $\nu_3$  &14.00456\,$\pm$\,0.00040 &0.590\,$\pm$\,0.021 &0.4984\,$\pm$\,0.0058  & 14 & $\nu_3$ &17.04825\,$\pm$\,0.00047 &0.501\,$\pm$\,0.022  &0.8969\,$\pm$\,0.0069  & 13 \\
$\nu_4$ &20.91822\,$\pm$\,0.00001  &0.100\,$\pm$\,0.005 &0.9778\,$\pm$\,0.0073  & 17 & $\nu_4$  &30.70567\,$\pm$\,0.00044 &0.532\,$\pm$\,0.021 &0.7363\,$\pm$\,0.0064  & 11 & $\nu_4$ &5.25878\,$\pm$\,0.00058  &0.405\,$\pm$\,0.022  &0.8720\,$\pm$\,0.0086  & 10 \\
$\nu_5$ &24.82455\,$\pm$\,0.00001  &0.101\,$\pm$\,0.005 &0.5483\,$\pm$\,0.0073  & 14 & $\nu_5$  &14.04931\,$\pm$\,0.00066 &0.458\,$\pm$\,0.021 &0.5547\,$\pm$\,0.0096  & 11 & $\nu_5$ &15.14783\,$\pm$\,0.00067 &0.351\,$\pm$\,0.022  &0.7665\,$\pm$\,0.0099  & 10 \\
 \hline
\end{tabular}
 \end{scriptsize}
\end{table}
 \end{landscape}

\setcounter{table}{3}
\begin{landscape}

\begin{table}
 \begin{scriptsize}
\label{tab:apt}
\caption{Continuation.}\label{tableap}
\begin{tabular}{llccc|llccc|llccc} \toprule
\hline\noalign{\smallskip}
   & Frequency  &  Amplitude  & Phase  &S/N     &    &Frequency  &  Amplitude   & Phase  &S/N & &Frequency  &  Amplitude   & Phase  &S/N\\
   & (d$^{-1}$) & (mmag)      & (rad)  &        &    &(d$^{-1}$) & (mmag)       & (rad)  &    & &(d$^{-1}$) & (mmag)       & (rad)  &    \\
   \hline\noalign{\smallskip}
\multicolumn{5}{c}{\hrulefill TIC\,440003271\hrulefill} & \multicolumn{5}{c}{\hrulefill TIC\,456905229\hrulefill} & \multicolumn{5}{c}{\hrulefill TIC\,467354611\hrulefill} \\ 
   \hline\noalign{\smallskip}
$\nu_1$  &18.71653\,$\pm$\,0.00031 &1.204\,$\pm$\,0.016 &0.1117\,$\pm$\,0.0021  & 36 & $\nu_1$ &15.16456\,$\pm$\,0.00003 &2.646\,$\pm$\,0.007  &0.7014\,$\pm$\,0.0004  & 167 & $\nu_1$  &14.90235\,$\pm$\,0.00001 &3.905\,$\pm$\,0.008 &0.5206\,$\pm$\,0.0004  & 166\\
$\nu_2$  &15.61934\,$\pm$\,0.00032 &1.192\,$\pm$\,0.016 &0.8186\,$\pm$\,0.0021  & 36 & $\nu_2$ &13.64681\,$\pm$\,0.00005 &1.468\,$\pm$\,0.007  &0.1107\,$\pm$\,0.0008  & 82 & $\nu_2$  &13.61211\,$\pm$\,0.00001 &3.039\,$\pm$\,0.007 &0.9023\,$\pm$\,0.0005  & 152 \\
$\nu_3$  &14.94367\,$\pm$\,0.00040 &0.940\,$\pm$\,0.016 &0.4500\,$\pm$\,0.0027  & 29 & $\nu_3$ &21.13332\,$\pm$\,0.00006 &1.226\,$\pm$\,0.007  &0.7881\,$\pm$\,0.0009  & 58 & $\nu_3$  &14.13054\,$\pm$\,0.00001 &2.320\,$\pm$\,0.009 &0.5640\,$\pm$\,0.0005  & 97   \\
$\nu_4$  &16.02039\,$\pm$\,0.00054 &0.697\,$\pm$\,0.016 &0.9872\,$\pm$\,0.0036  & 21 & $\nu_4$ &19.93937\,$\pm$\,0.00009 &0.852\,$\pm$\,0.007  &0.5775\,$\pm$\,0.0013  & 46 & $\nu_4$  &14.09850\,$\pm$\,0.00001 &1.035\,$\pm$\,0.007 &0.4843\,$\pm$\,0.0011  & 43    \\
$\nu_5$  &20.54519\,$\pm$\,0.00058 &0.657\,$\pm$\,0.016 &0.0864\,$\pm$\,0.0038  & 20 & $\nu_5$ &28.24970\,$\pm$\,0.00009 &0.809\,$\pm$\,0.007  &0.4093\,$\pm$\,0.0014  & 46 & $\nu_5$  &12.21896\,$\pm$\,0.00001 &0.702\,$\pm$\,0.011 &0.5071\,$\pm$\,0.0019  & 51  \\
 \hline
\end{tabular}
 \end{scriptsize}
\end{table}
 \end{landscape}

\setcounter{figure}{4}

\begin{figure*}
 \begin{minipage}[b]{0.24\textwidth}
  \includegraphics[height=3.5cm, width=1.0\textwidth]{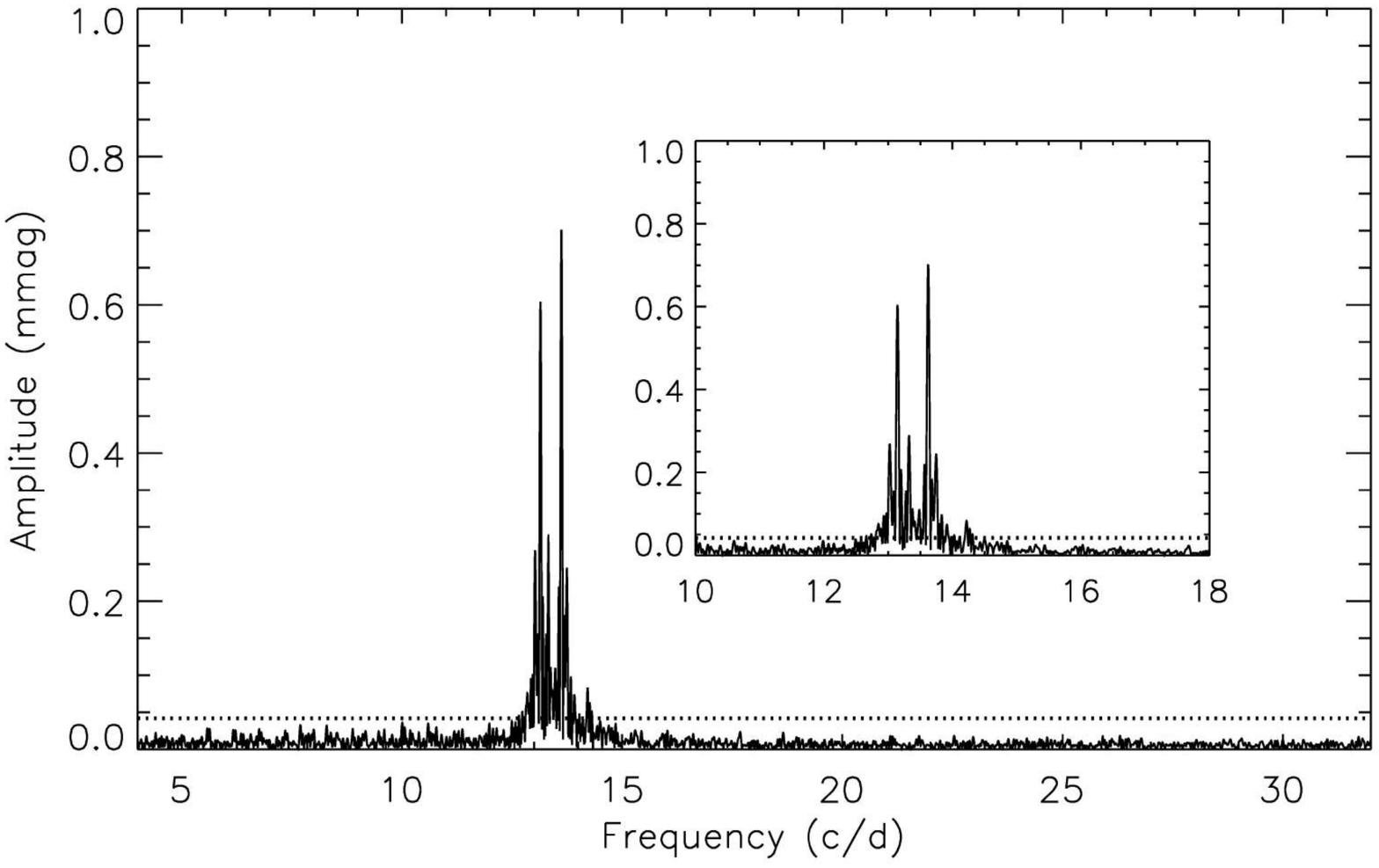}
  \end{minipage}
 \begin{minipage}[b]{0.24\textwidth}
  \includegraphics[height=3.5cm, width=1.0\textwidth]{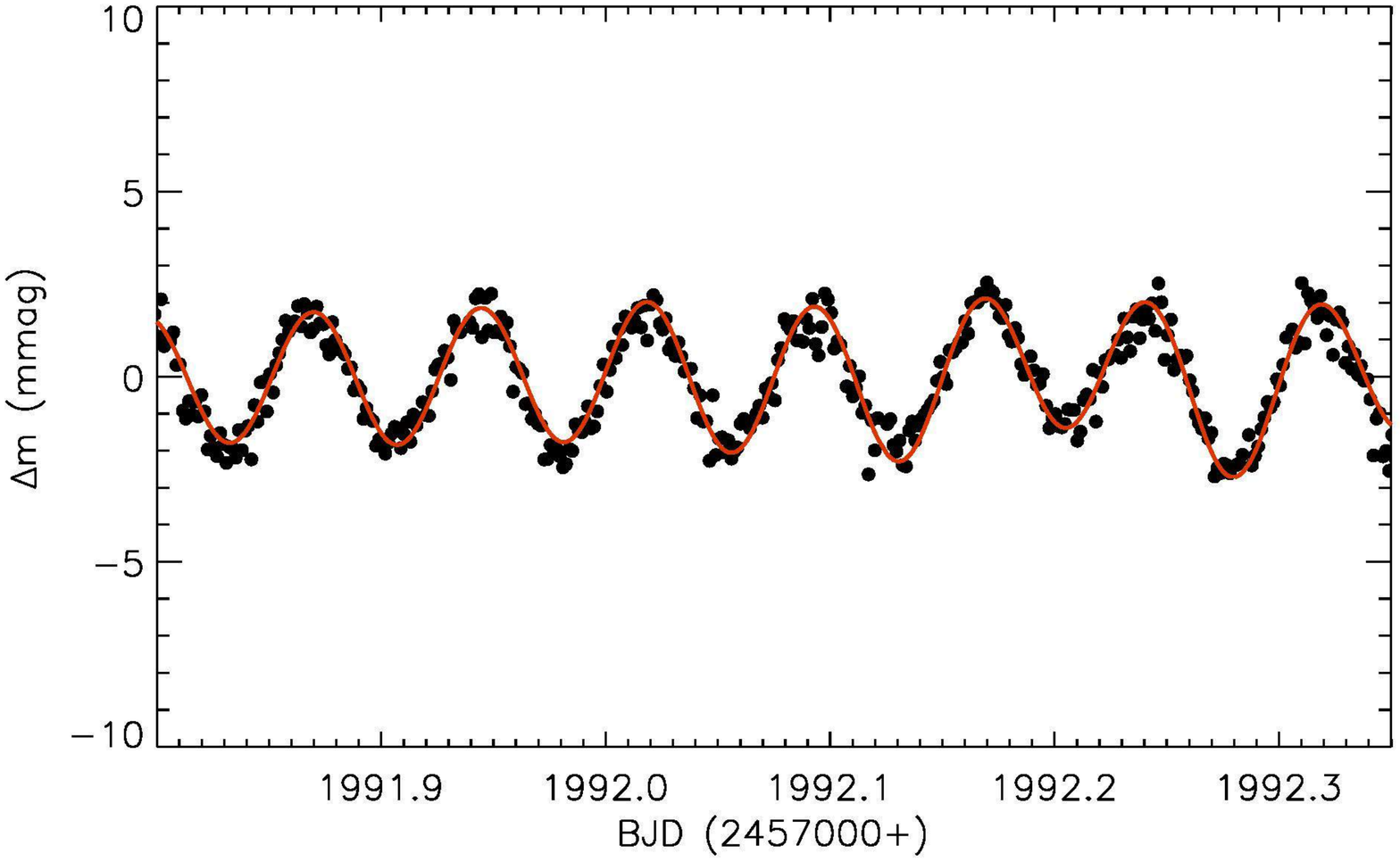}
 \end{minipage}
  \begin{minipage}[b]{0.24\textwidth}
  \includegraphics[height=3.5cm, width=1.0\textwidth]{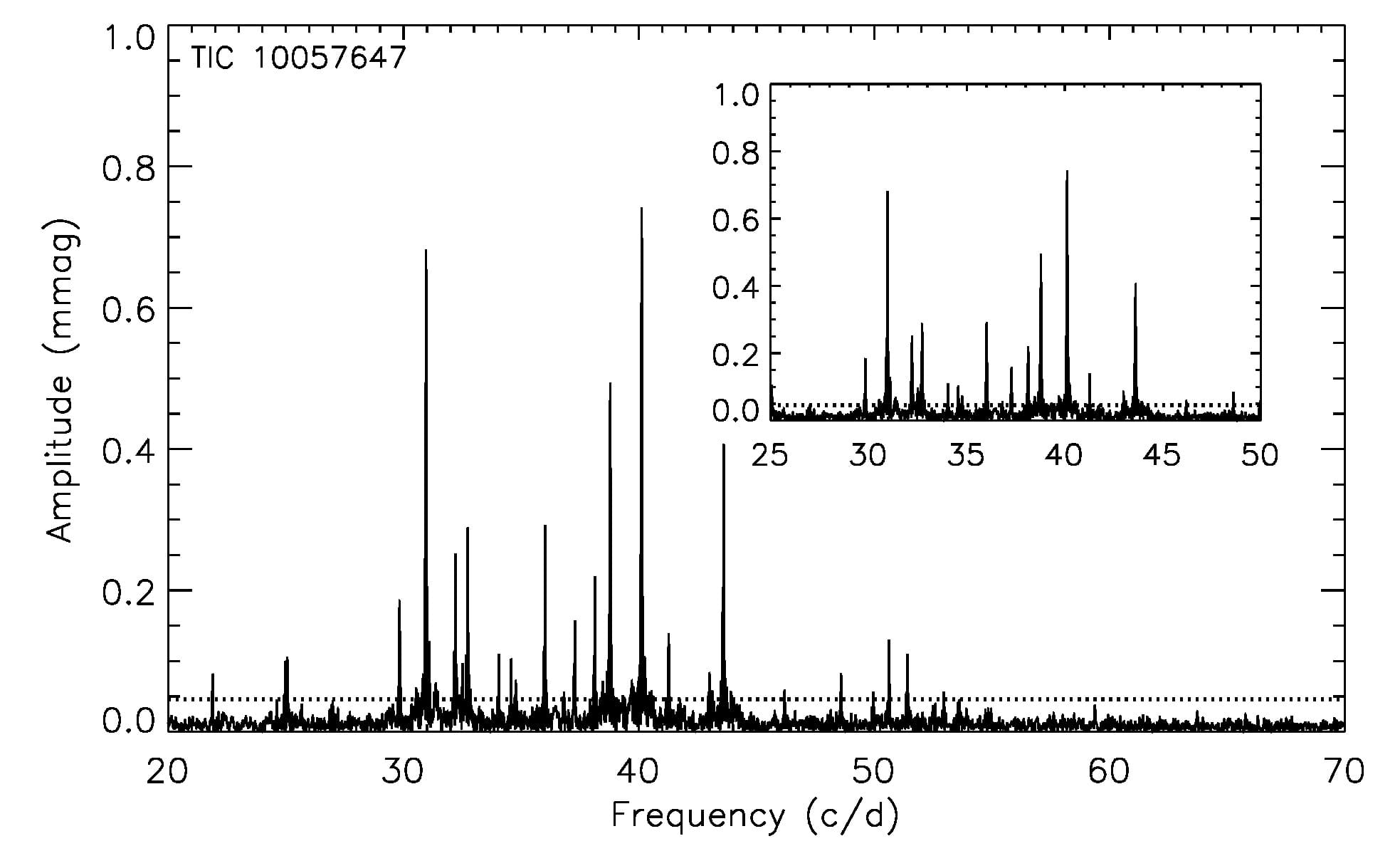}
  \end{minipage}
  \begin{minipage}[b]{0.24\textwidth}
 \includegraphics[height=3.5cm, width=1.0\textwidth]{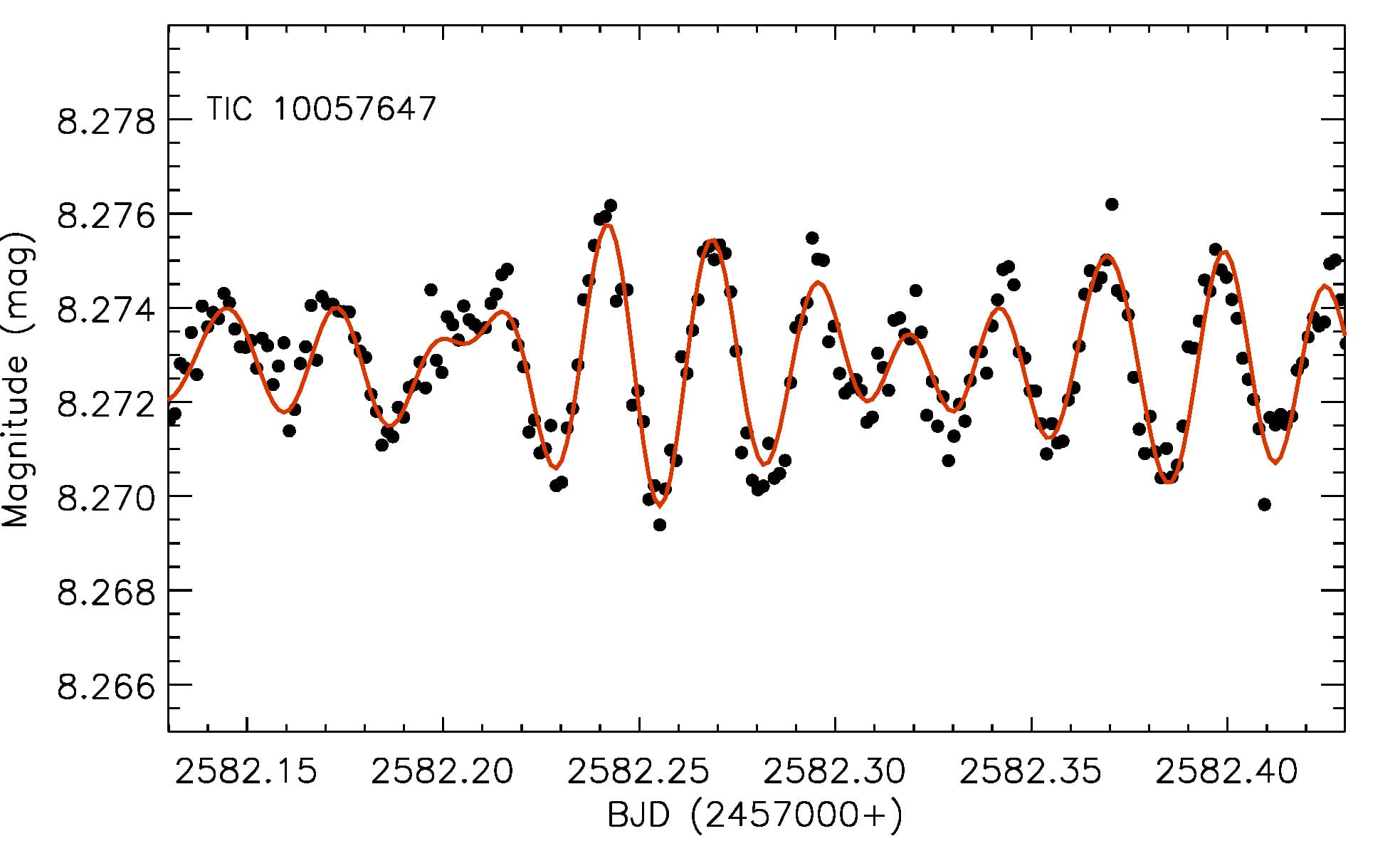}
 \end{minipage}
 \begin{minipage}[b]{0.24\textwidth}
  \includegraphics[height=3.5cm, width=1\textwidth]{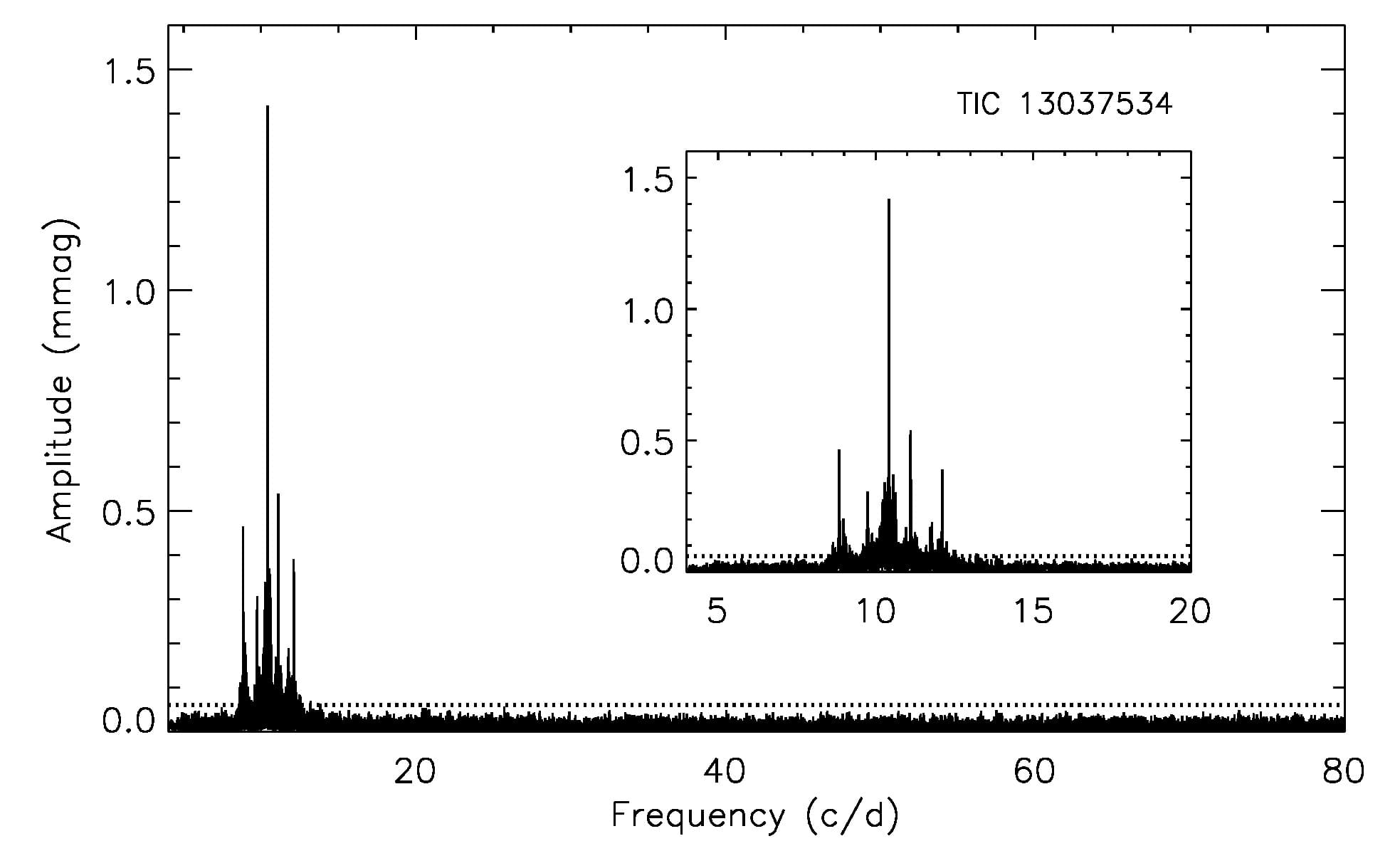}
  \end{minipage}
  \begin{minipage}[b]{0.24\textwidth}
  \includegraphics[height=3.5cm, width=1\textwidth]{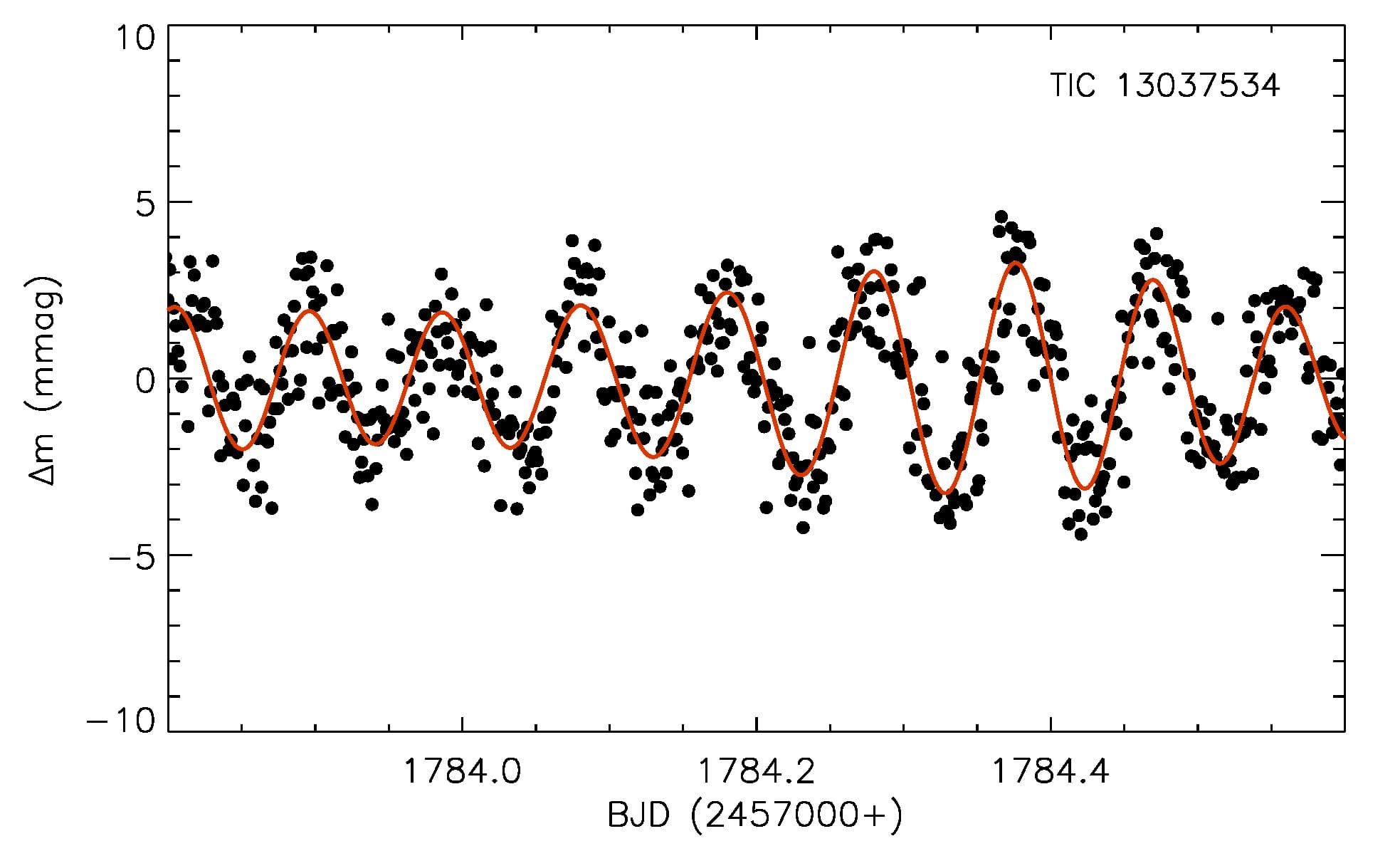}
  \end{minipage}
  \begin{minipage}[b]{0.24\textwidth}
  \includegraphics[height=3.5cm, width=1\textwidth]{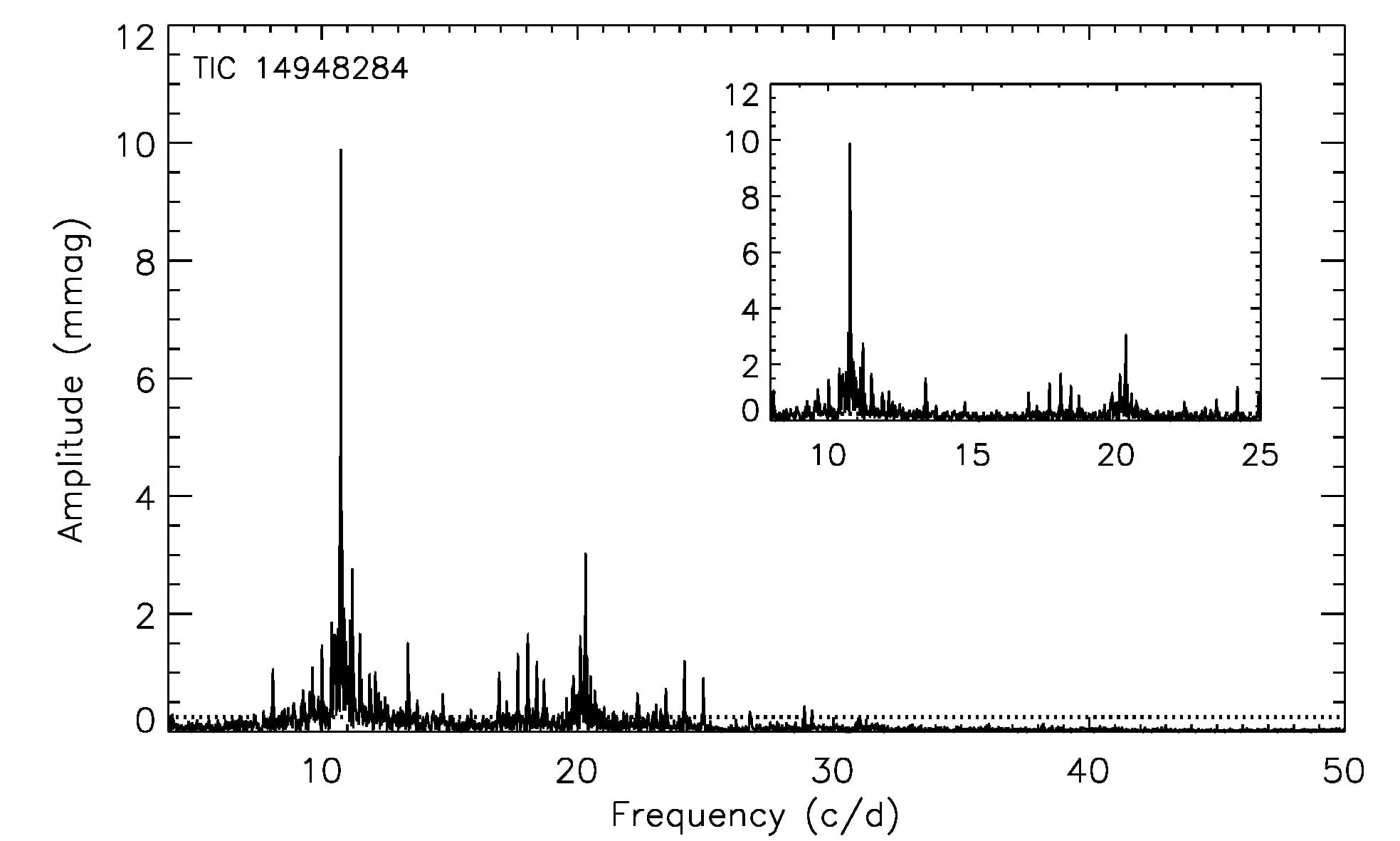}
 \end{minipage}
  \begin{minipage}[b]{0.24\textwidth}
  \includegraphics[height=3.5cm, width=1\textwidth]{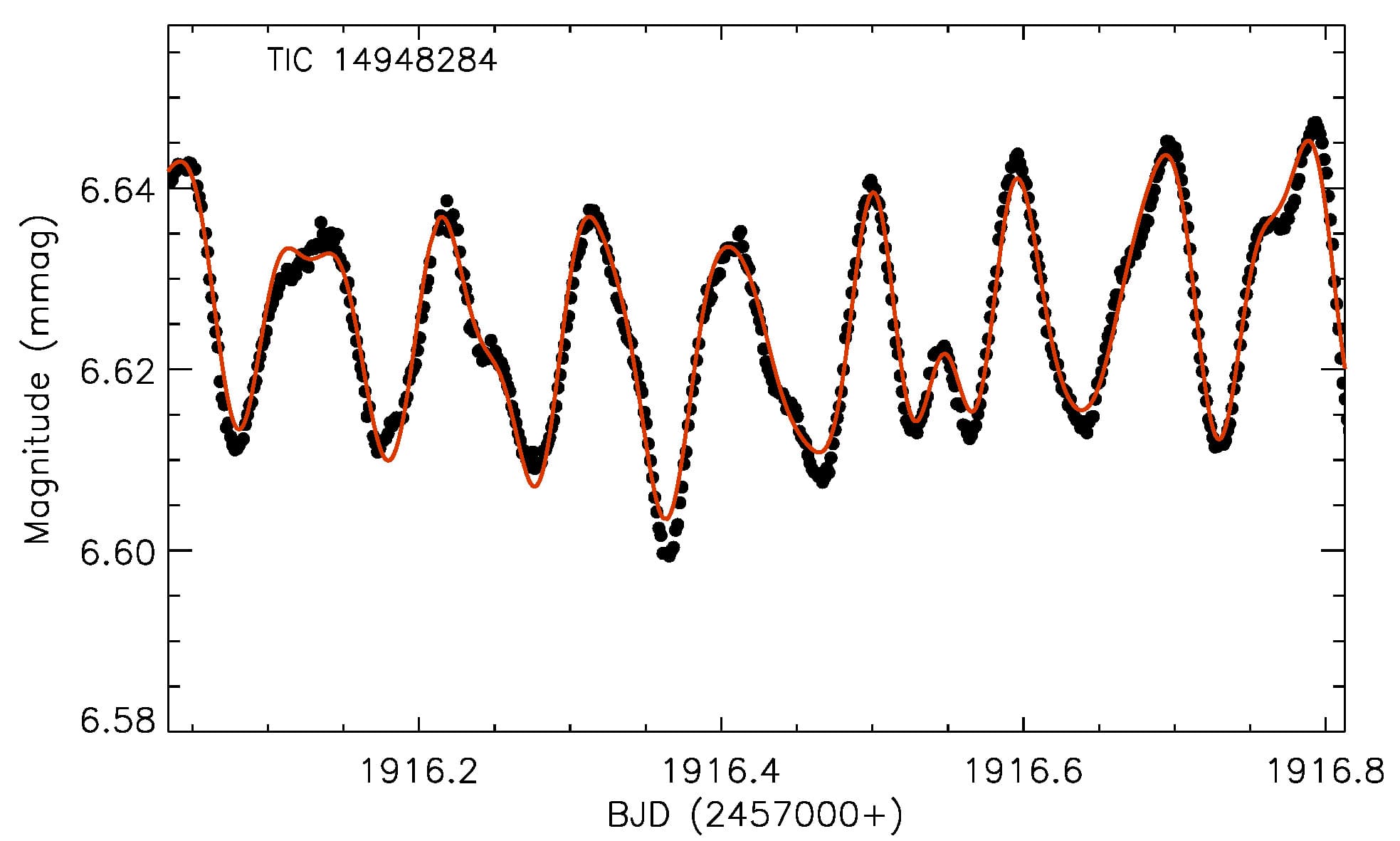}
 \end{minipage}
 \begin{minipage}[b]{0.24\textwidth}
  \includegraphics[height=3.5cm, width=1\textwidth]{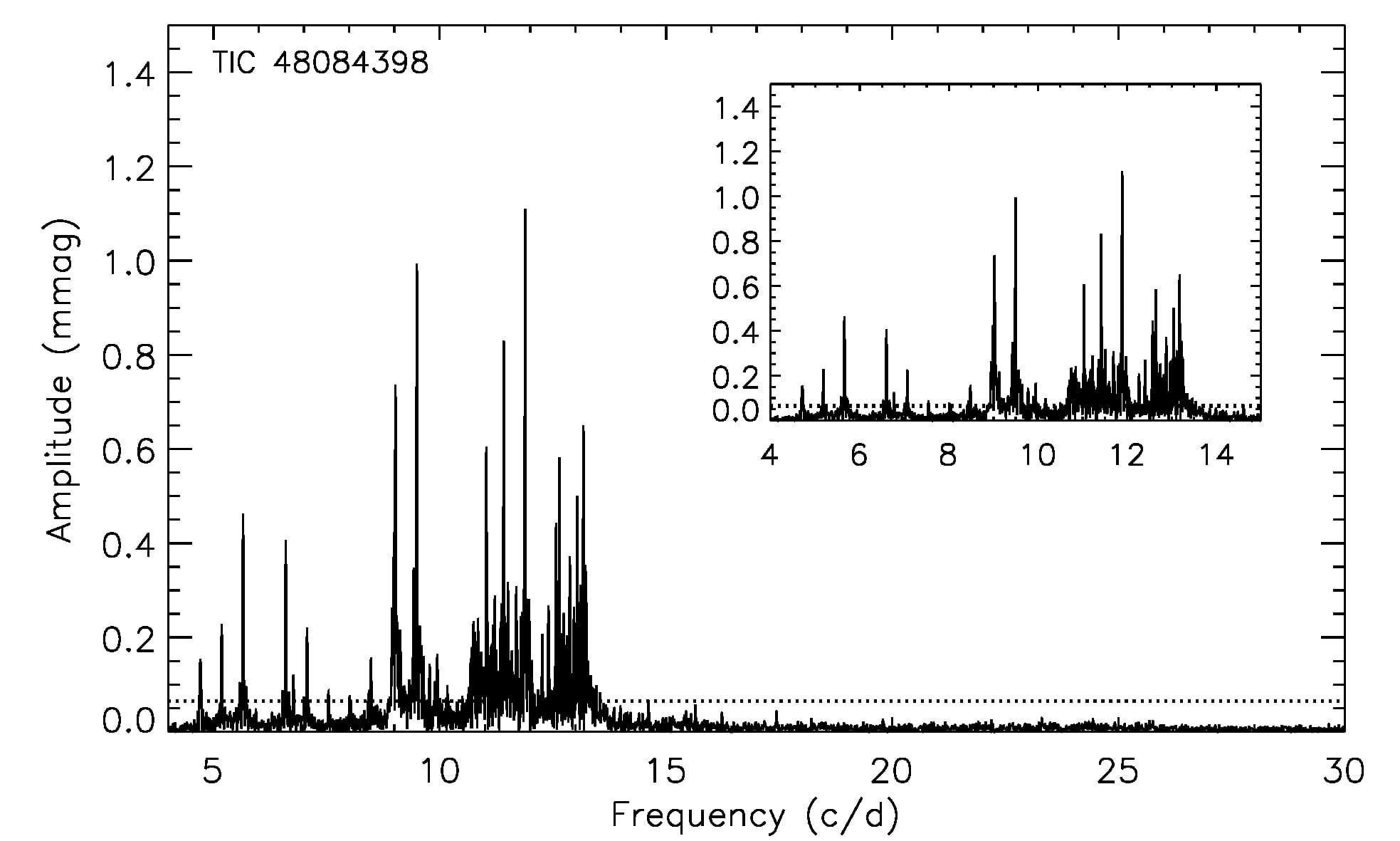}
  \end{minipage}
  \begin{minipage}[b]{0.24\textwidth}
  \includegraphics[height=3.5cm, width=1\textwidth]{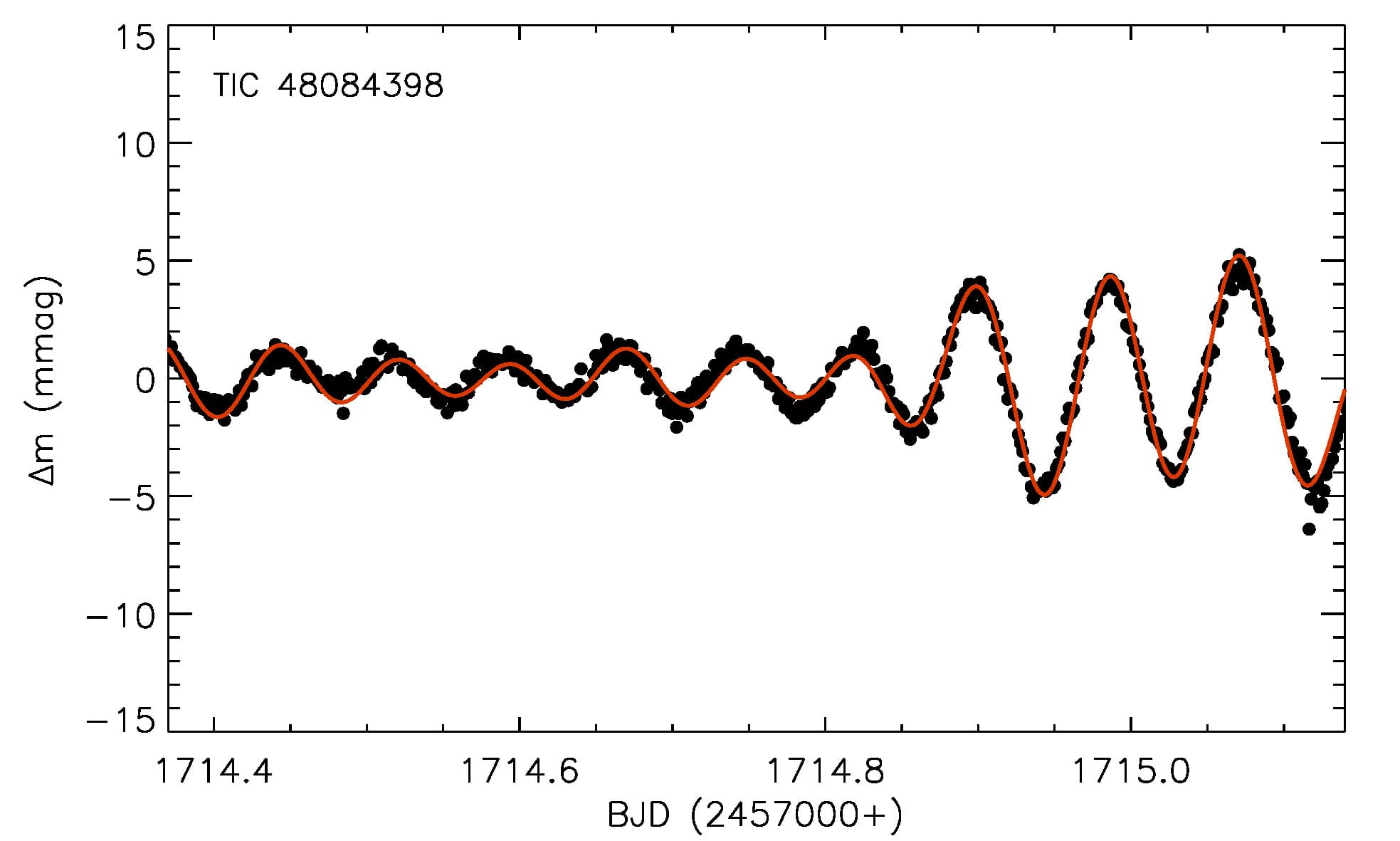}
  \end{minipage}
   \begin{minipage}[b]{0.24\textwidth}
  \includegraphics[height=3.5cm, width=1\textwidth]{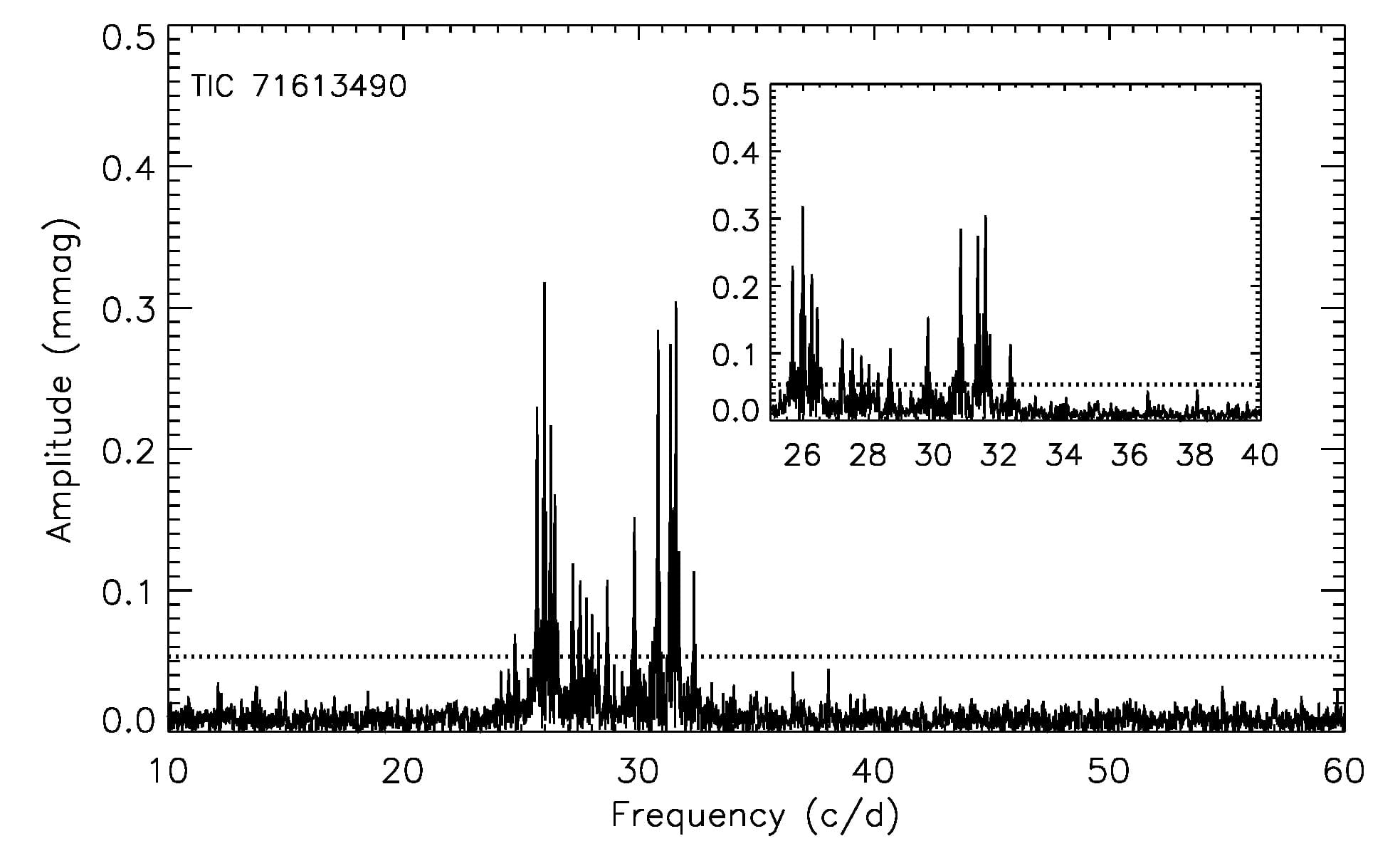}
 \end{minipage}
\begin{minipage}[b]{0.24\textwidth}
 \includegraphics[height=3.5cm, width=1\textwidth]{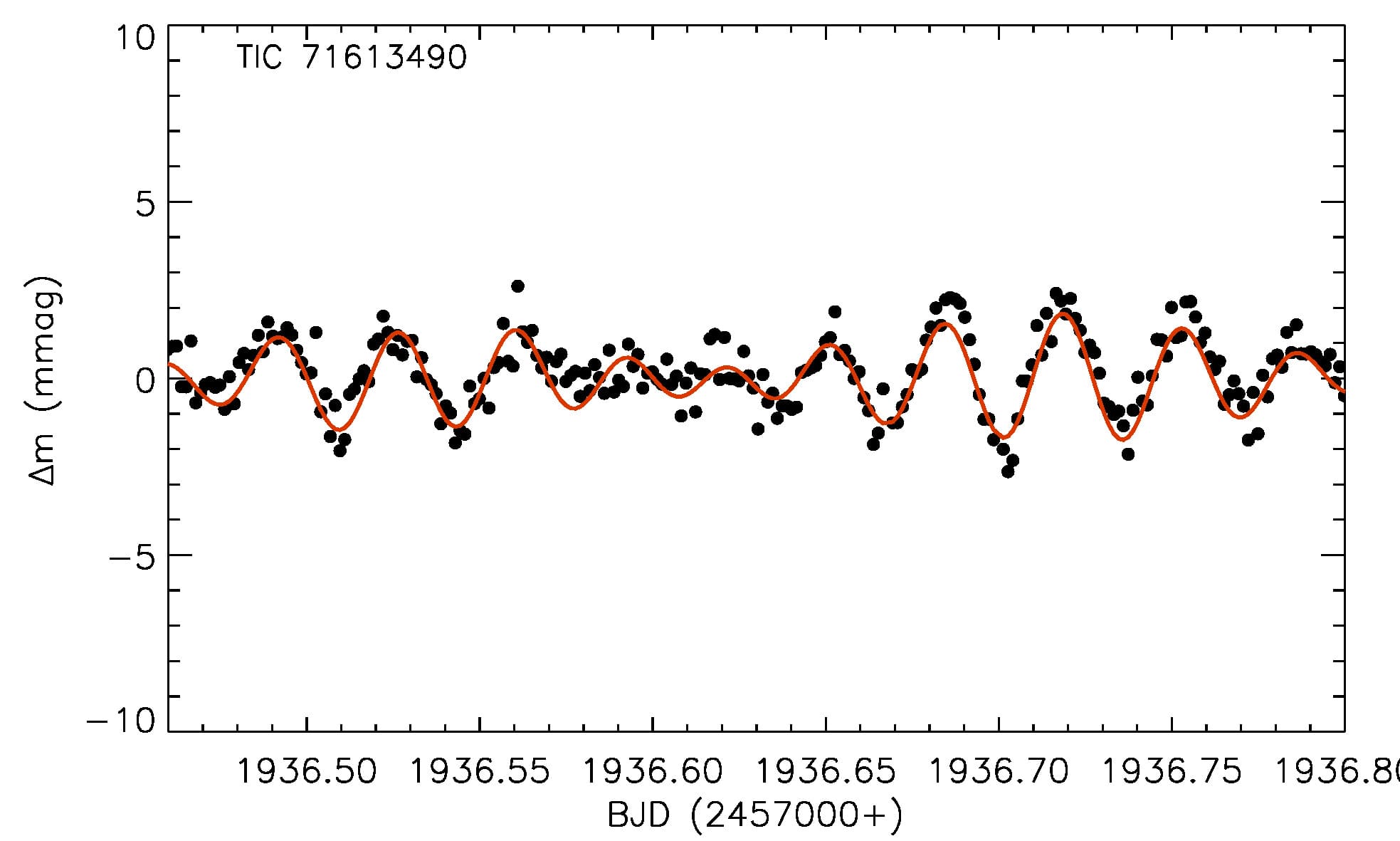}
 \end{minipage}
  \begin{minipage}[b]{0.24\textwidth}
  \includegraphics[height=3.5cm, width=1\textwidth]{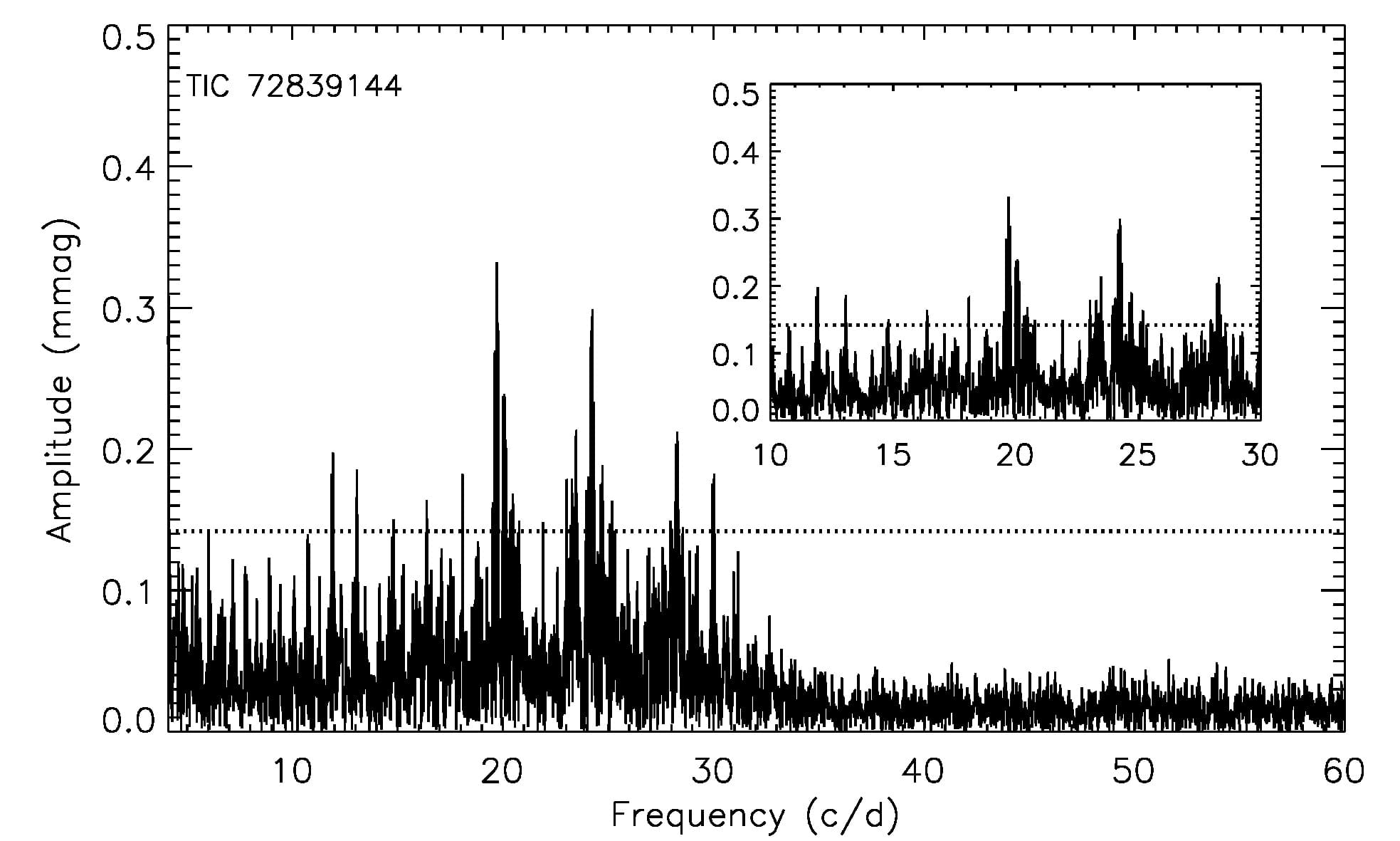}
 \end{minipage}
\begin{minipage}[b]{0.24\textwidth}
  \includegraphics[height=3.5cm, width=1\textwidth]{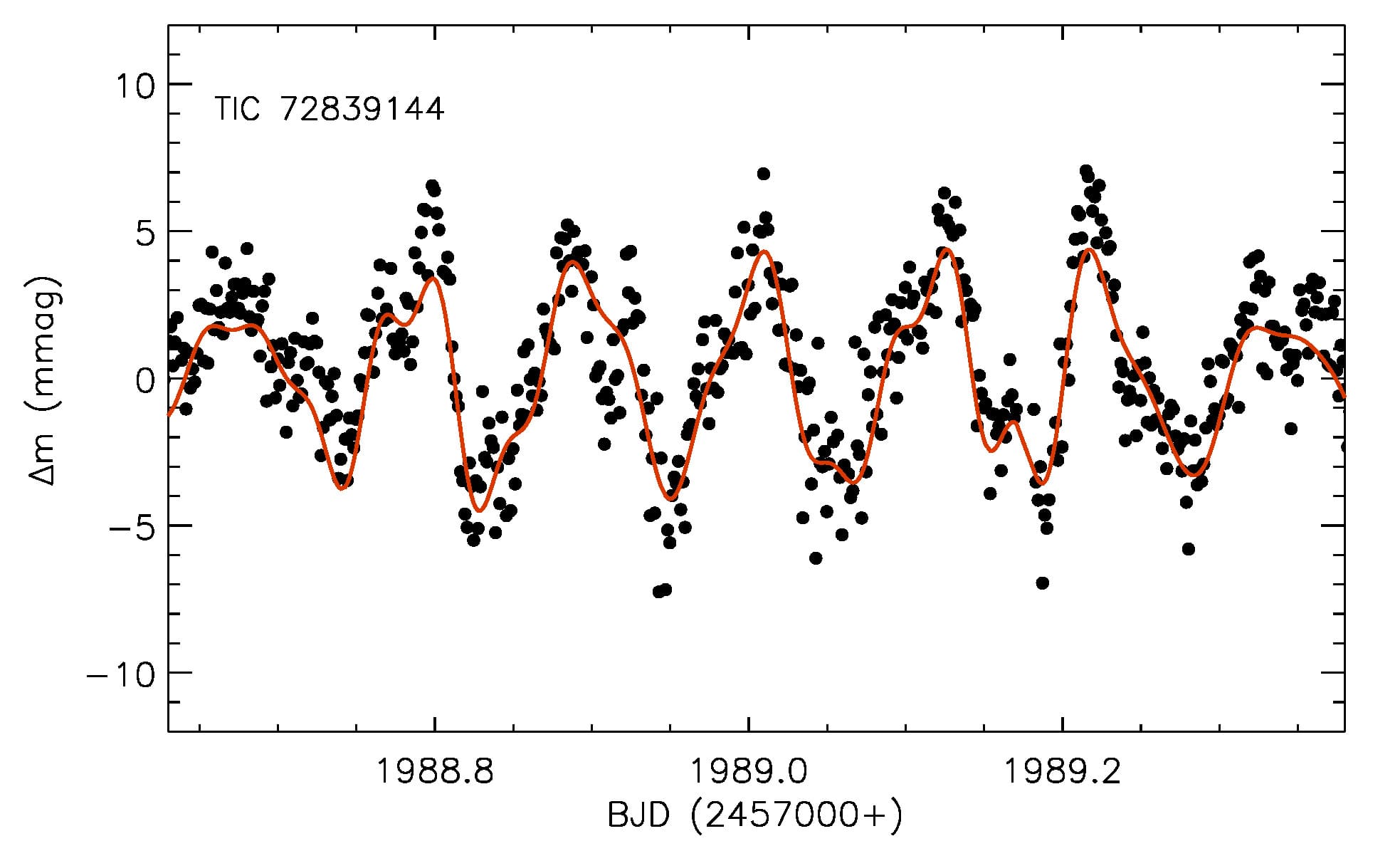}
  \end{minipage}
    \begin{minipage}[b]{0.24\textwidth}
  \includegraphics[height=3.5cm, width=1\textwidth]{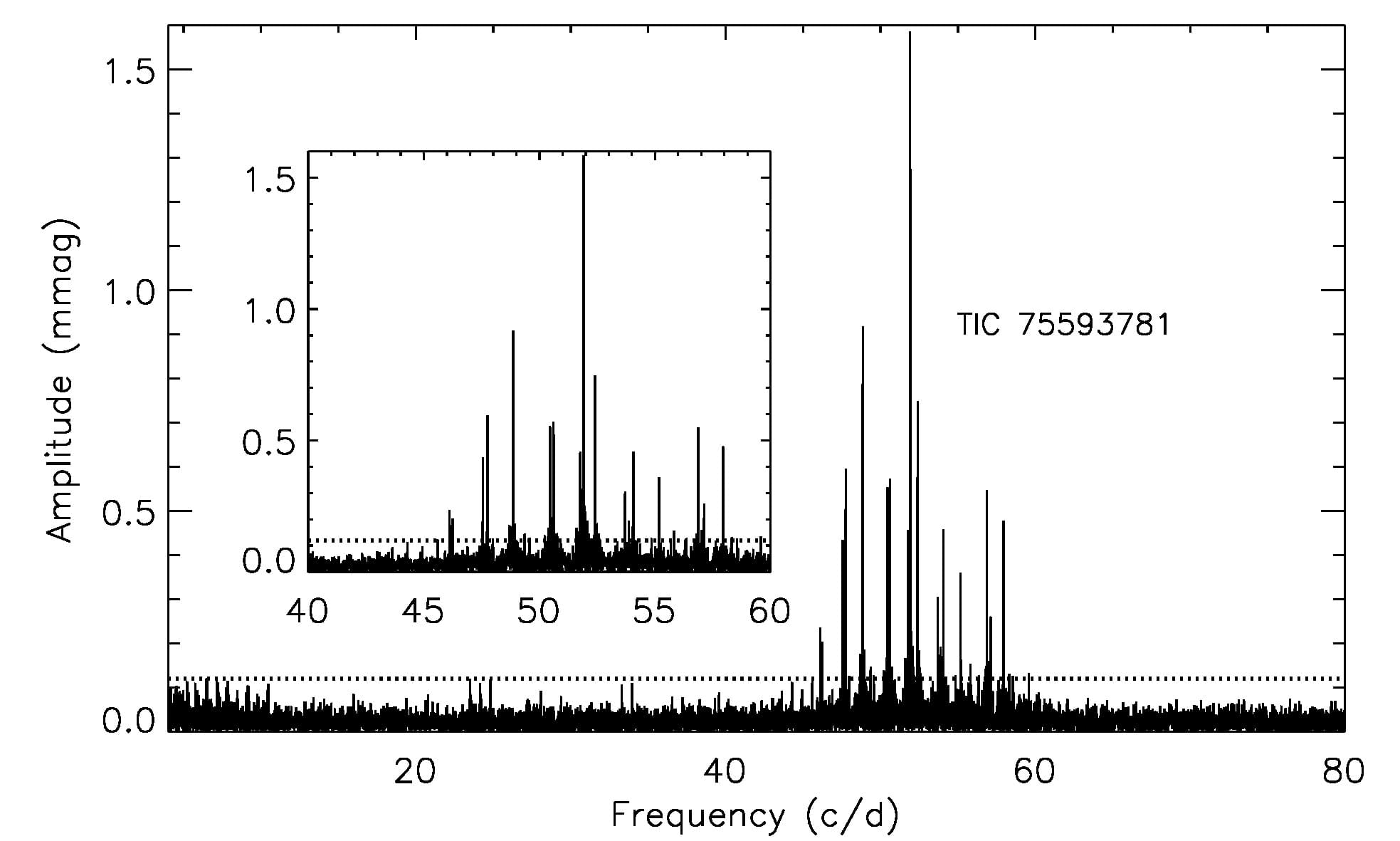}
  \end{minipage}
  \begin{minipage}[b]{0.24\textwidth}
  \includegraphics[height=3.5cm, width=1\textwidth]{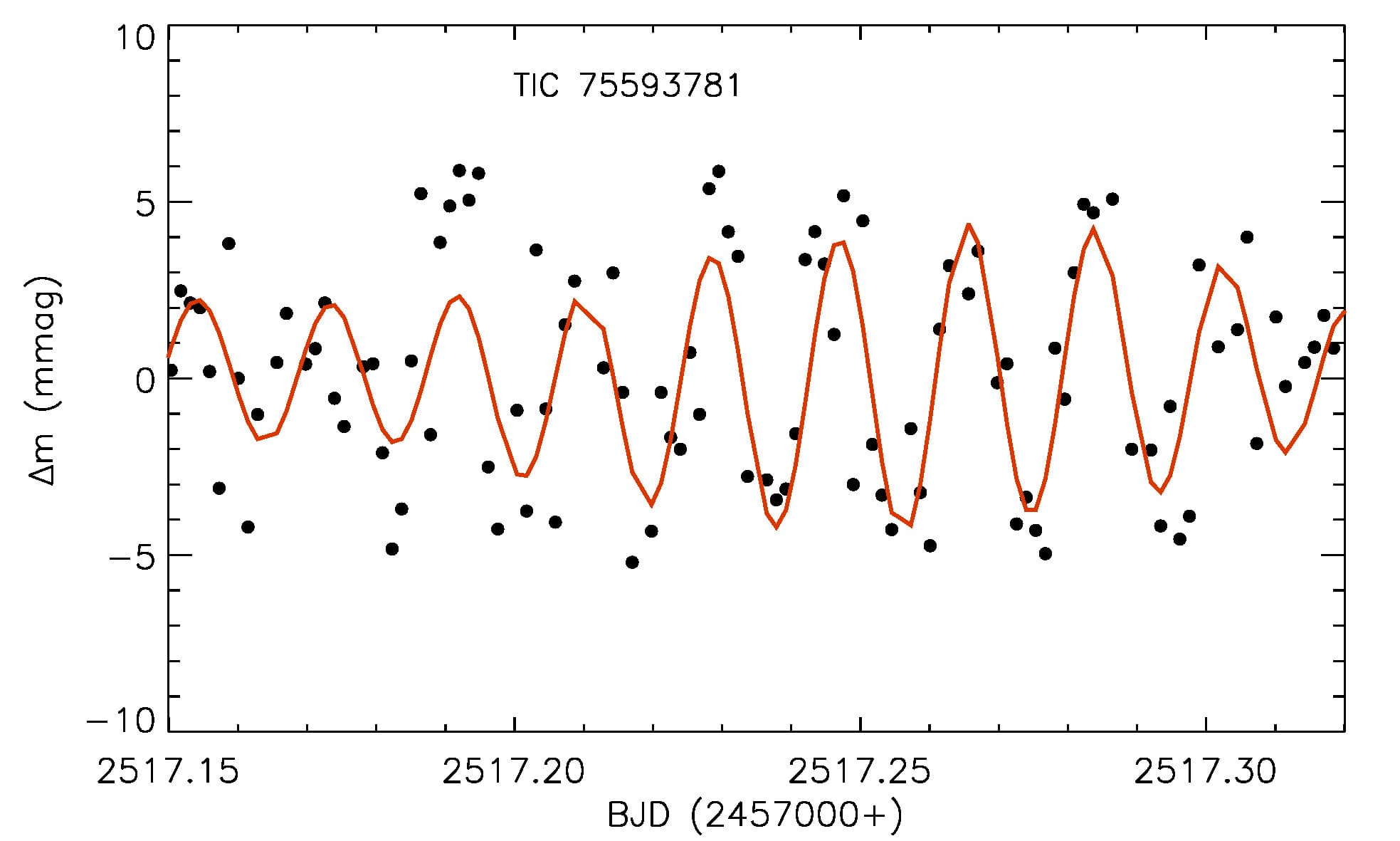}
  \end{minipage}
   \begin{minipage}[b]{0.24\textwidth}
  \includegraphics[height=3.5cm, width=1.0\textwidth]{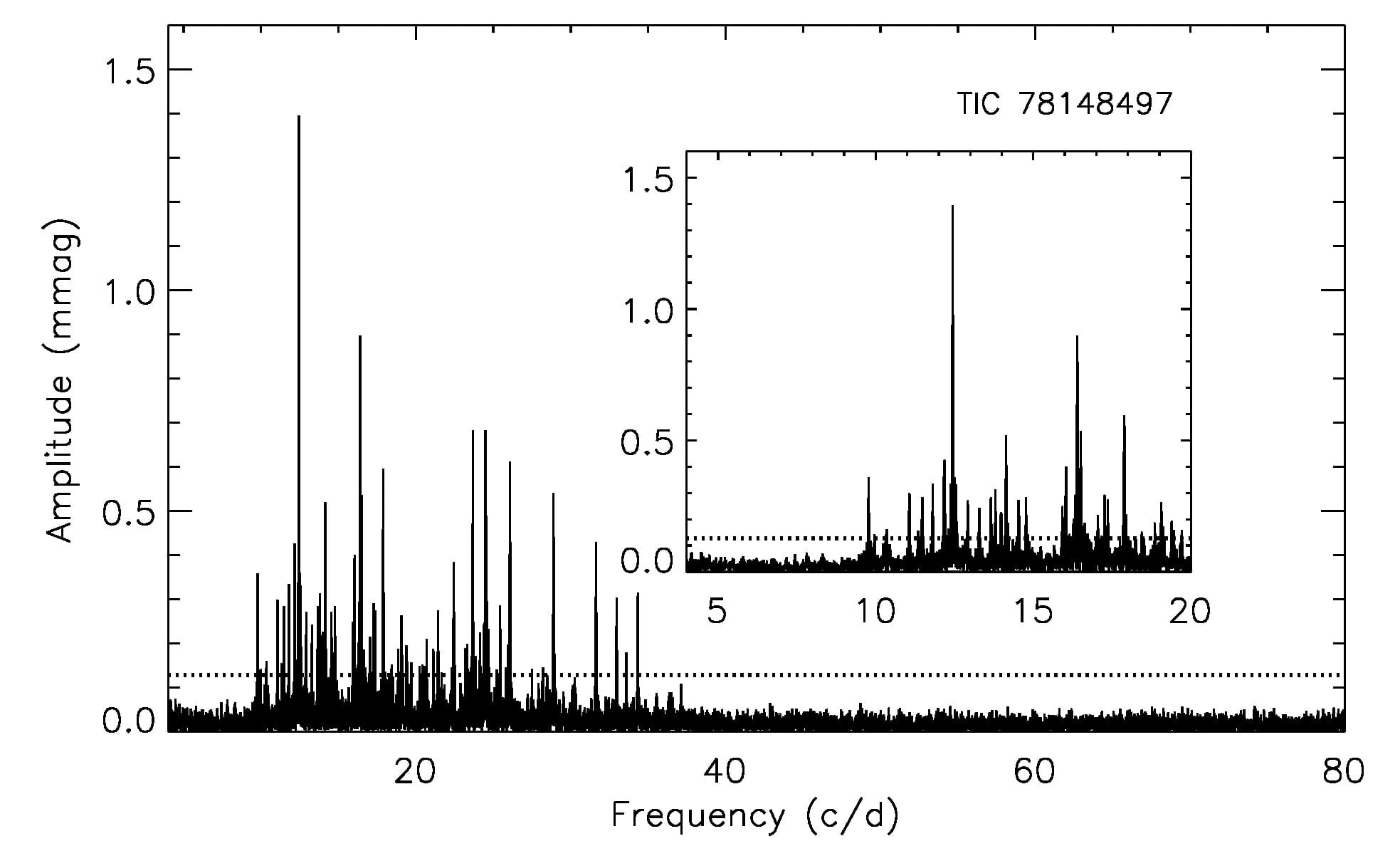}
  \end{minipage}
  \begin{minipage}[b]{0.24\textwidth}
  \includegraphics[height=3.5cm, width=1.0\textwidth]{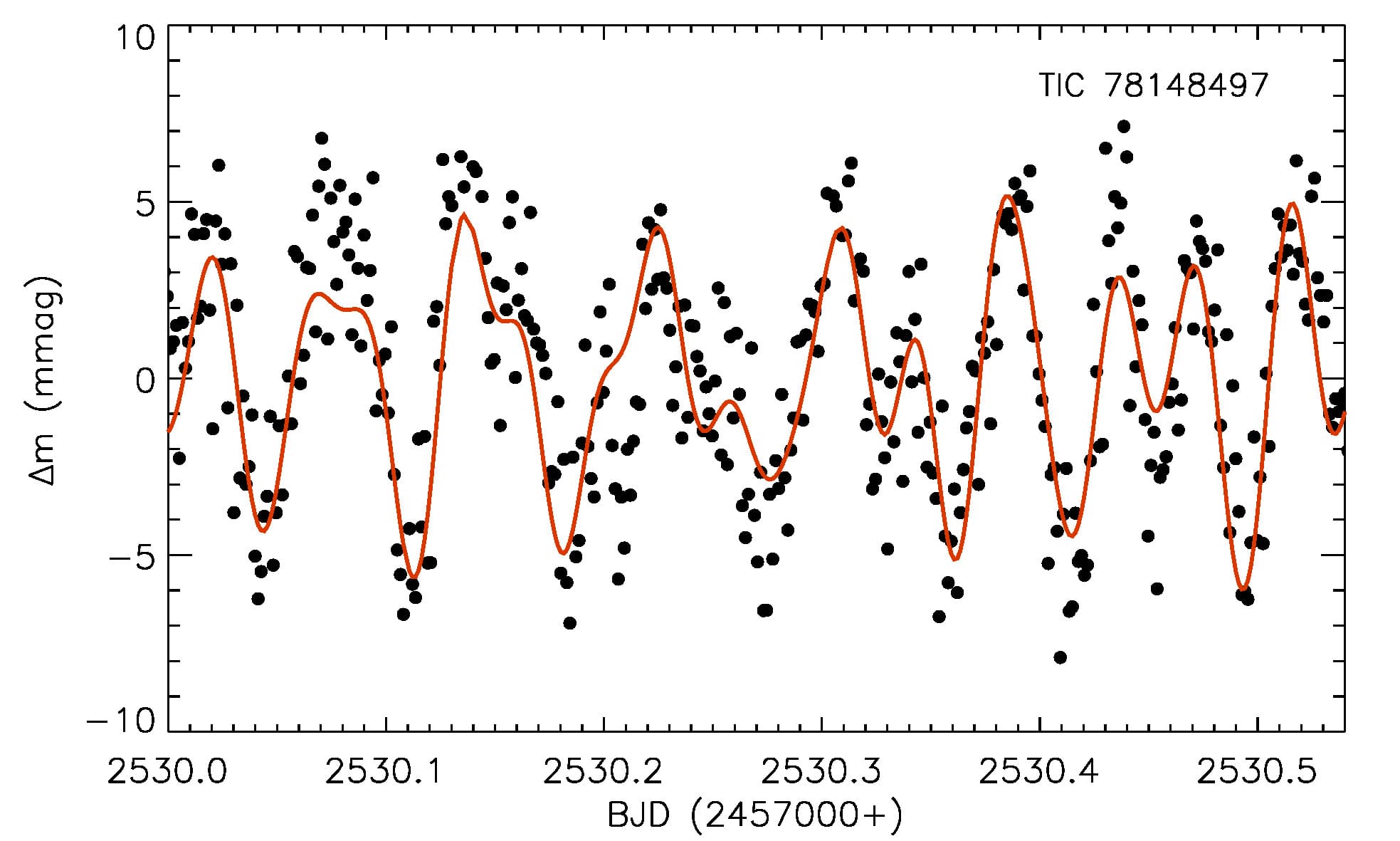}
 \end{minipage}
   \begin{minipage}[b]{0.24\textwidth}
  \includegraphics[height=3.5cm, width=1\textwidth]{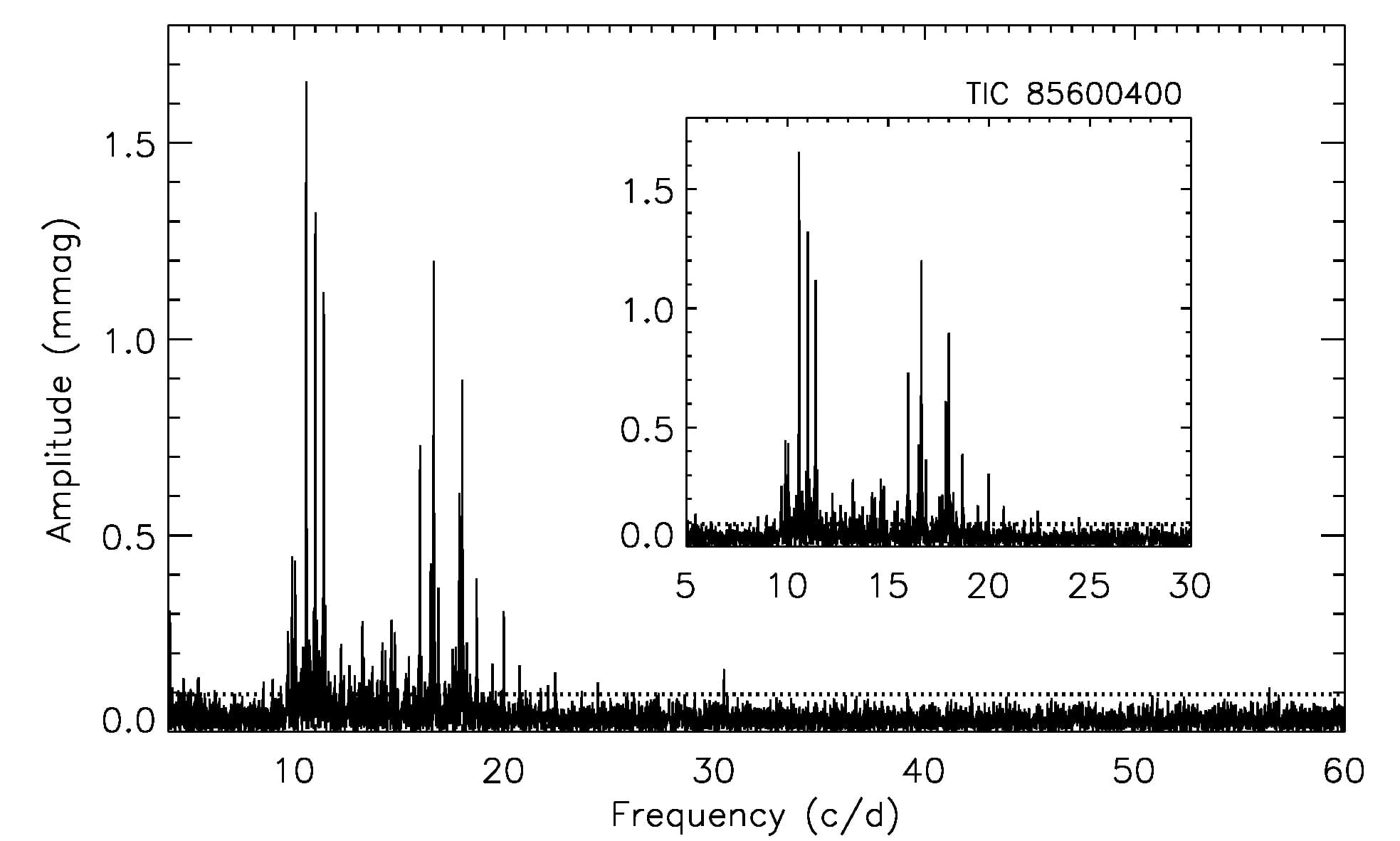}
  \end{minipage}
 \begin{minipage}[b]{0.24\textwidth}
 \includegraphics[height=3.5cm, width=1\textwidth]{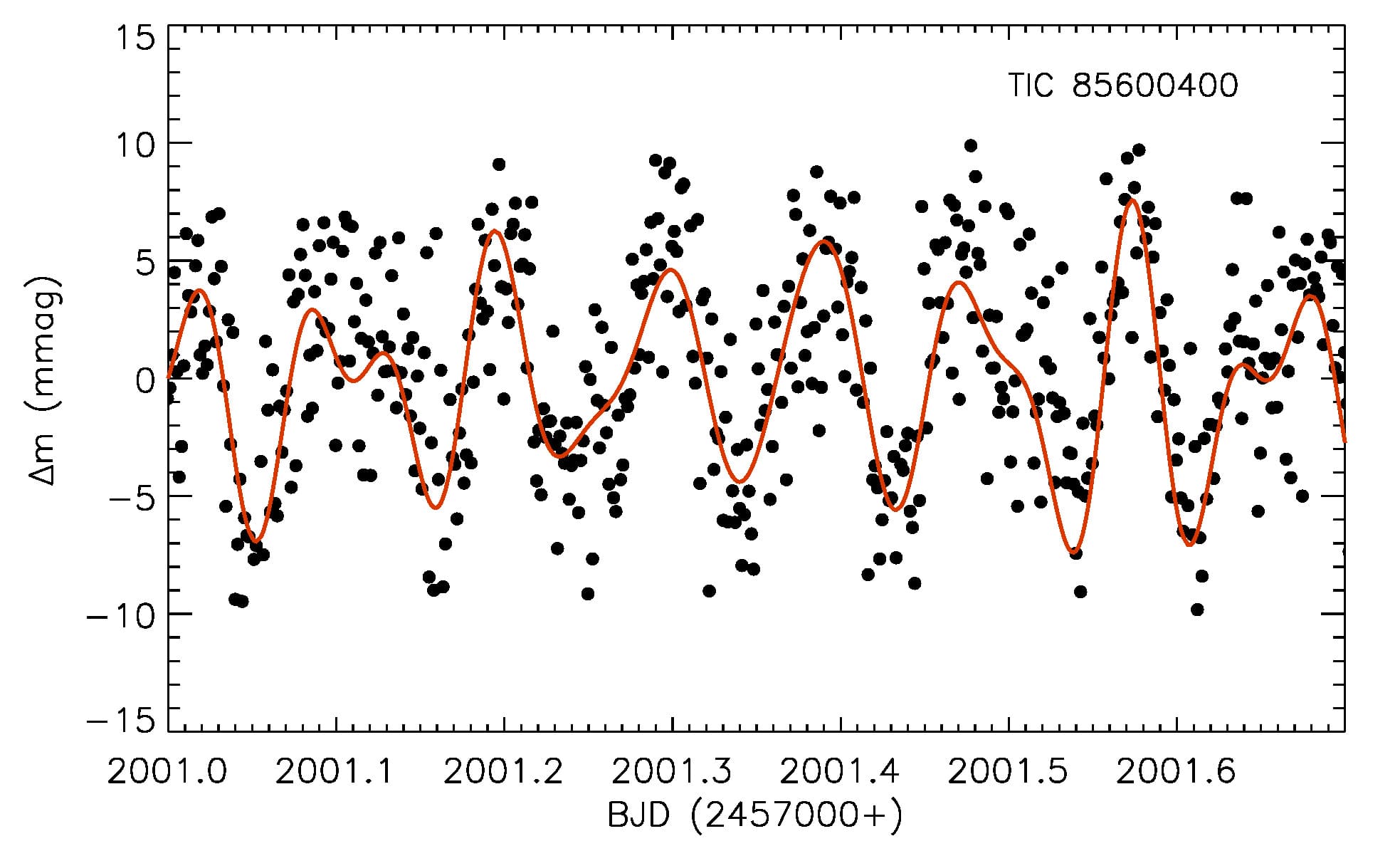}
 \end{minipage}
 \begin{minipage}[b]{0.24\textwidth}
  \includegraphics[height=3.5cm, width=1\textwidth]{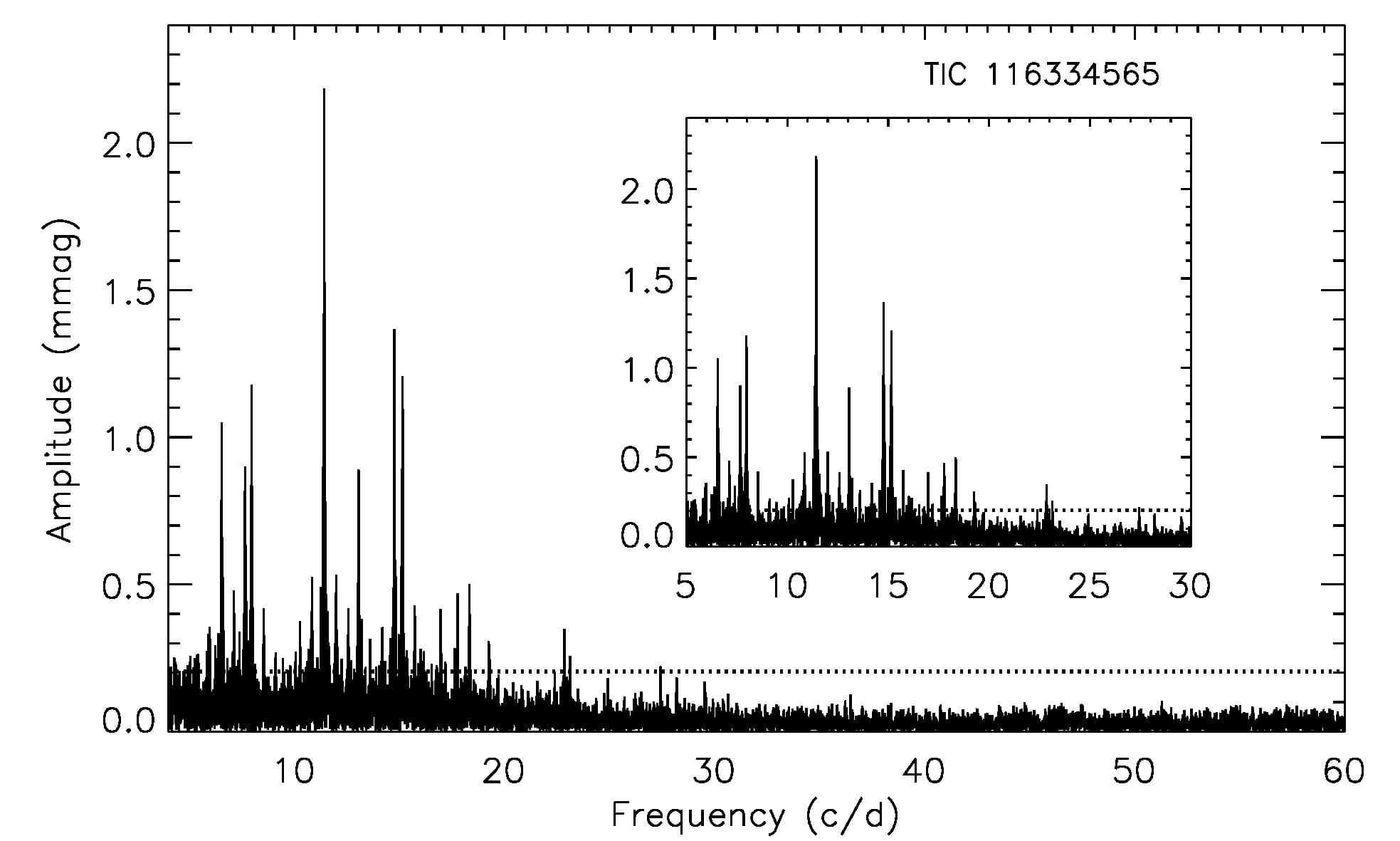}
 \end{minipage}
\begin{minipage}[b]{0.24\textwidth}
 \includegraphics[height=3.5cm, width=1\textwidth]{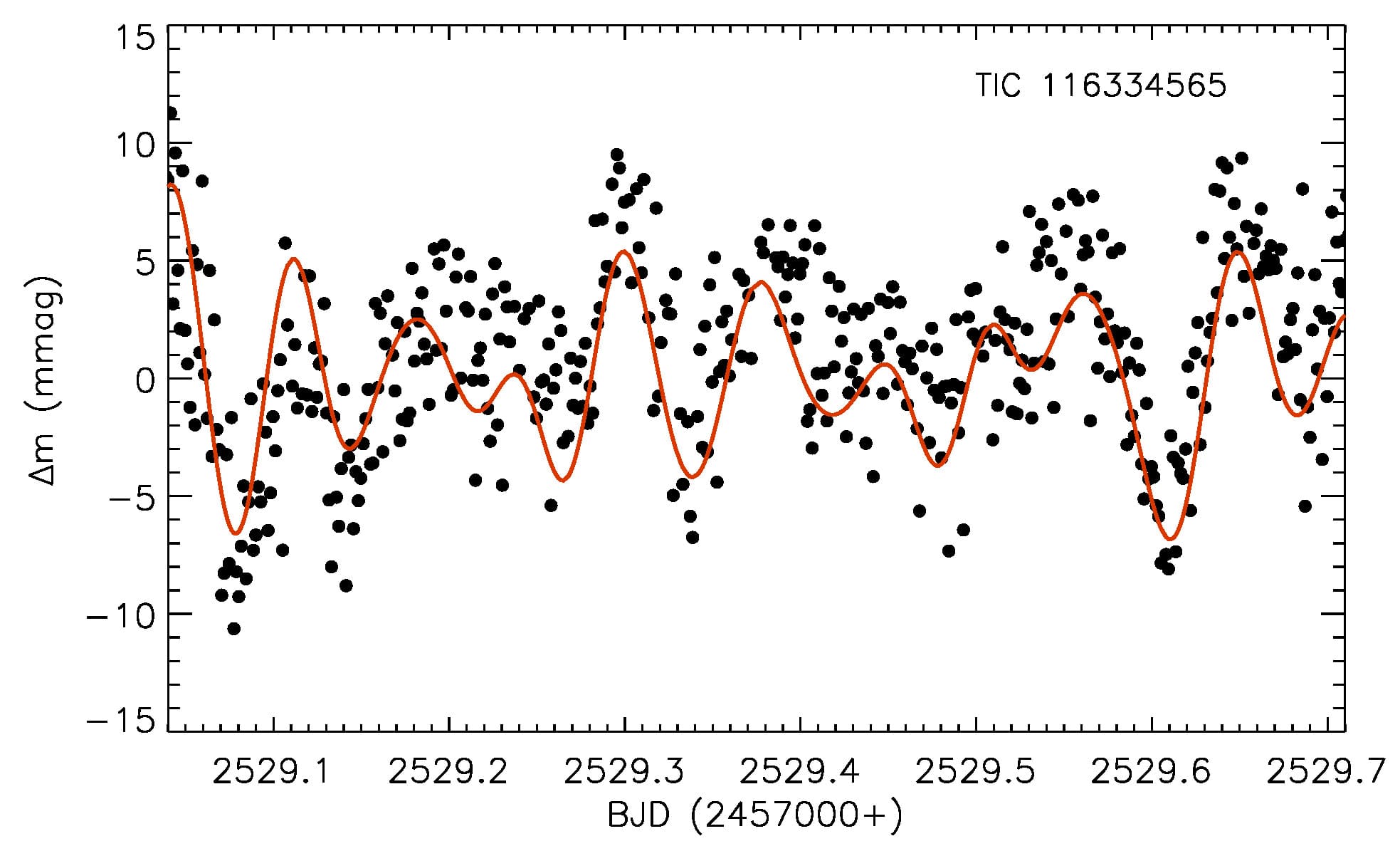}
\end{minipage}
 \begin{minipage}[b]{0.24\textwidth}
  \includegraphics[height=3.5cm, width=1\textwidth]{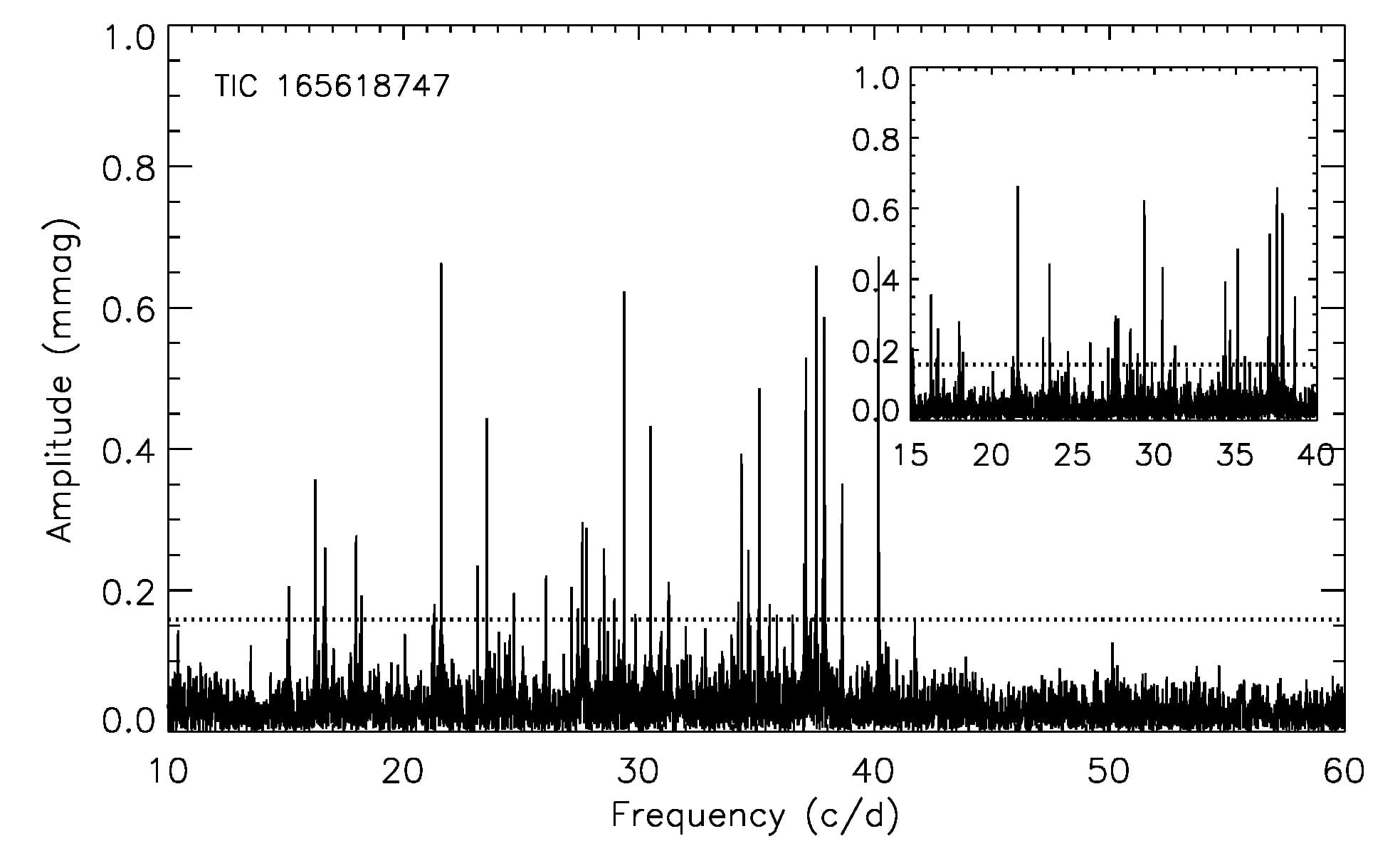}
  \end{minipage}
  \begin{minipage}[b]{0.24\textwidth}
  \includegraphics[height=3.5cm, width=1\textwidth]{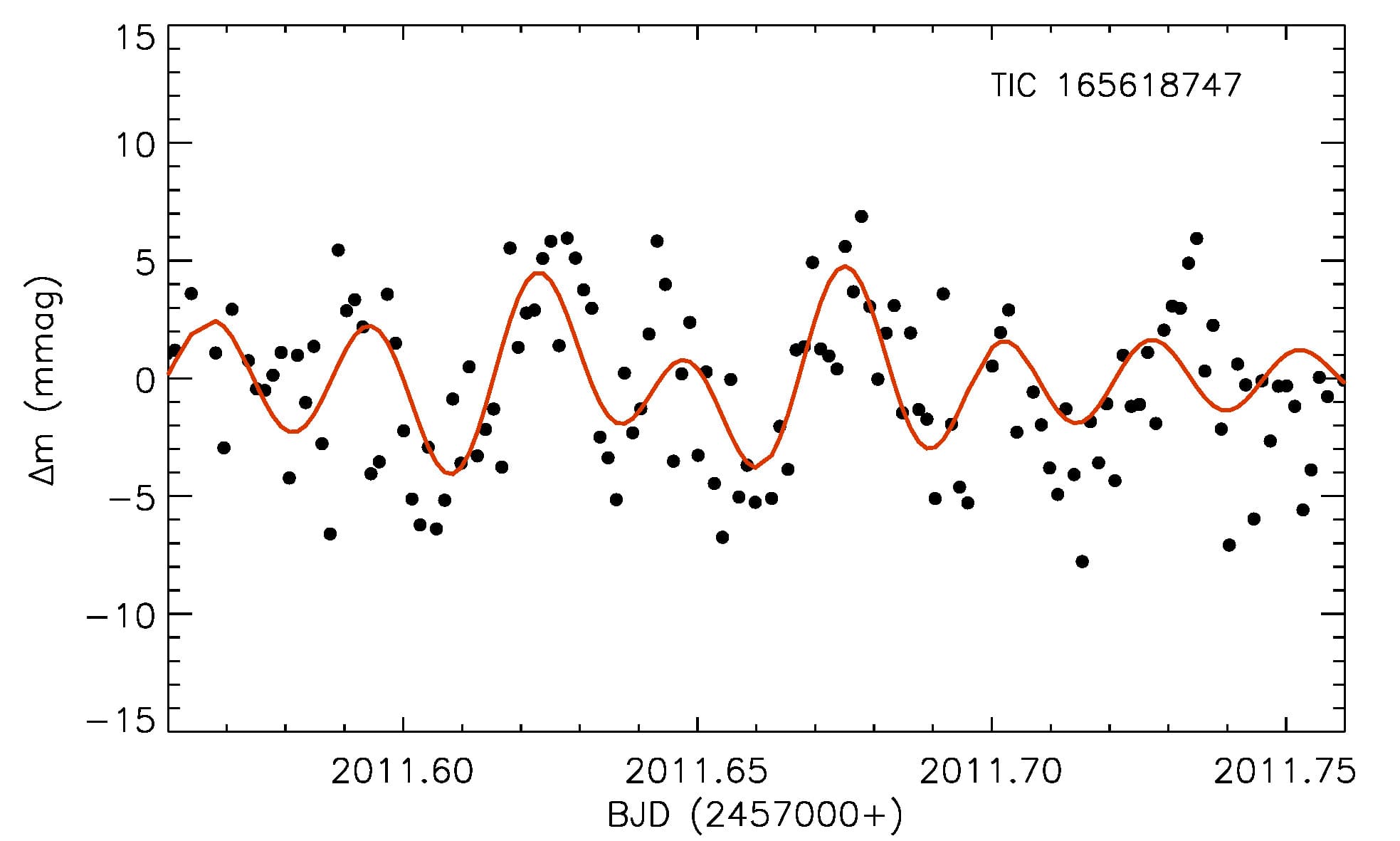}
  \end{minipage}
  \begin{minipage}[b]{0.24\textwidth}
  \includegraphics[height=3.5cm, width=1\textwidth]{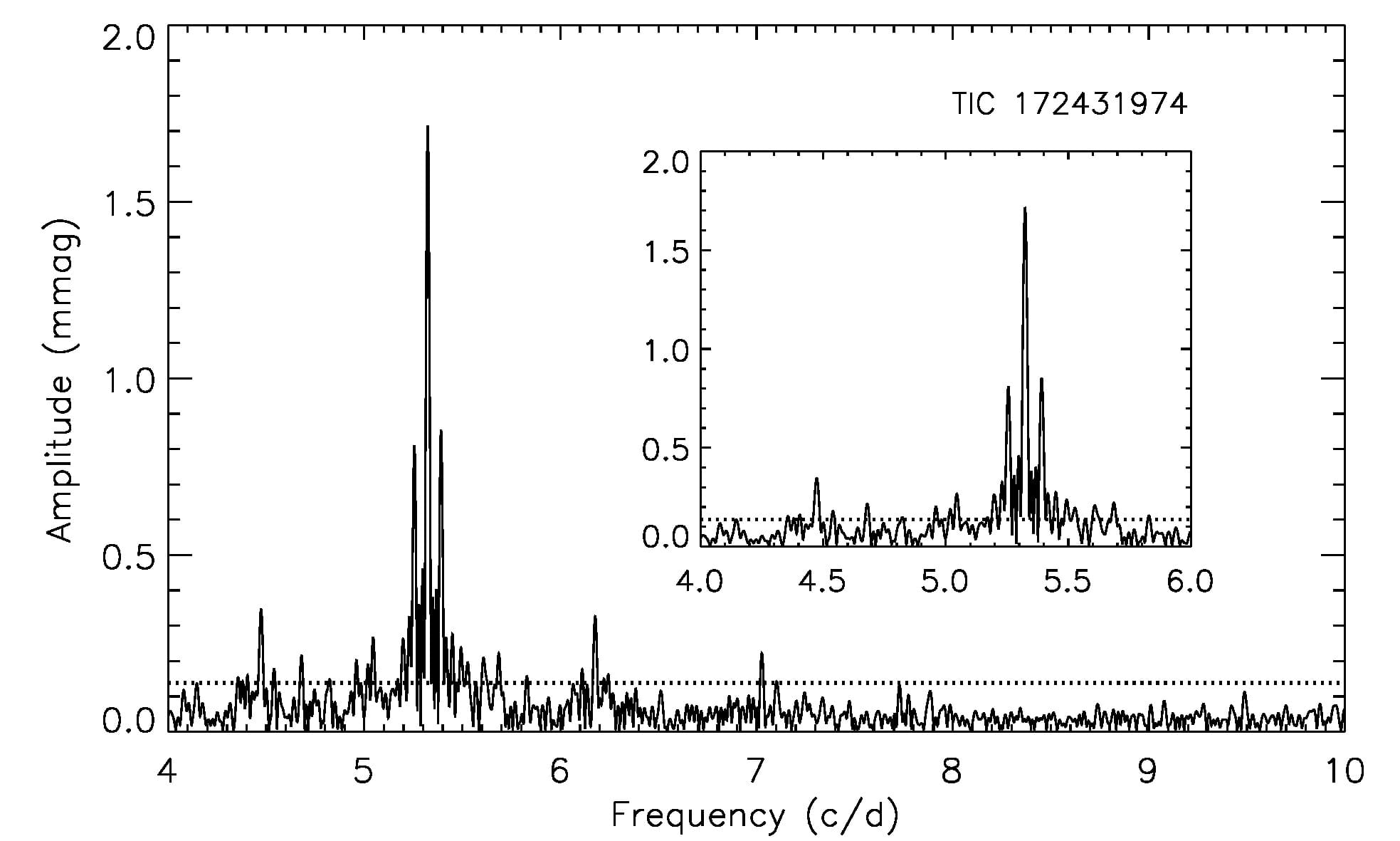}
 \end{minipage}
\begin{minipage}[b]{0.24\textwidth}
  \includegraphics[height=3.5cm, width=1\textwidth]{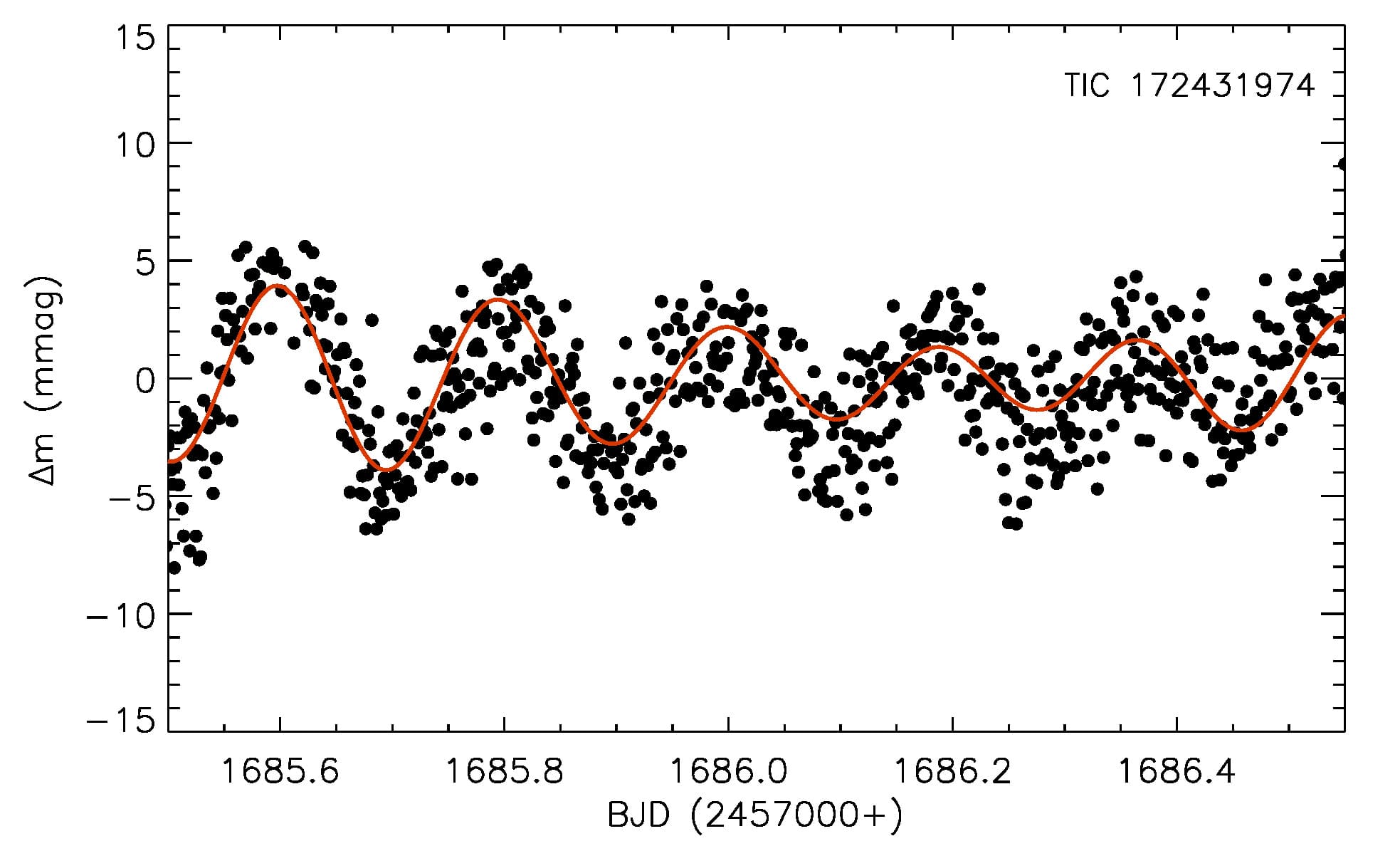}
 \end{minipage}
   \begin{minipage}[b]{0.24\textwidth}
  \includegraphics[height=3.5cm, width=1\textwidth]{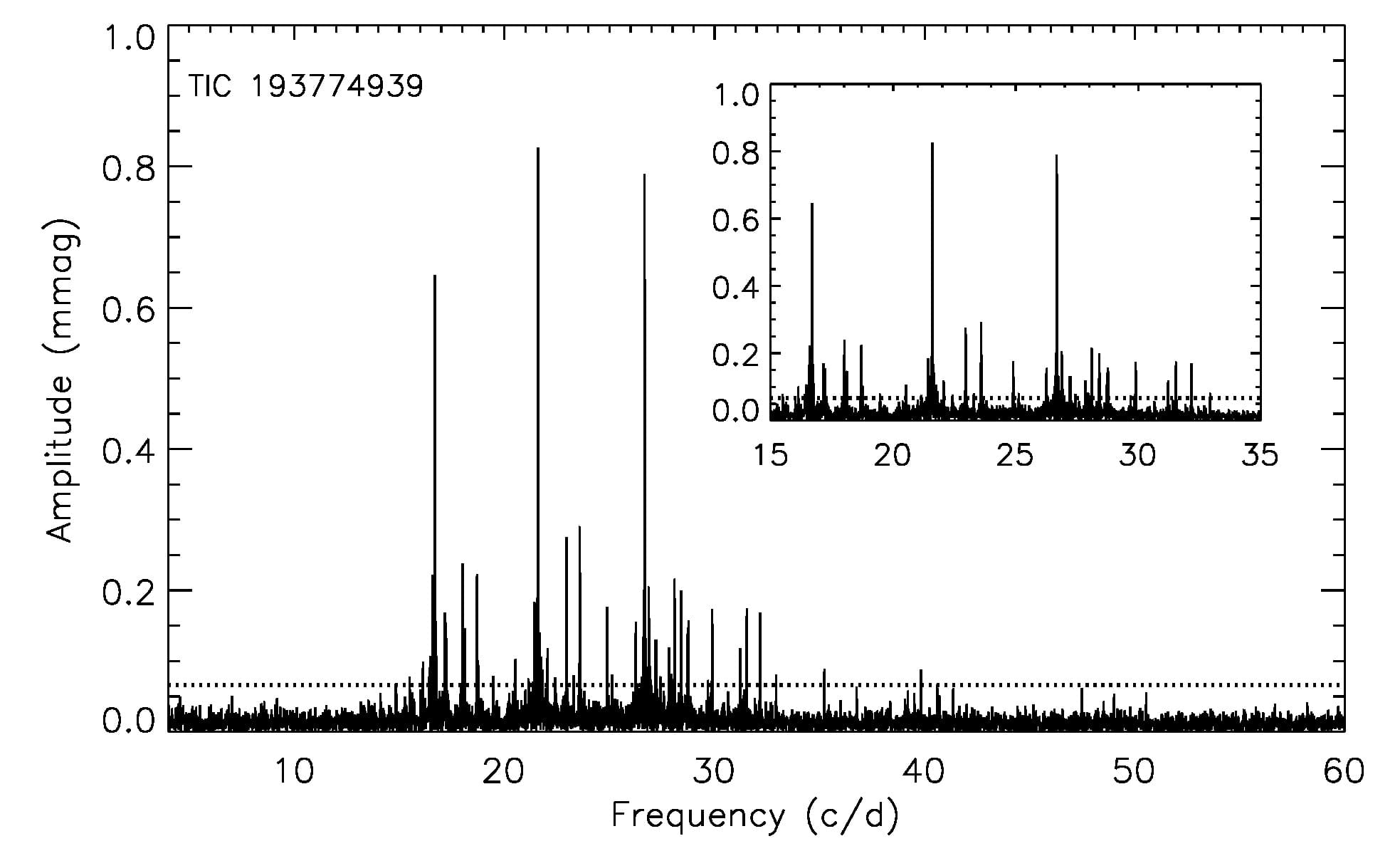}
 \end{minipage}
\begin{minipage}[b]{0.24\textwidth}
 \includegraphics[height=3.5cm, width=1\textwidth]{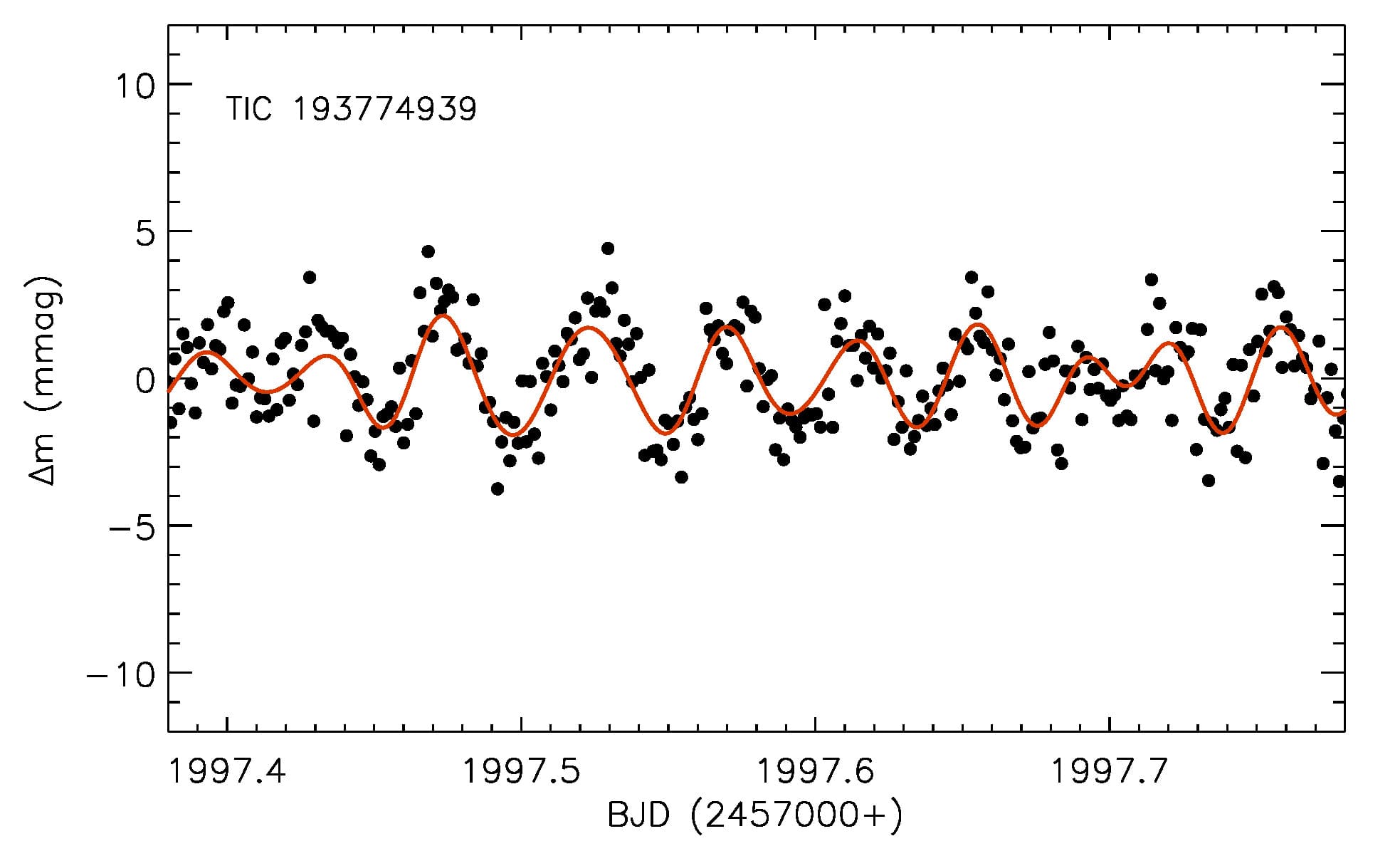}
 \end{minipage}
    \caption{Amplitude spectra of the targets and the theoretical frequency fit (red solid lines) to the observations. Dotted lines represent the 4.5-$\sigma$ level.}\label{spec2}
\end{figure*}

\setcounter{figure}{4}

\begin{figure*}
  \begin{minipage}[b]{0.24\textwidth}
  \includegraphics[height=3.5cm, width=1\textwidth]{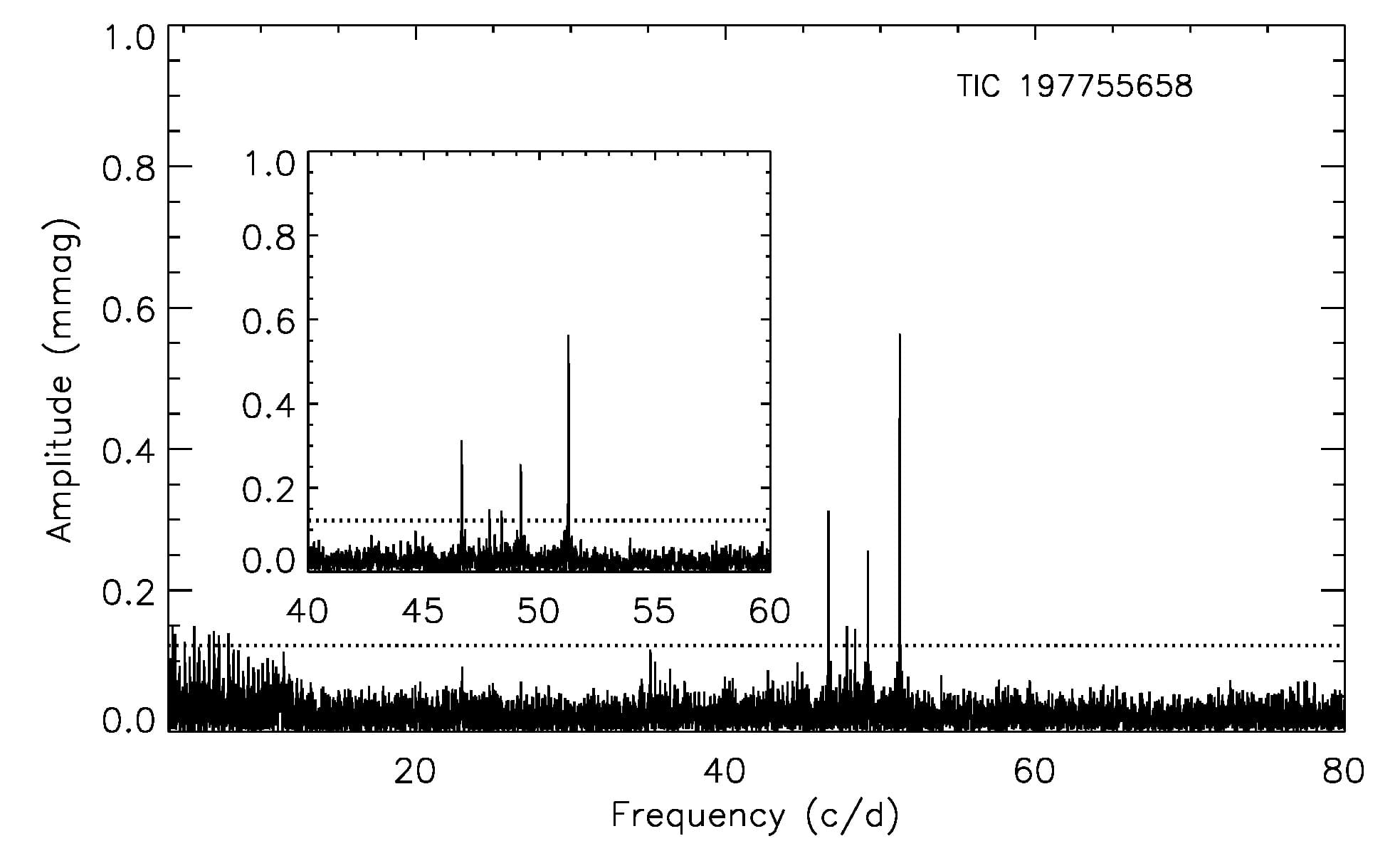}
  \end{minipage}
  \begin{minipage}[b]{0.24\textwidth}
  \includegraphics[height=3.5cm, width=1\textwidth]{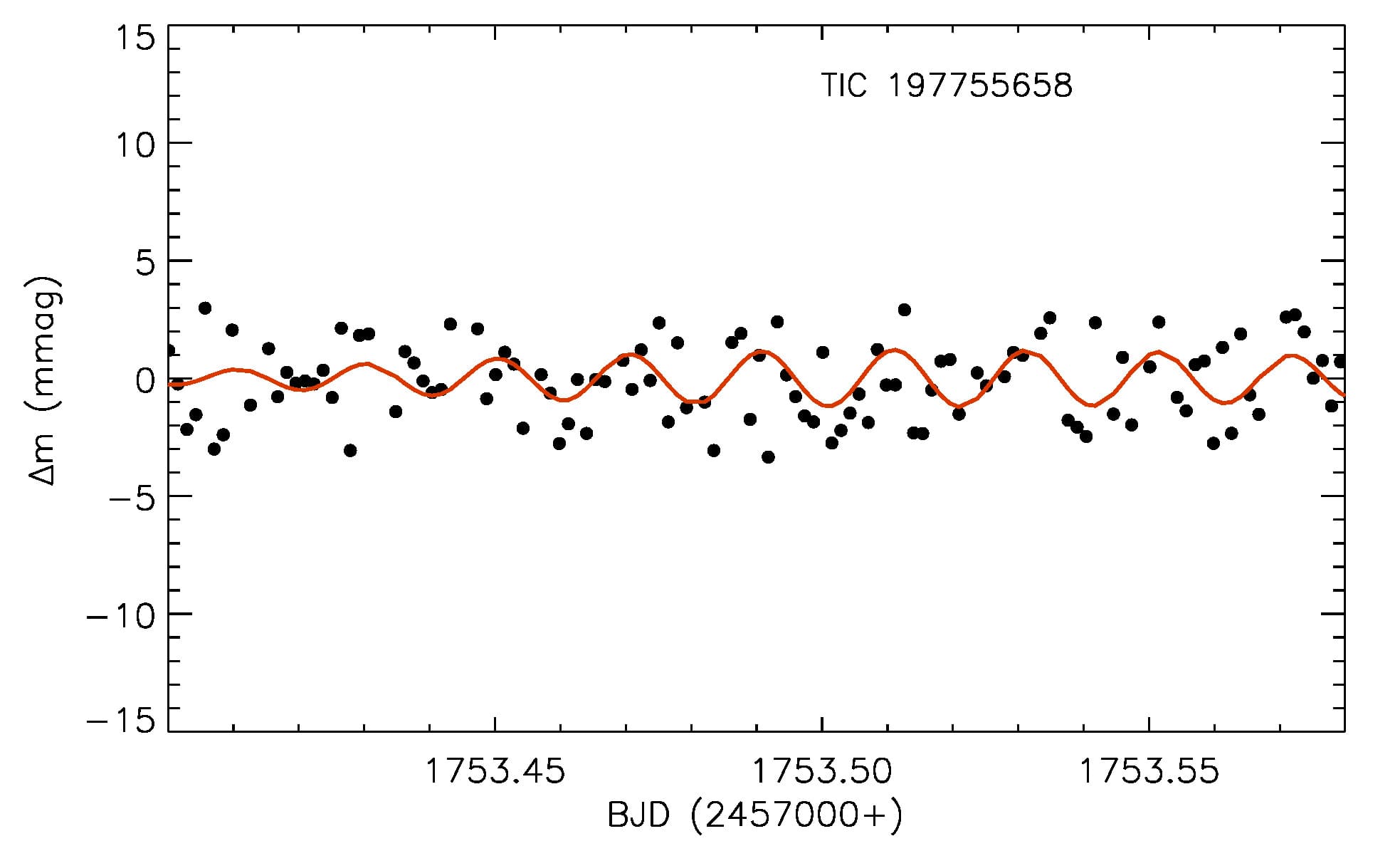}
  \end{minipage}
 \begin{minipage}[b]{0.24\textwidth}
  \includegraphics[height=3.5cm, width=1.0\textwidth]{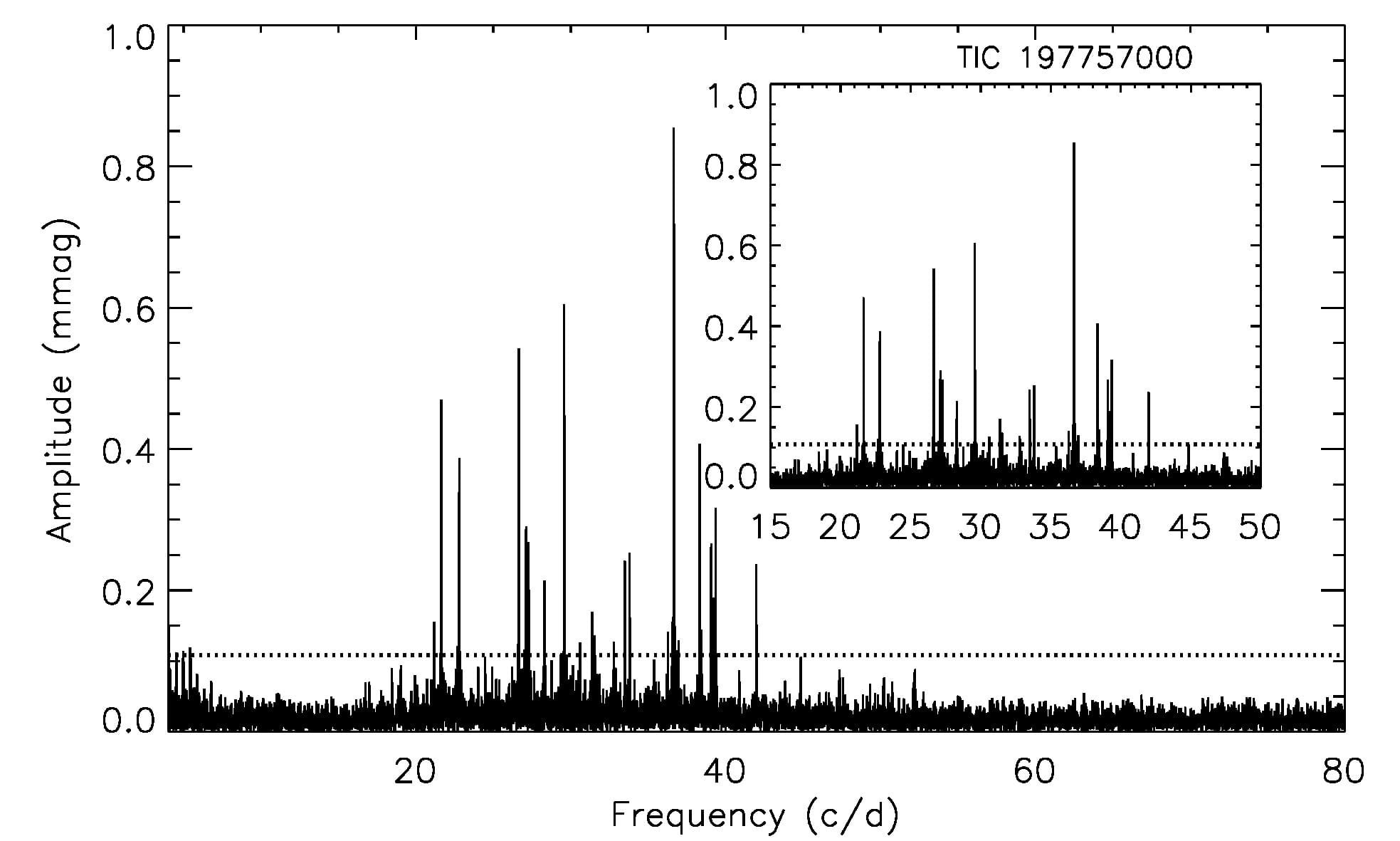}
 \end{minipage}
\begin{minipage}[b]{0.24\textwidth}
  \includegraphics[height=3.5cm, width=1.0\textwidth]{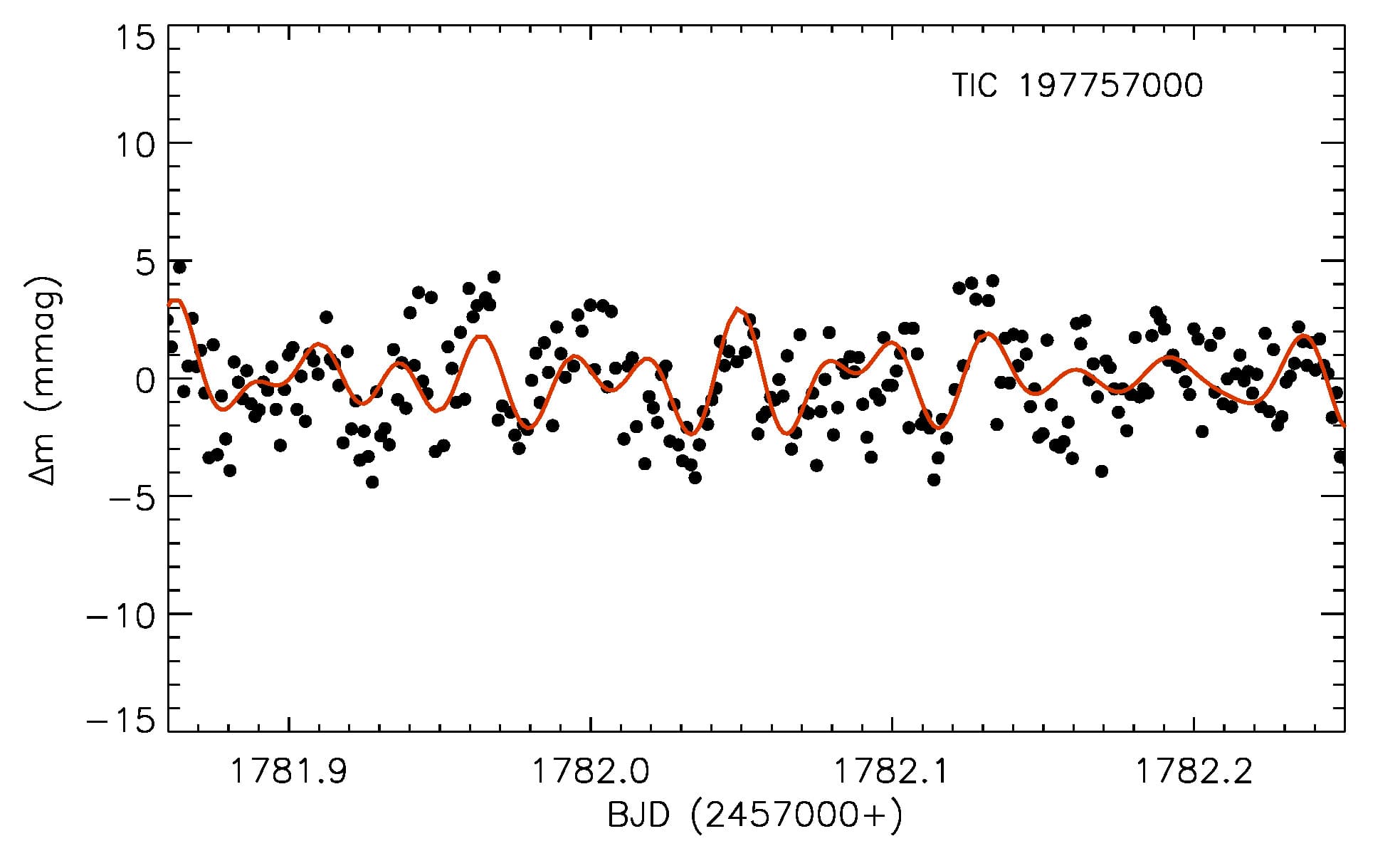}
 \end{minipage}
 \begin{minipage}[b]{0.24\textwidth}
  \includegraphics[height=3.5cm, width=1.0\textwidth]{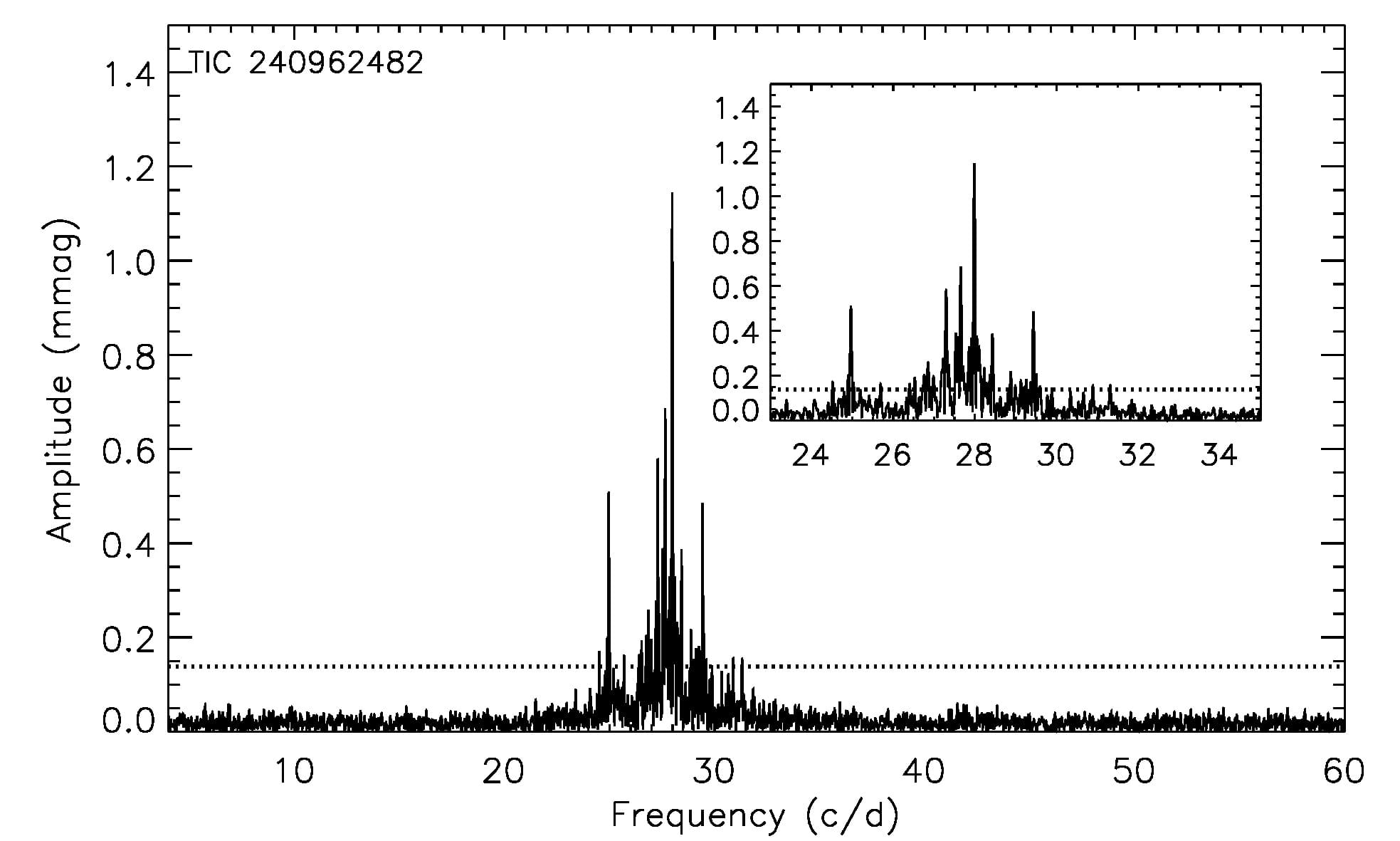}
  \end{minipage}
 \begin{minipage}[b]{0.24\textwidth}
 \includegraphics[height=3.5cm, width=1.0\textwidth]{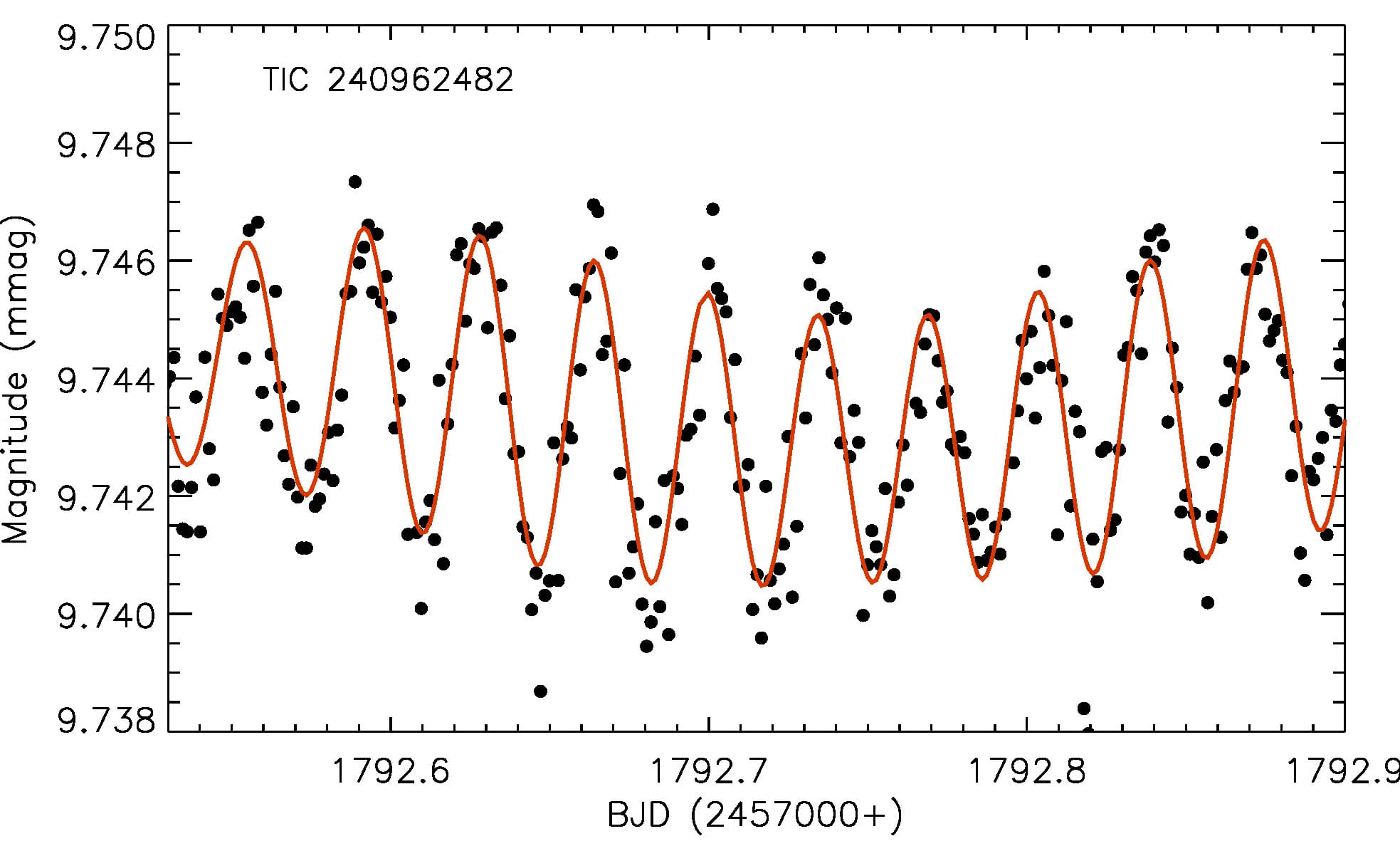}
 \end{minipage}
 \begin{minipage}[b]{0.24\textwidth}
  \includegraphics[height=3.5cm, width=1\textwidth]{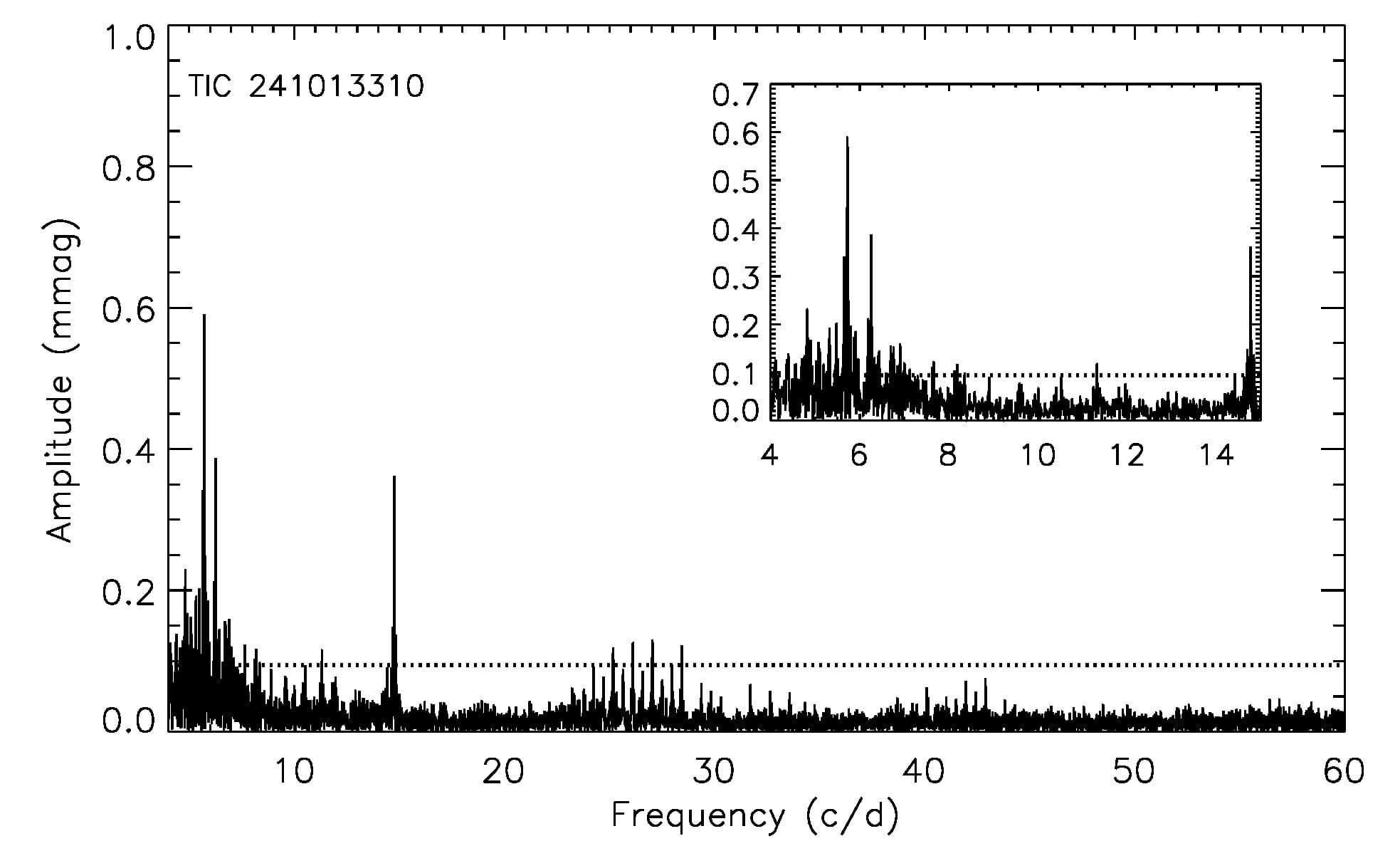}
  \end{minipage}
 \begin{minipage}[b]{0.24\textwidth}
 \includegraphics[height=3.5cm, width=1\textwidth]{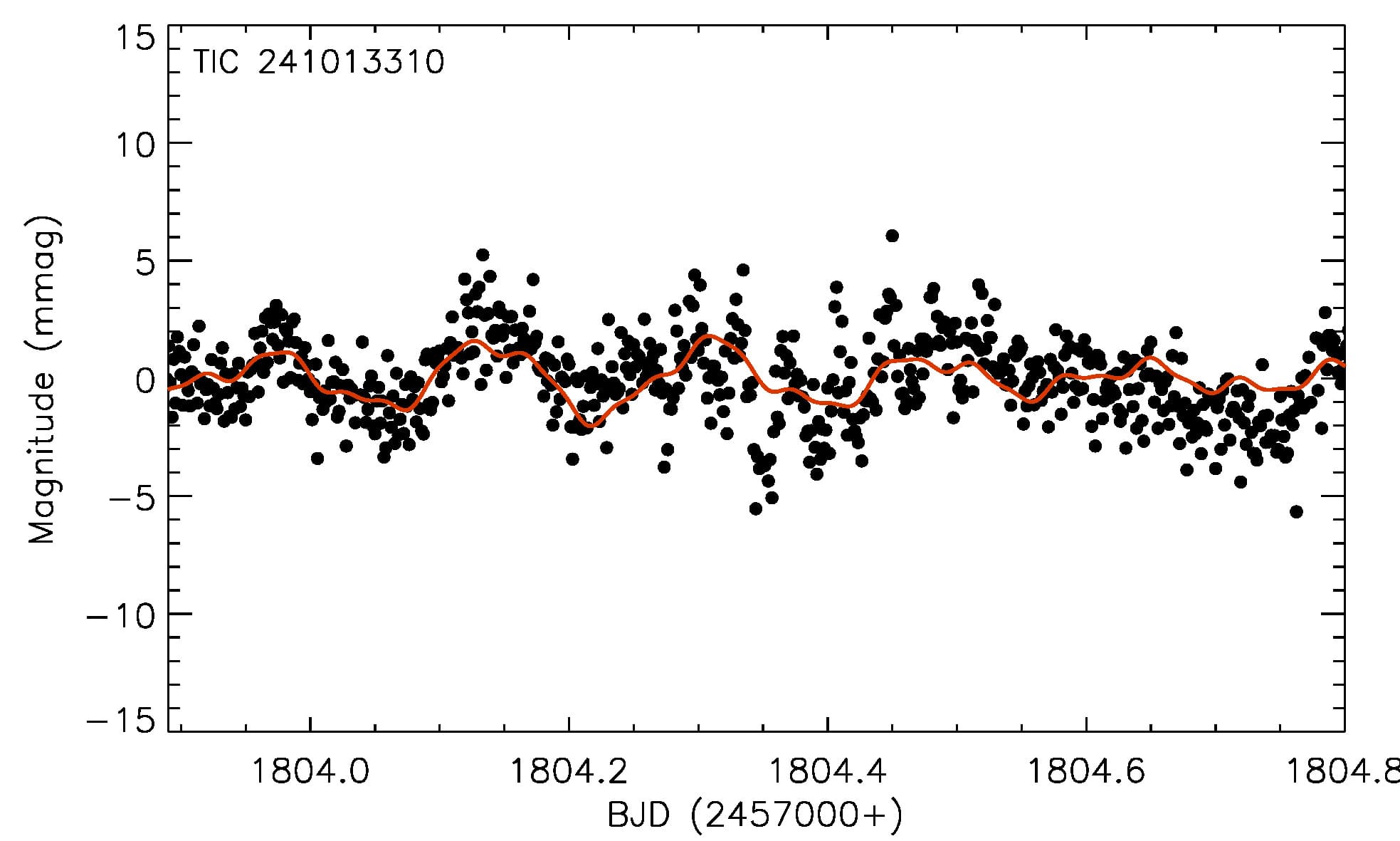}
 \end{minipage}
  \begin{minipage}[b]{0.24\textwidth}
  \includegraphics[height=3.5cm, width=1\textwidth]{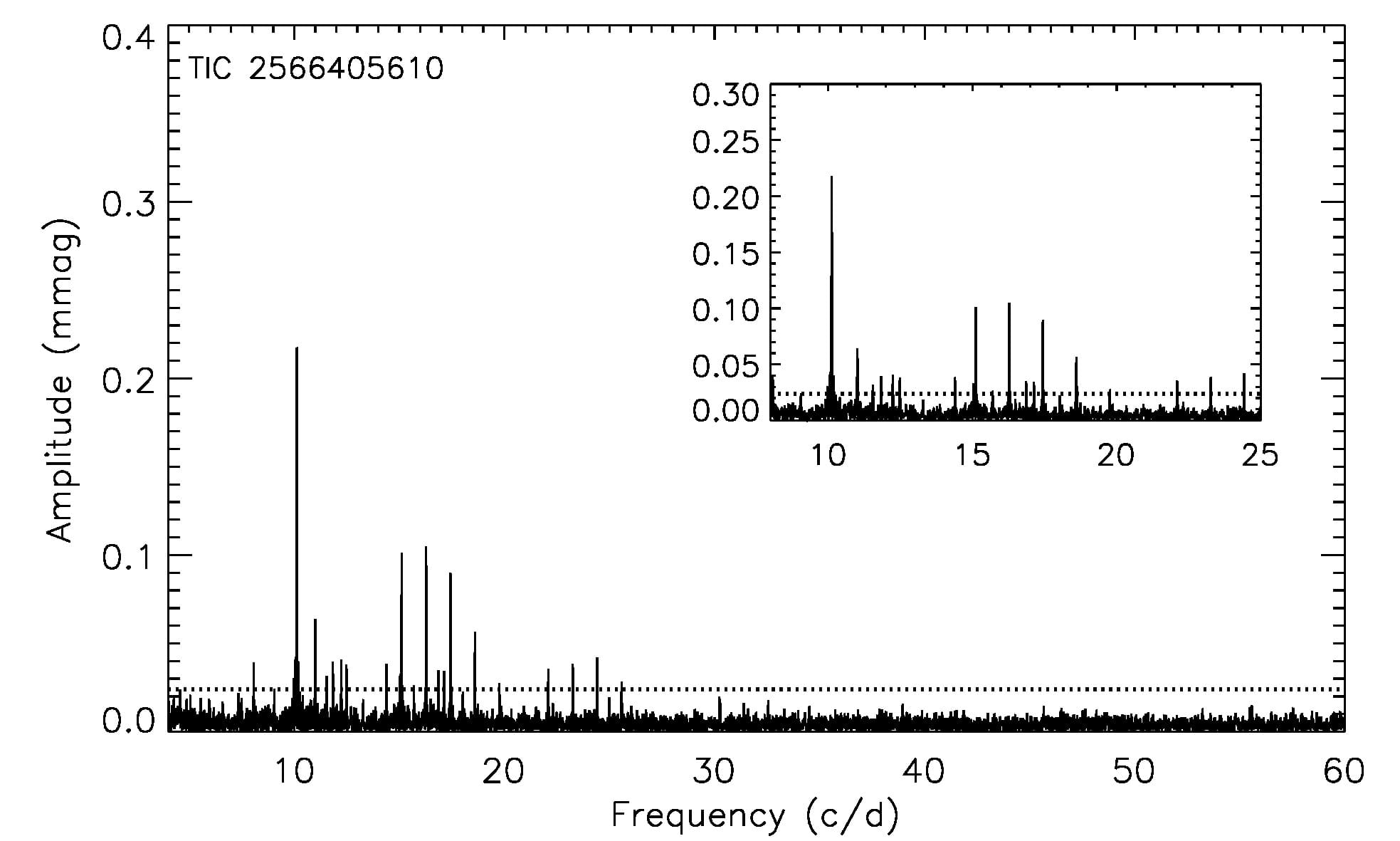}
 \end{minipage}
 \begin{minipage}[b]{0.24\textwidth}
 \includegraphics[height=3.5cm, width=1\textwidth]{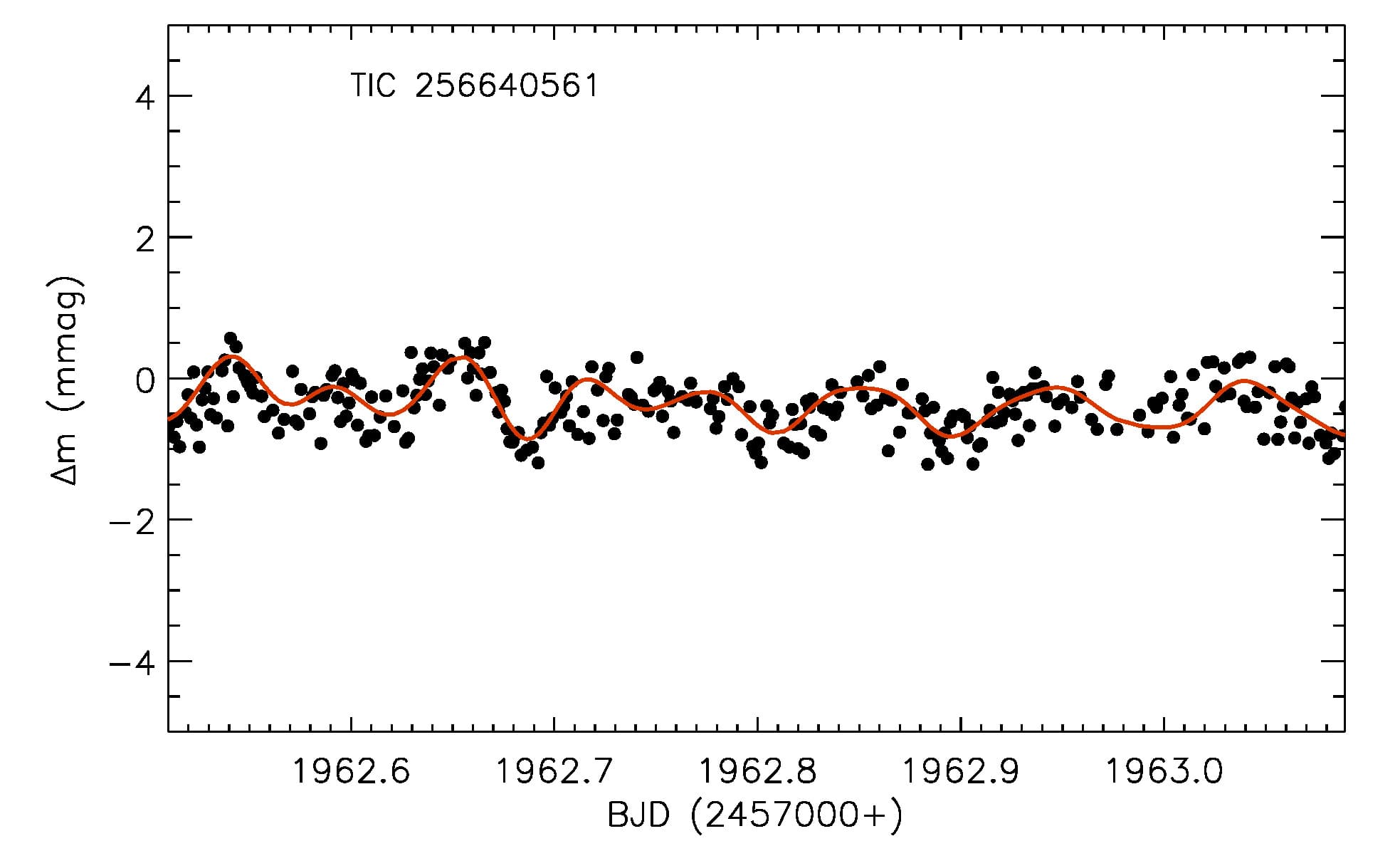}
 \end{minipage}
  \begin{minipage}[b]{0.24\textwidth}
  \includegraphics[height=3.5cm, width=1\textwidth]{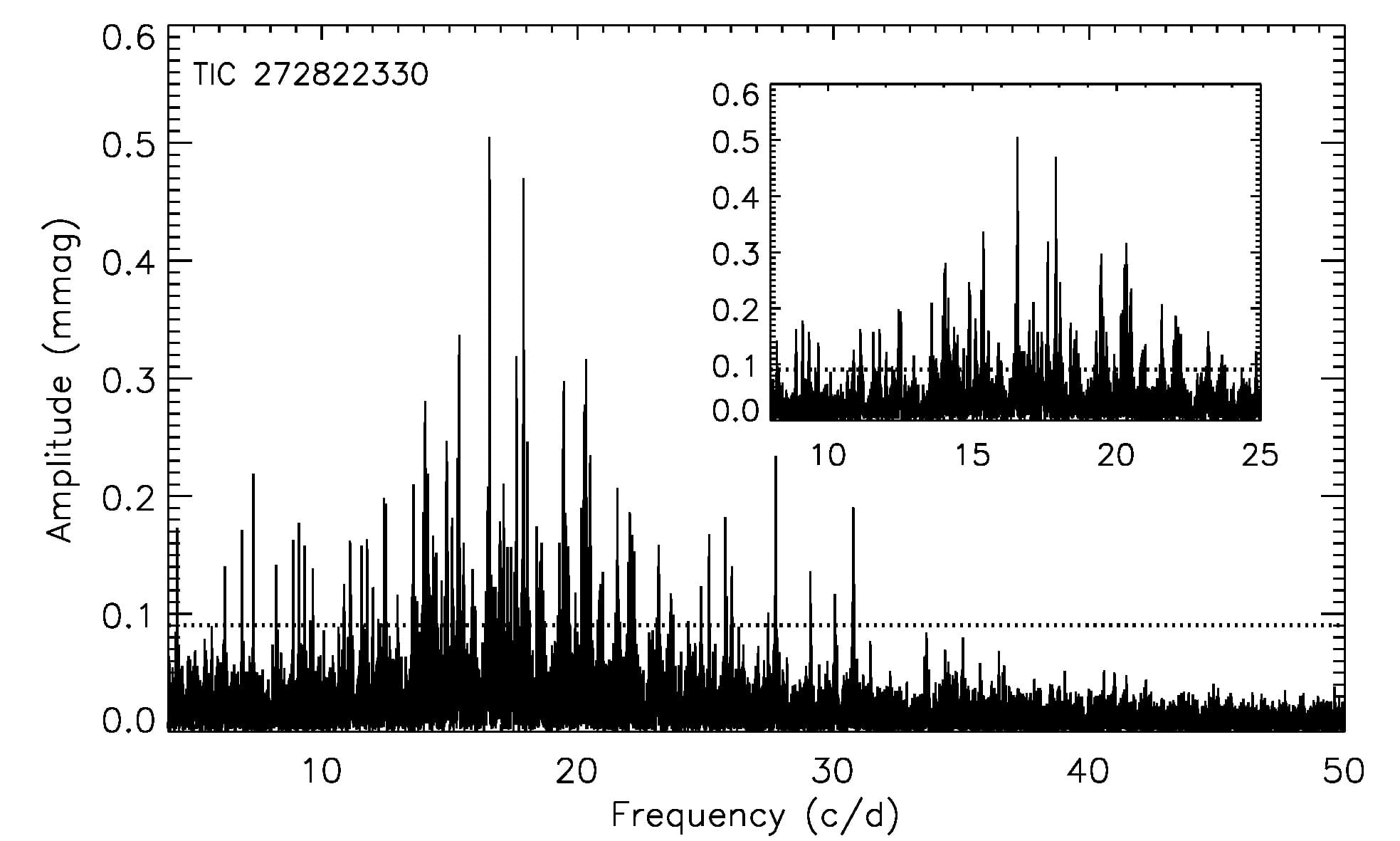}
  \end{minipage}
 \begin{minipage}[b]{0.24\textwidth}
 \includegraphics[height=3.5cm, width=1\textwidth]{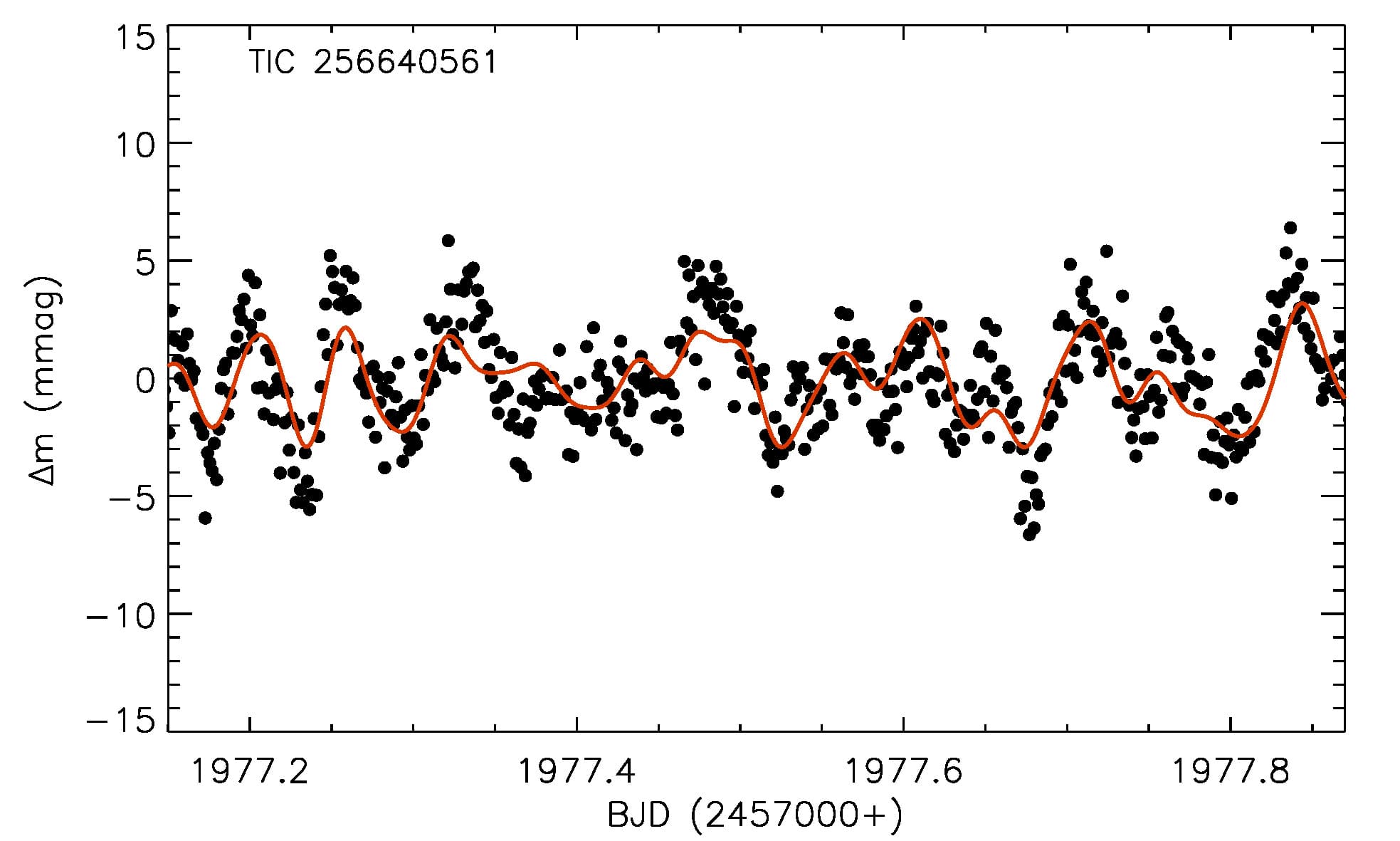}
 \end{minipage}
  \begin{minipage}[b]{0.24\textwidth}
  \includegraphics[height=3.5cm, width=1\textwidth]{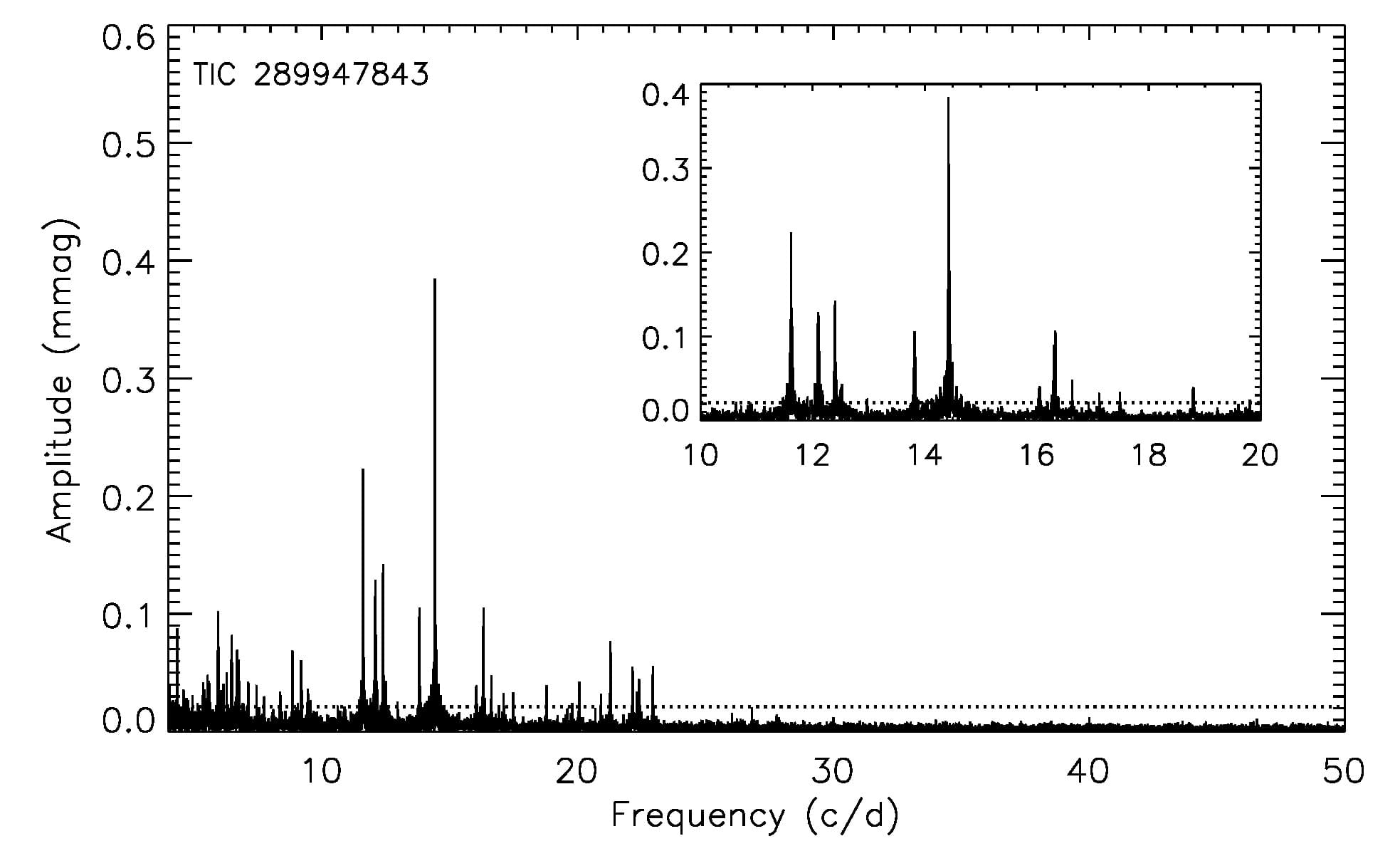}
 \end{minipage}
 \begin{minipage}[b]{0.24\textwidth}
  \includegraphics[height=3.5cm, width=1\textwidth]{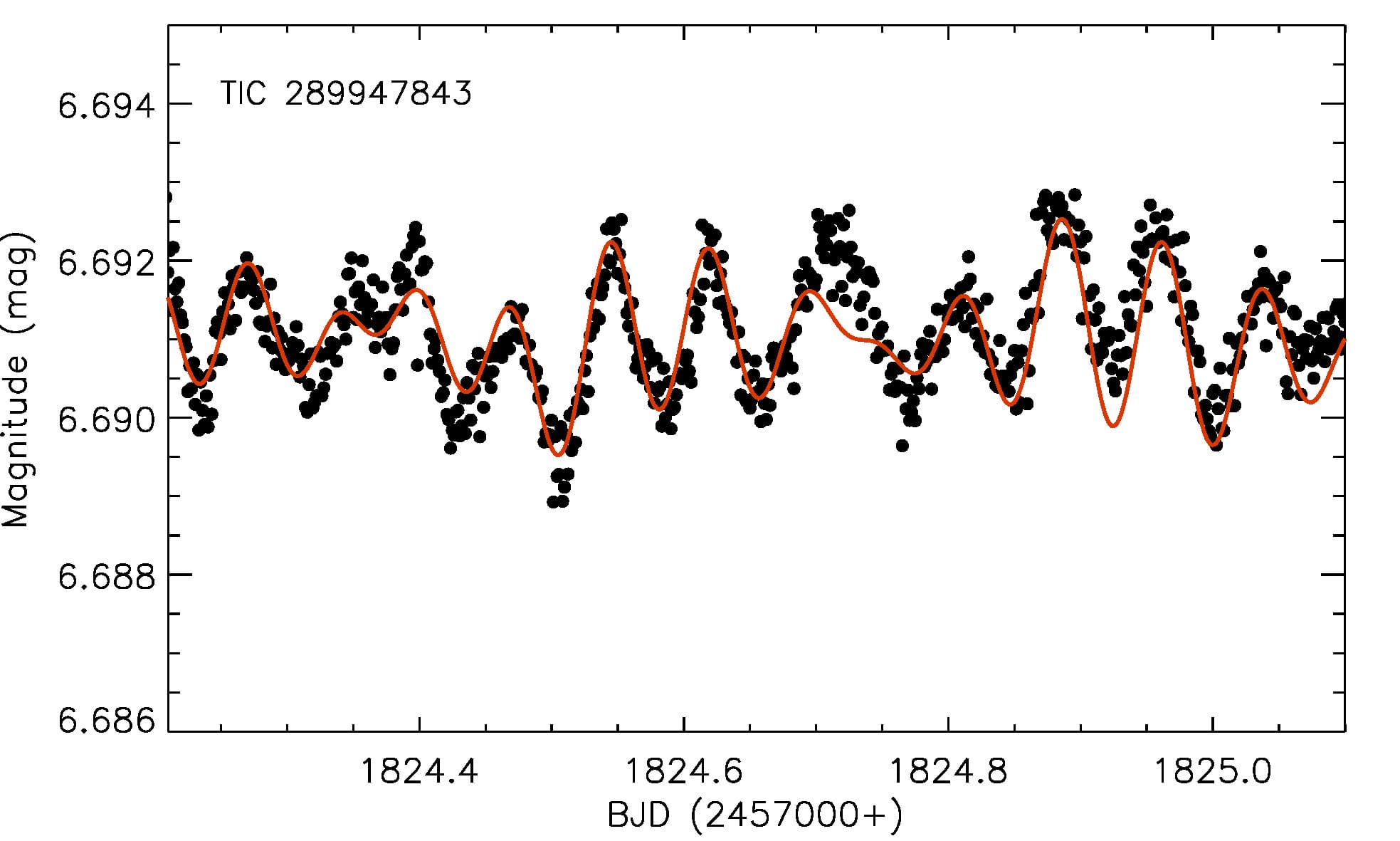}
 \end{minipage}
  \begin{minipage}[b]{0.24\textwidth}
  \includegraphics[height=3.5cm, width=1\textwidth]{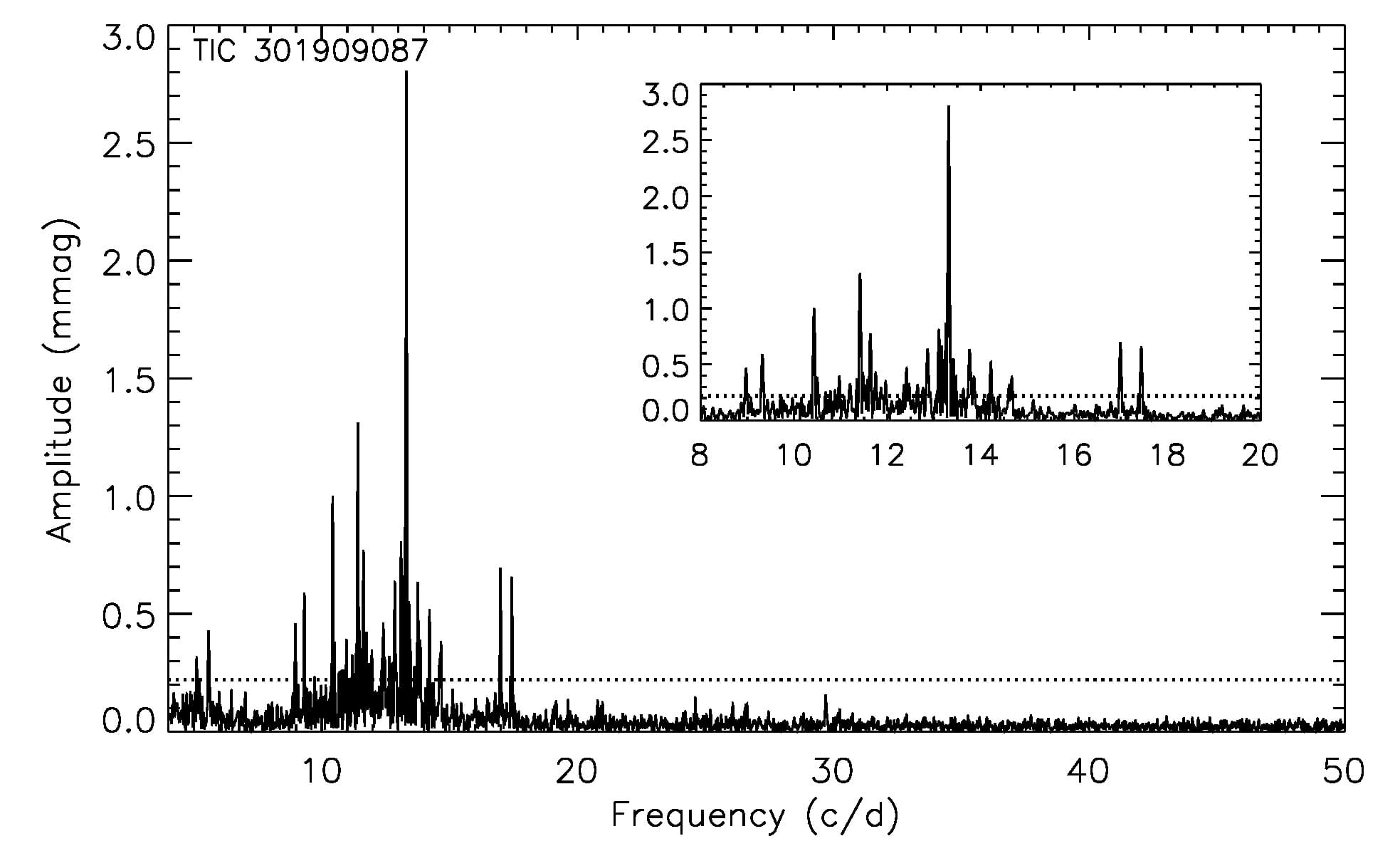}
 \end{minipage}
\begin{minipage}[b]{0.24\textwidth}
 \includegraphics[height=3.5cm, width=1\textwidth]{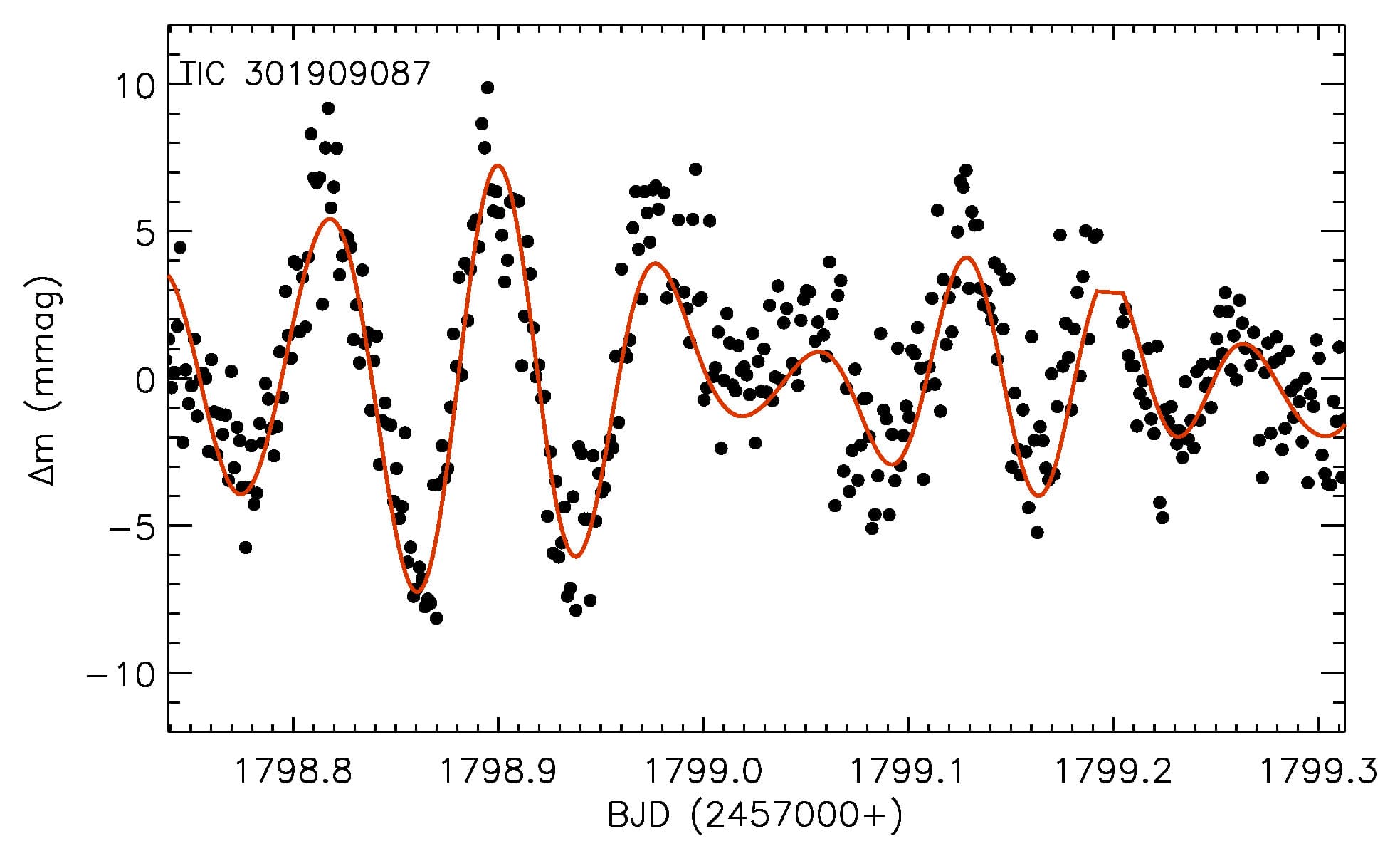}
 \end{minipage}
  \begin{minipage}[b]{0.24\textwidth}
  \includegraphics[height=3.5cm, width=1\textwidth]{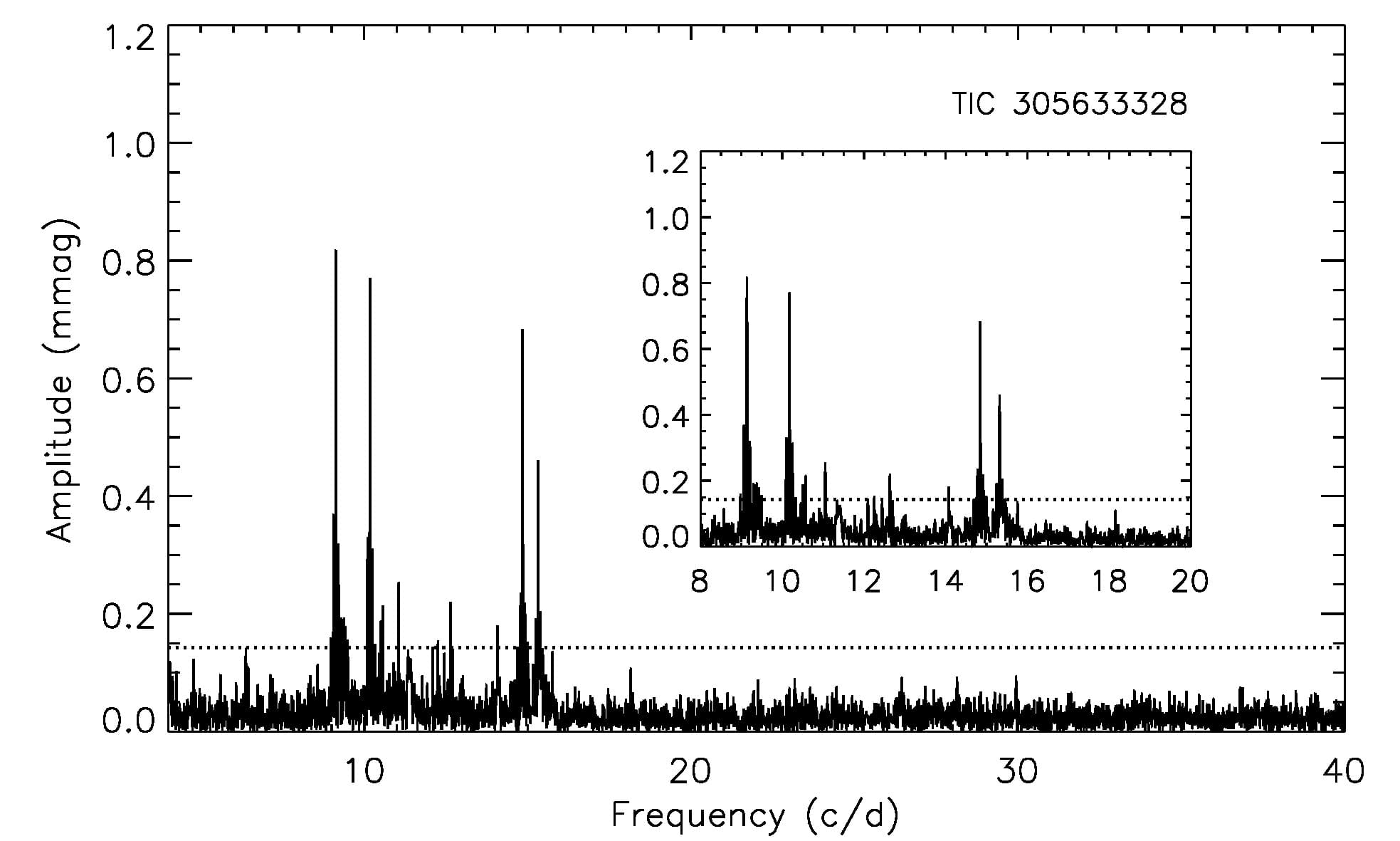}
  \end{minipage}
 \begin{minipage}[b]{0.24\textwidth}
 \includegraphics[height=3.5cm, width=1\textwidth]{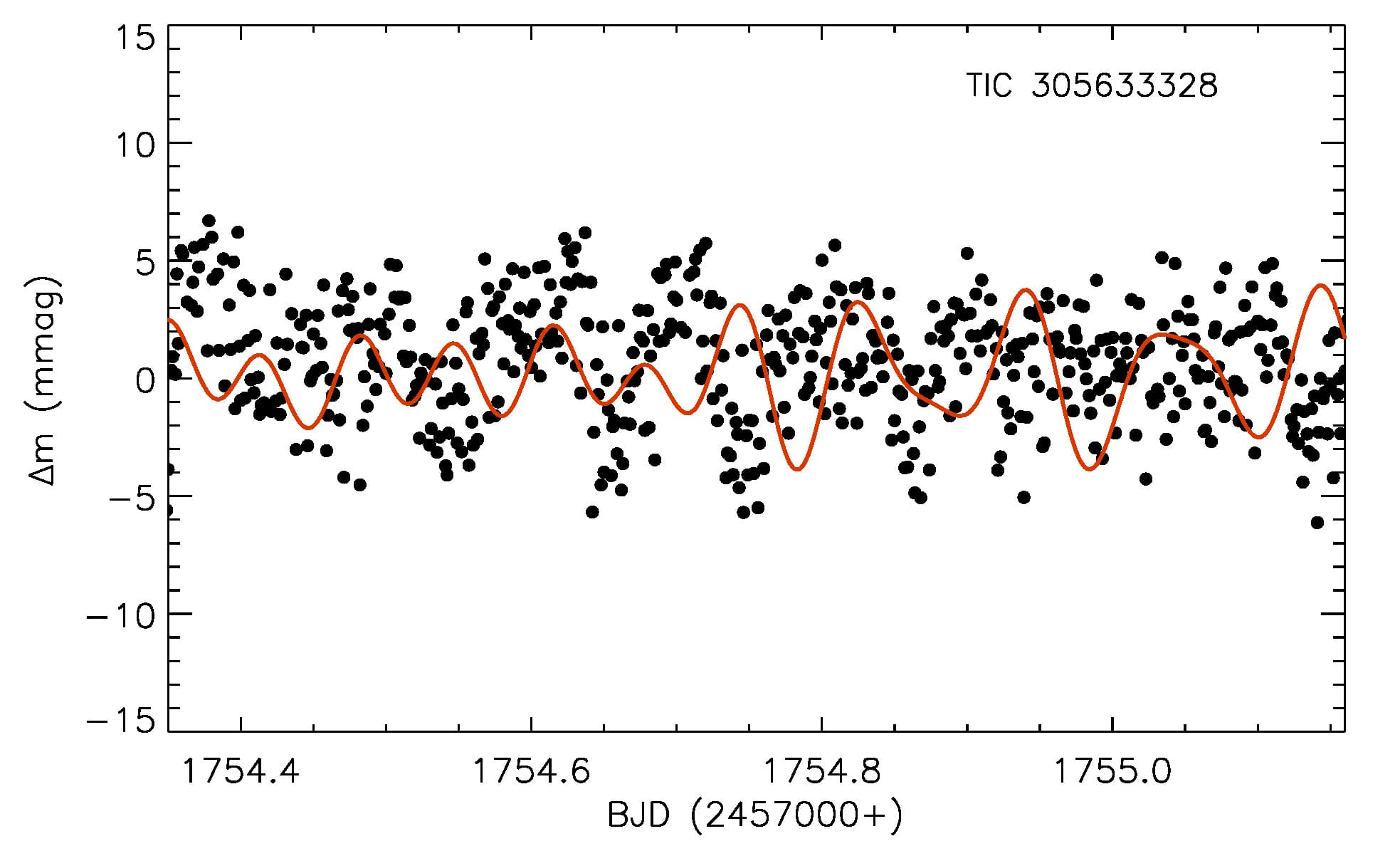}
 \end{minipage}
  \begin{minipage}[b]{0.24\textwidth}
  \includegraphics[height=3.5cm, width=1.0\textwidth]{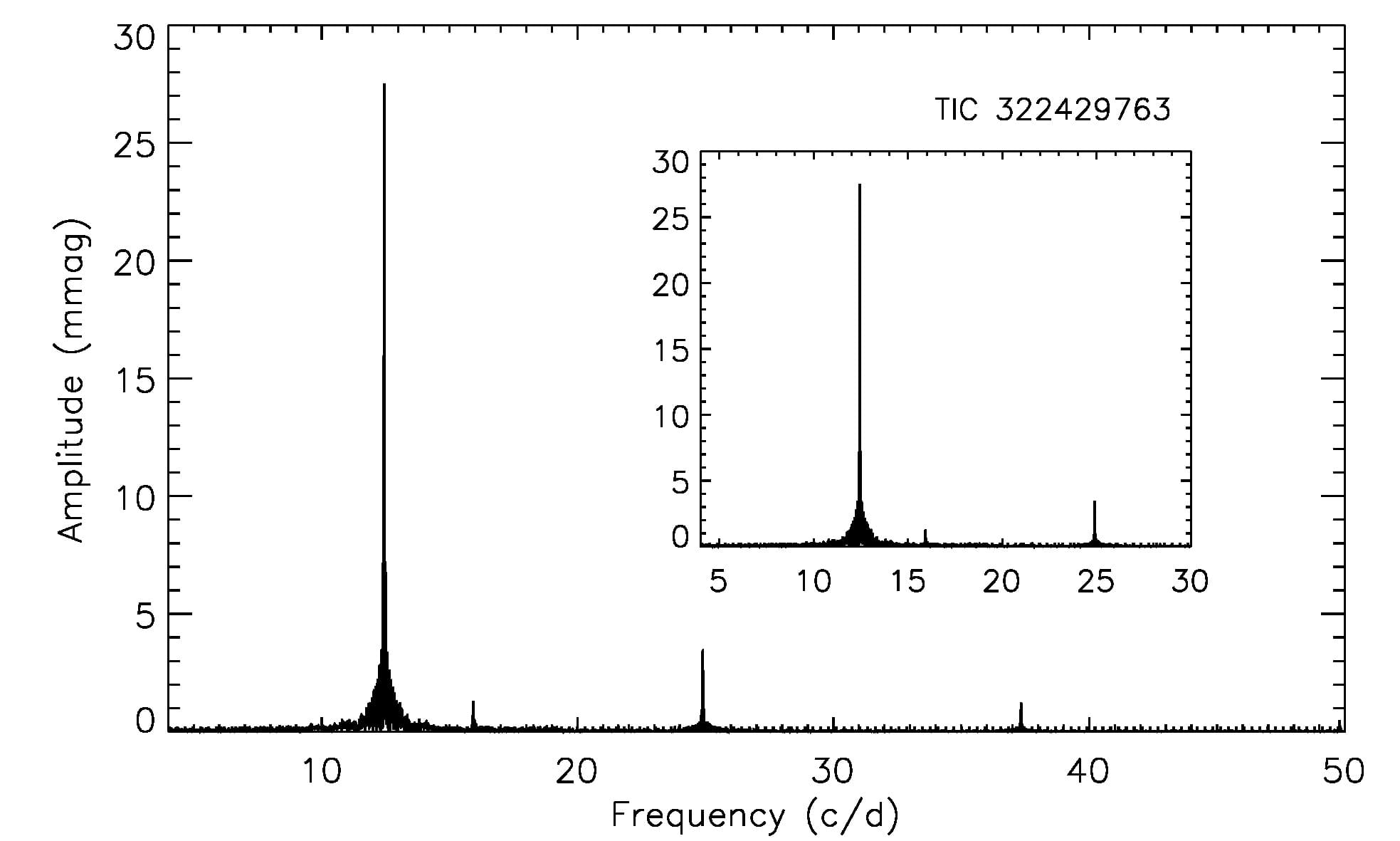}
 \end{minipage}
 \begin{minipage}[b]{0.24\textwidth}
 \includegraphics[height=3.5cm, width=1.0\textwidth]{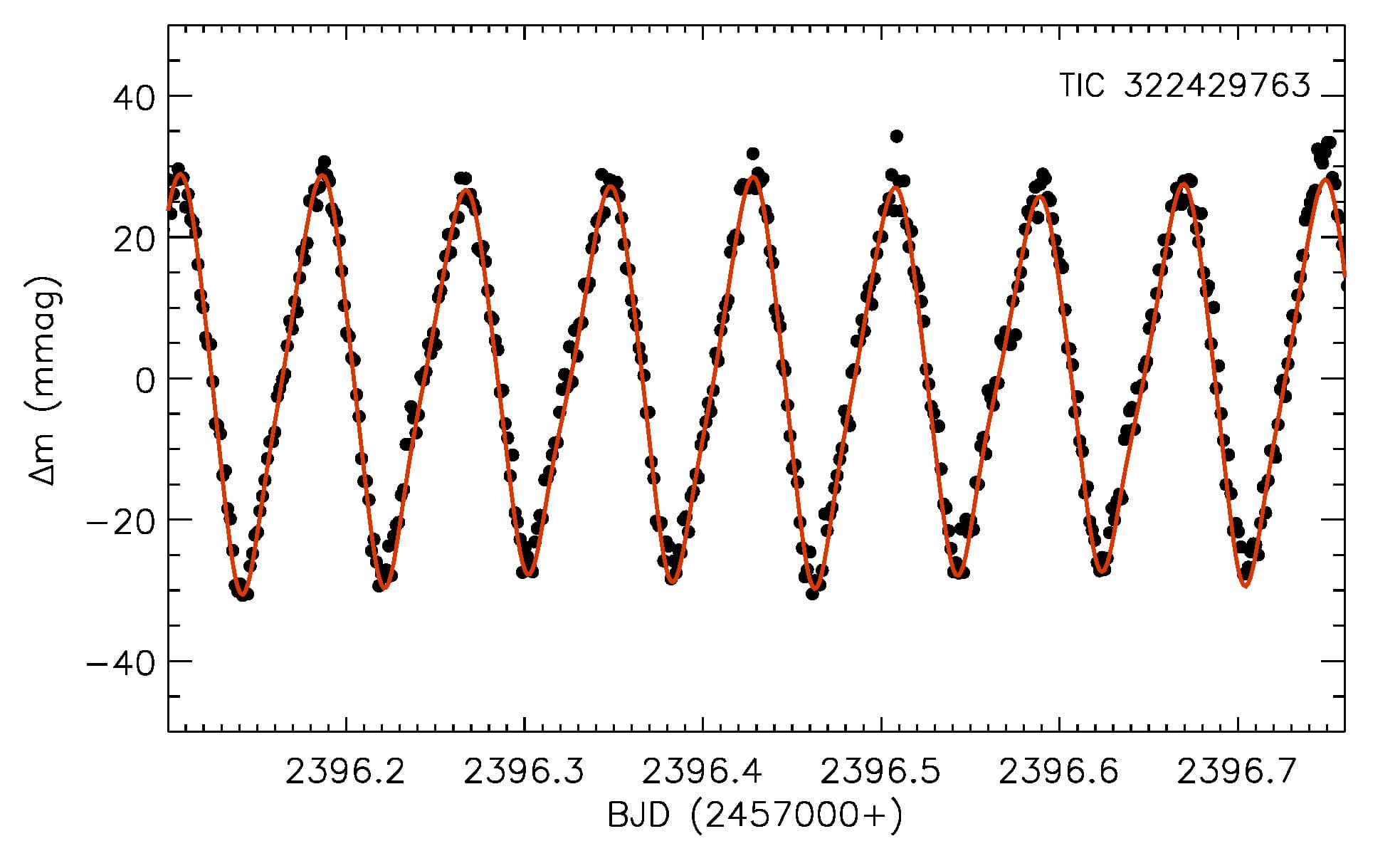}
 \end{minipage}
 \begin{minipage}[b]{0.24\textwidth}
  \includegraphics[height=3.5cm, width=1.0\textwidth]{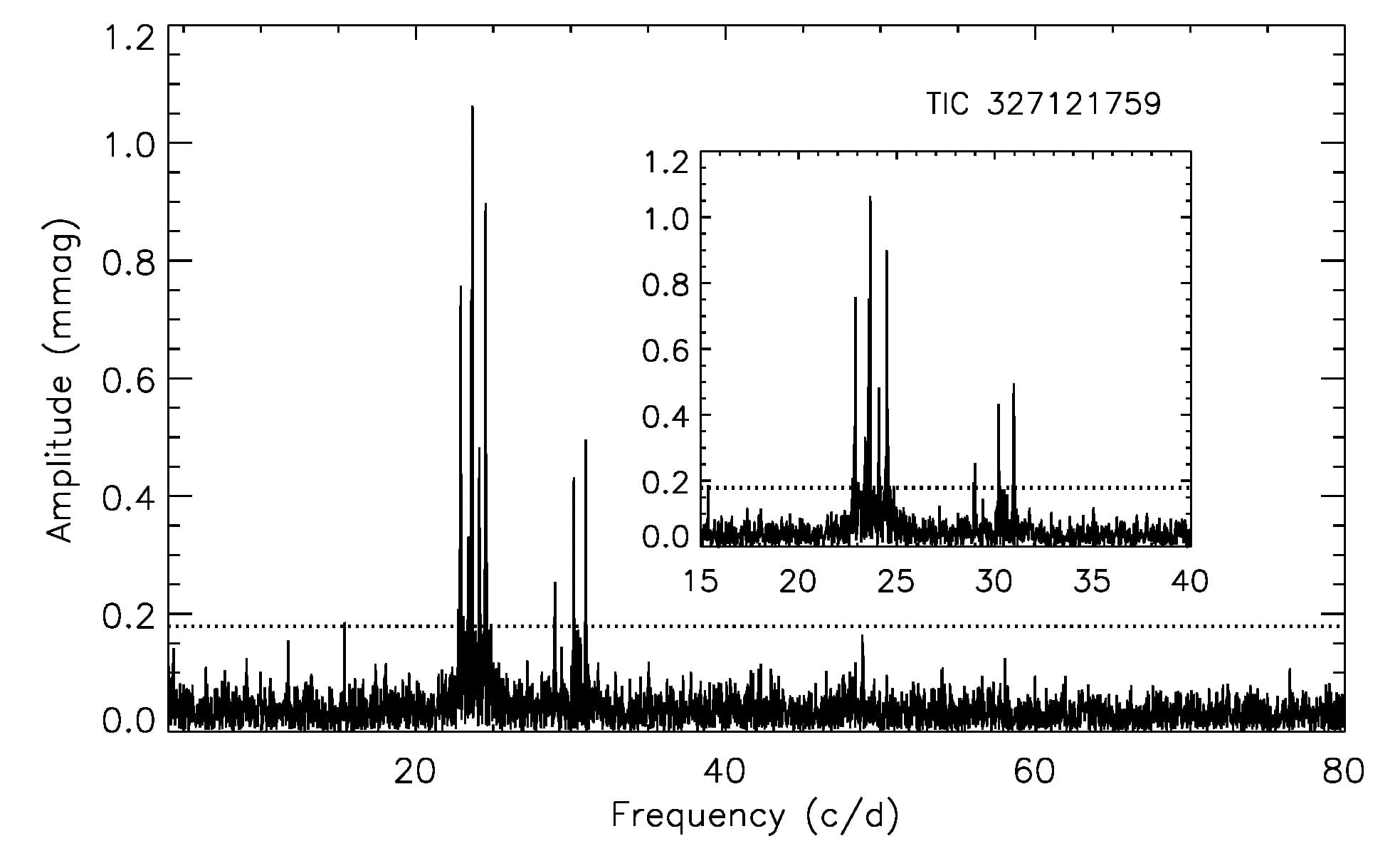}
  \end{minipage}
  \begin{minipage}[b]{0.24\textwidth}
  \includegraphics[height=3.5cm, width=1.0\textwidth]{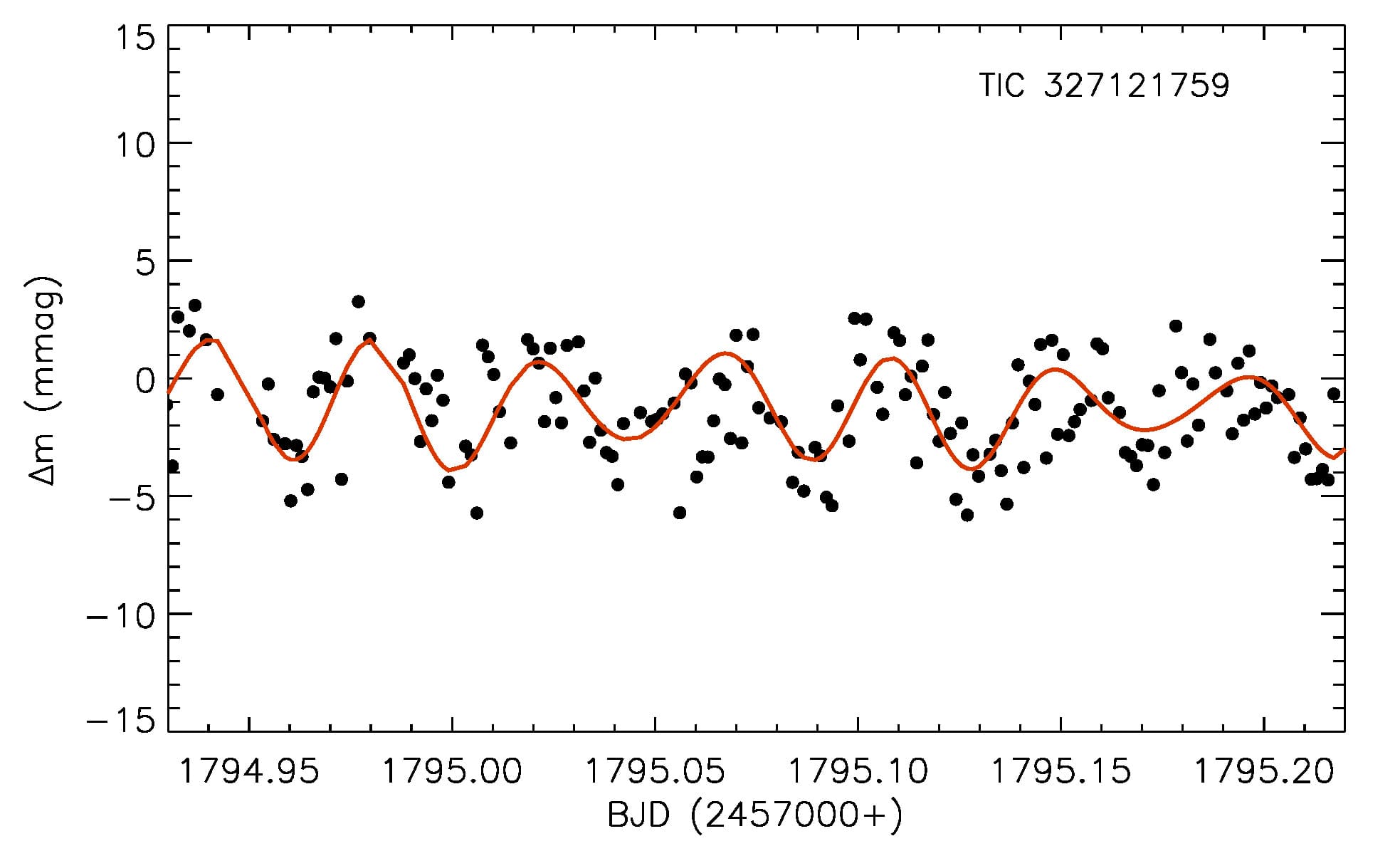}
  \end{minipage}
\begin{minipage}[b]{0.24\textwidth}
  \includegraphics[height=3.5cm, width=1\textwidth]{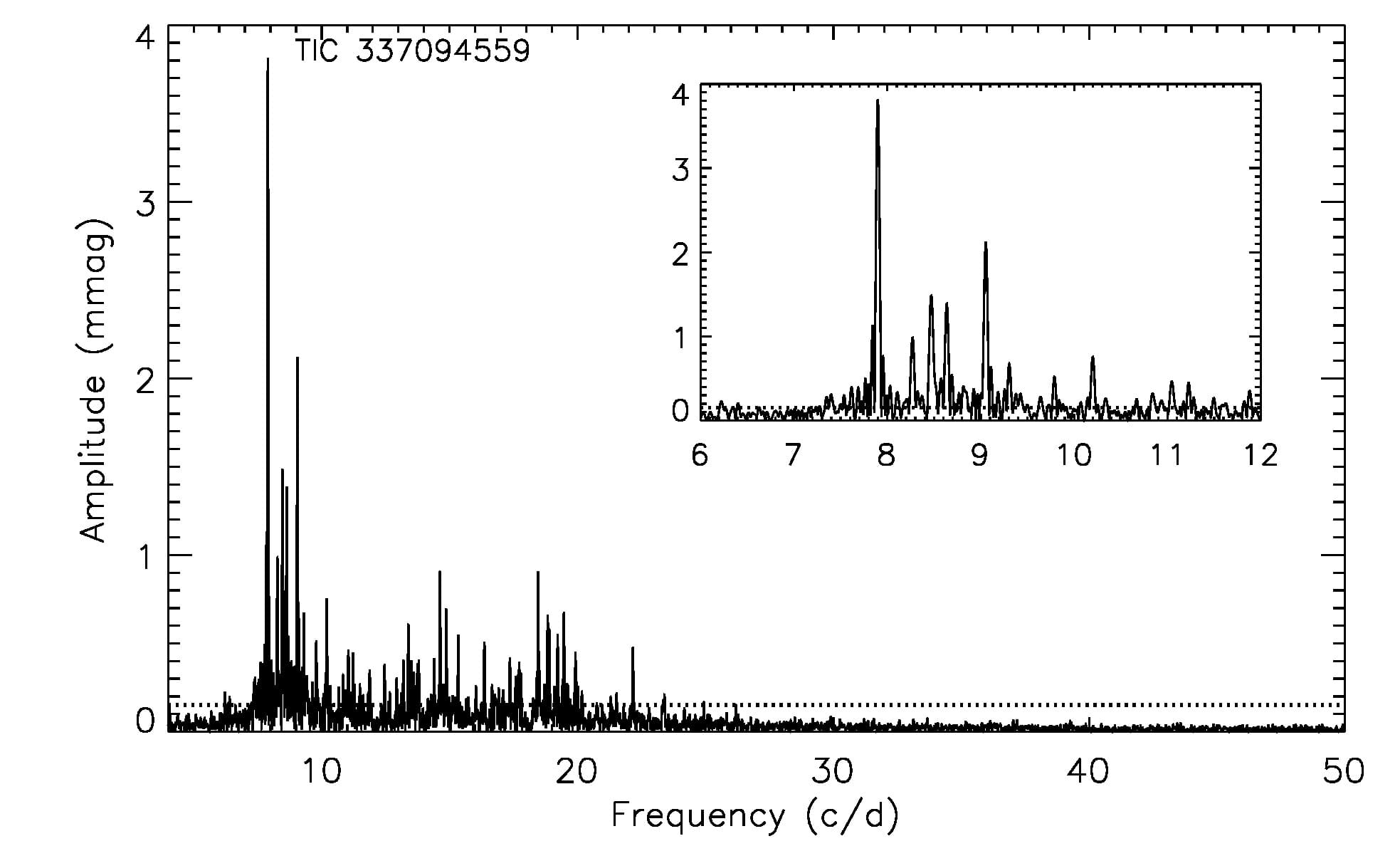}
  \end{minipage}
  \begin{minipage}[b]{0.24\textwidth}
  \includegraphics[height=3.5cm, width=1\textwidth]{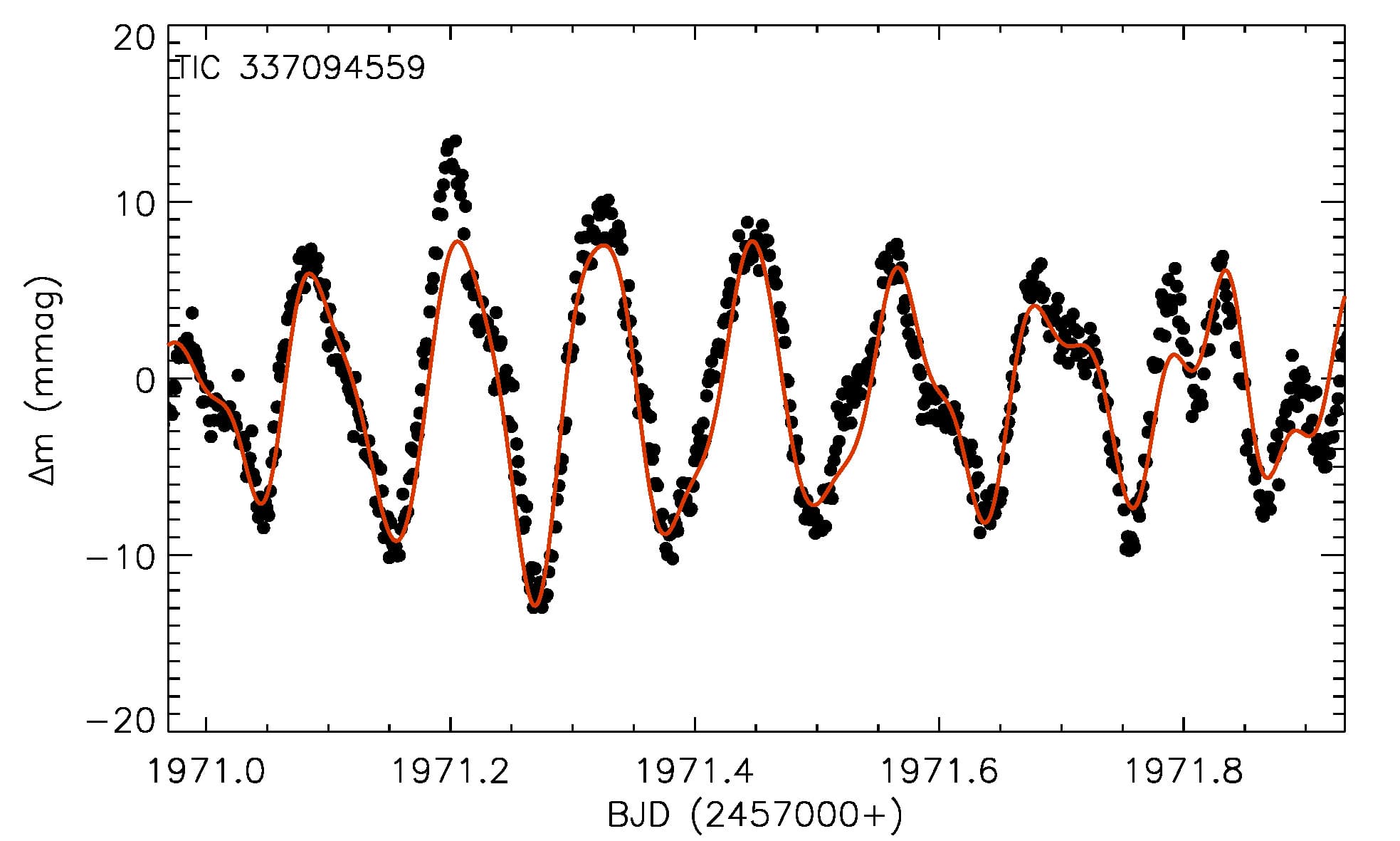}
  \end{minipage}
 \begin{minipage}[b]{0.24\textwidth}
  \includegraphics[height=3.5cm, width=1\textwidth]{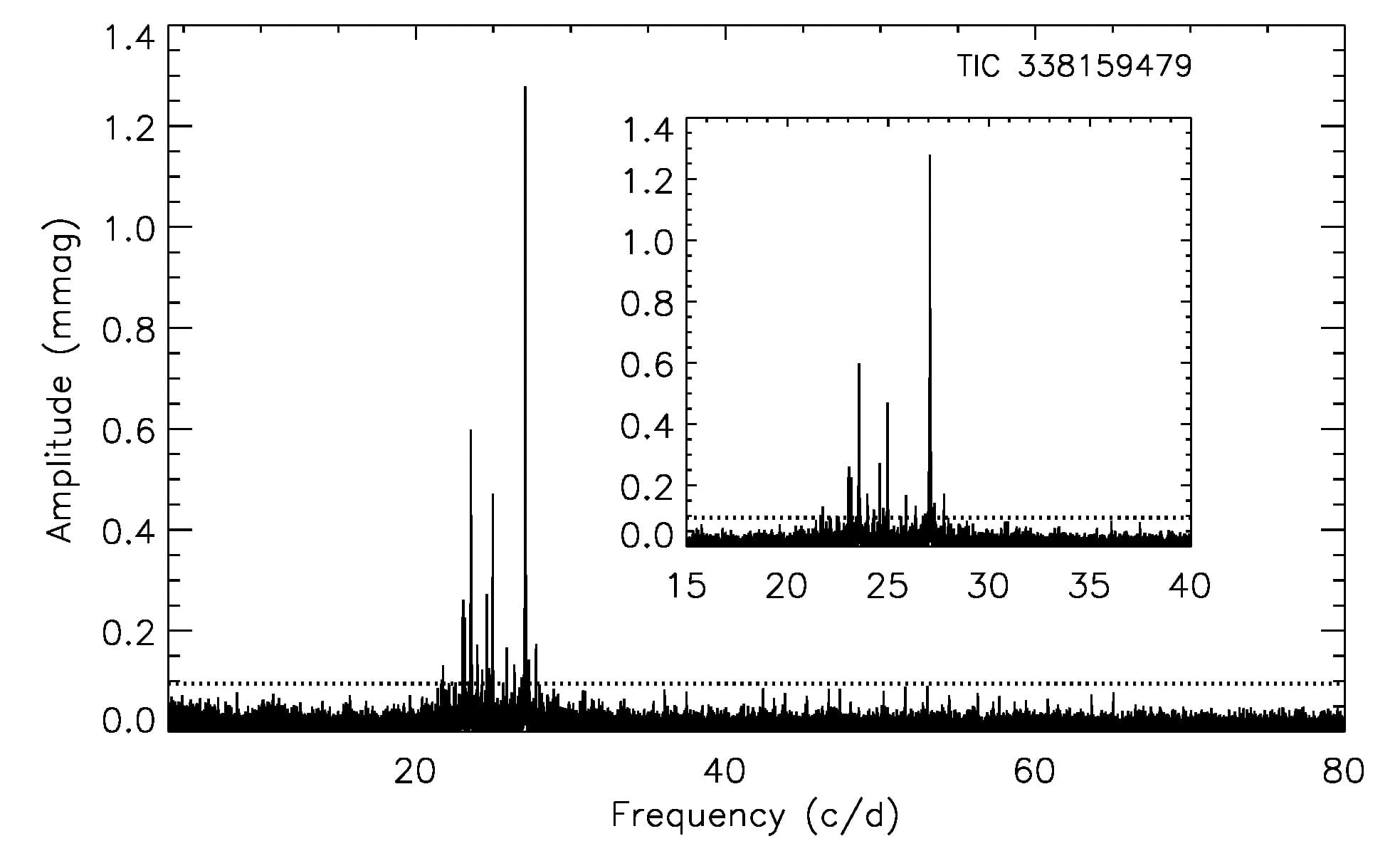}
 \end{minipage}
 \begin{minipage}[b]{0.24\textwidth}
  \includegraphics[height=3.5cm, width=1\textwidth]{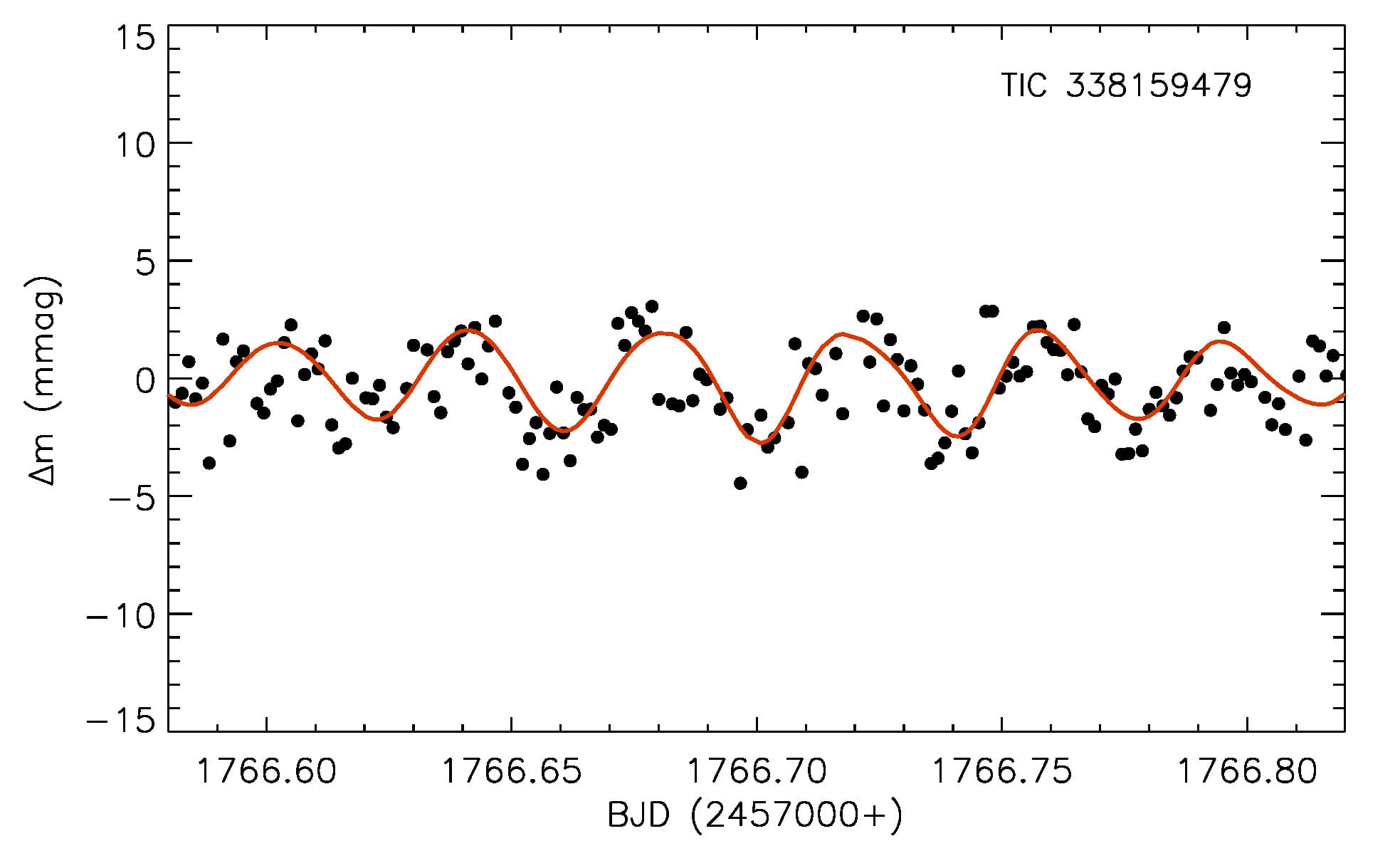}
 \end{minipage}
 \begin{minipage}[b]{0.24\textwidth}
  \includegraphics[height=3.5cm, width=1\textwidth]{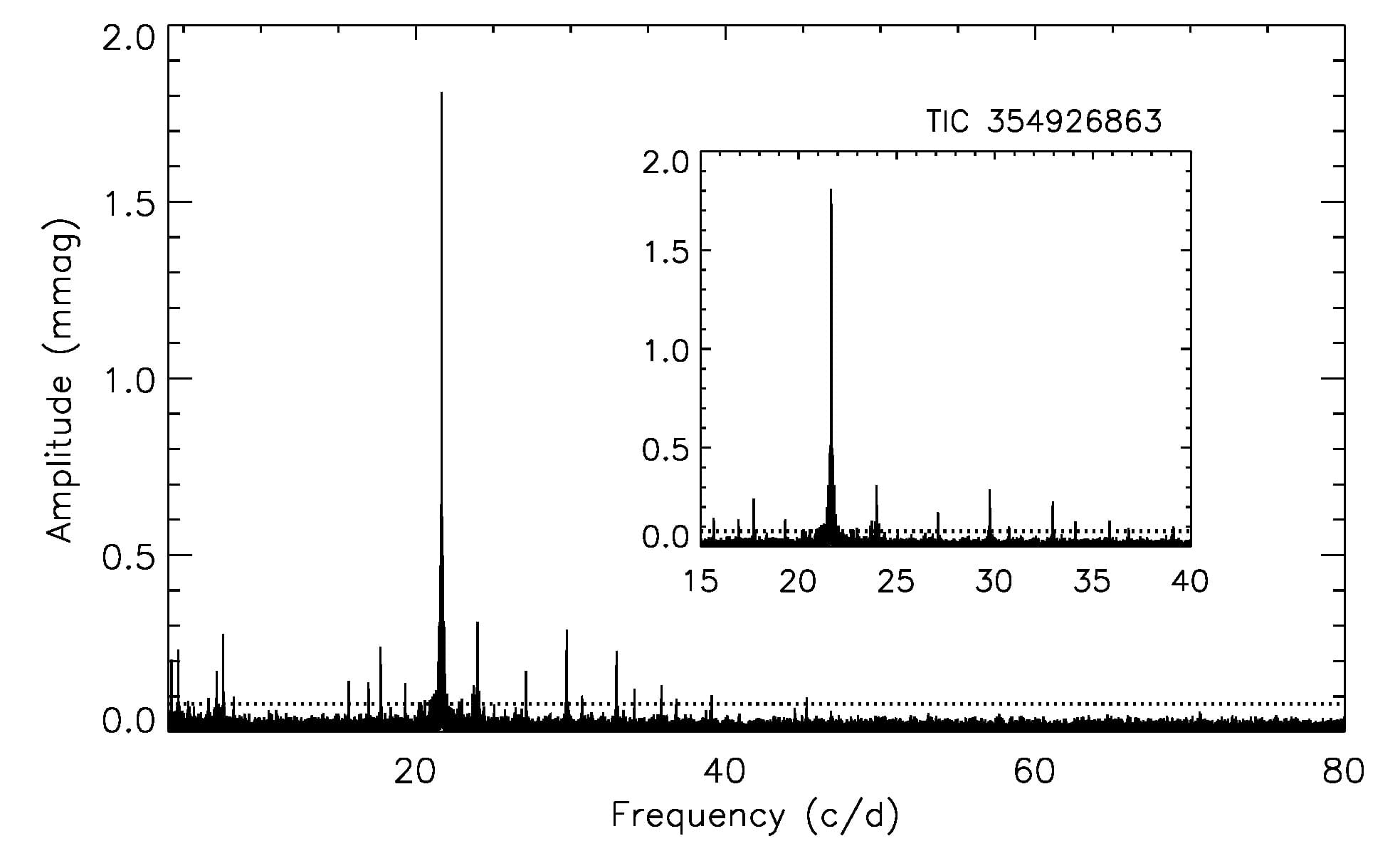}
  \end{minipage}
  \begin{minipage}[b]{0.24\textwidth}
  \includegraphics[height=3.5cm, width=1\textwidth]{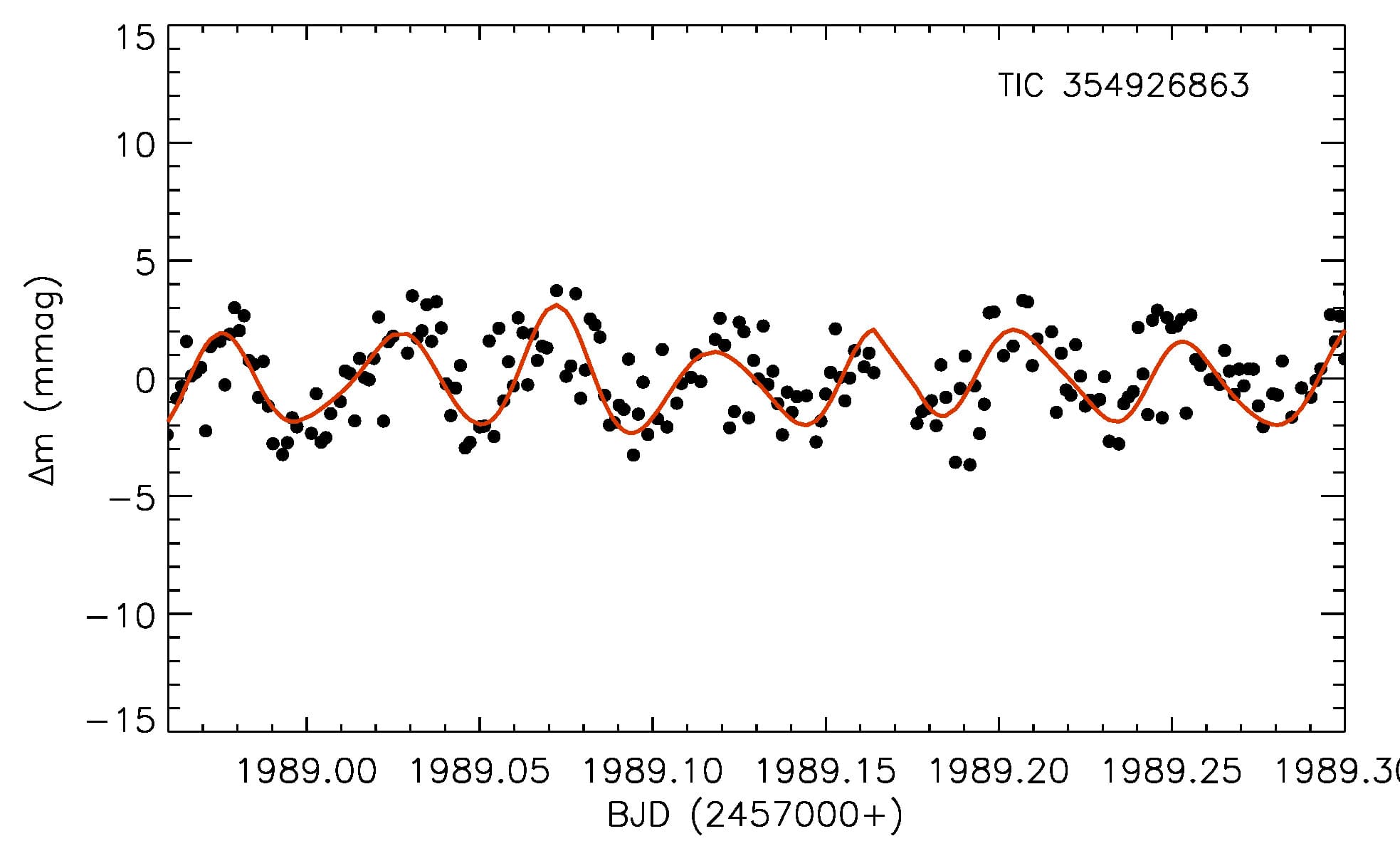}
  \end{minipage}
     \caption{Continuation.}\label{spec2}
\end{figure*}

\setcounter{figure}{4}

\begin{figure*}
 \begin{minipage}[b]{0.24\textwidth}
  \includegraphics[height=3.5cm, width=1\textwidth]{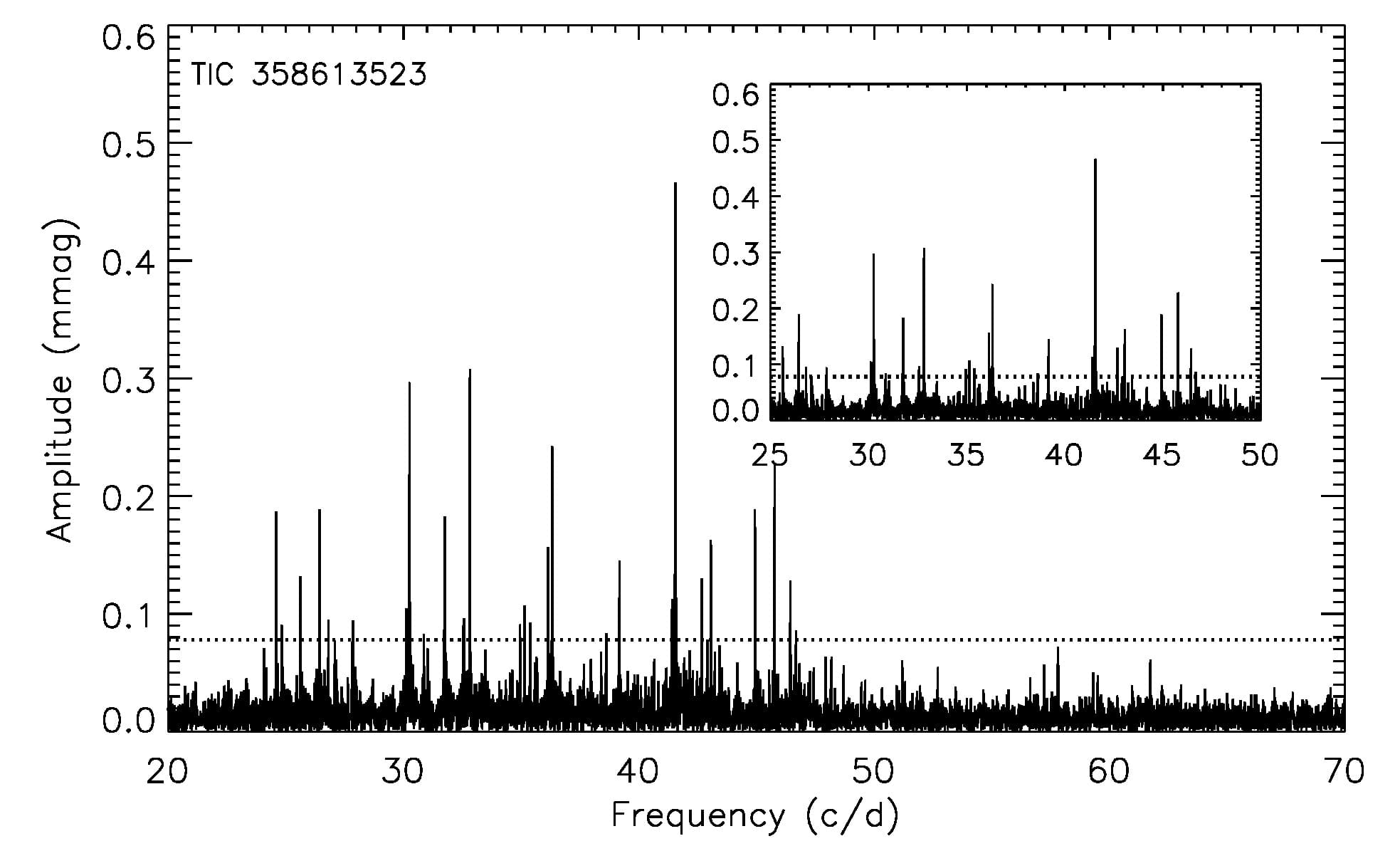}
 \end{minipage}
\begin{minipage}[b]{0.24\textwidth}
 \includegraphics[height=3.5cm, width=1\textwidth]{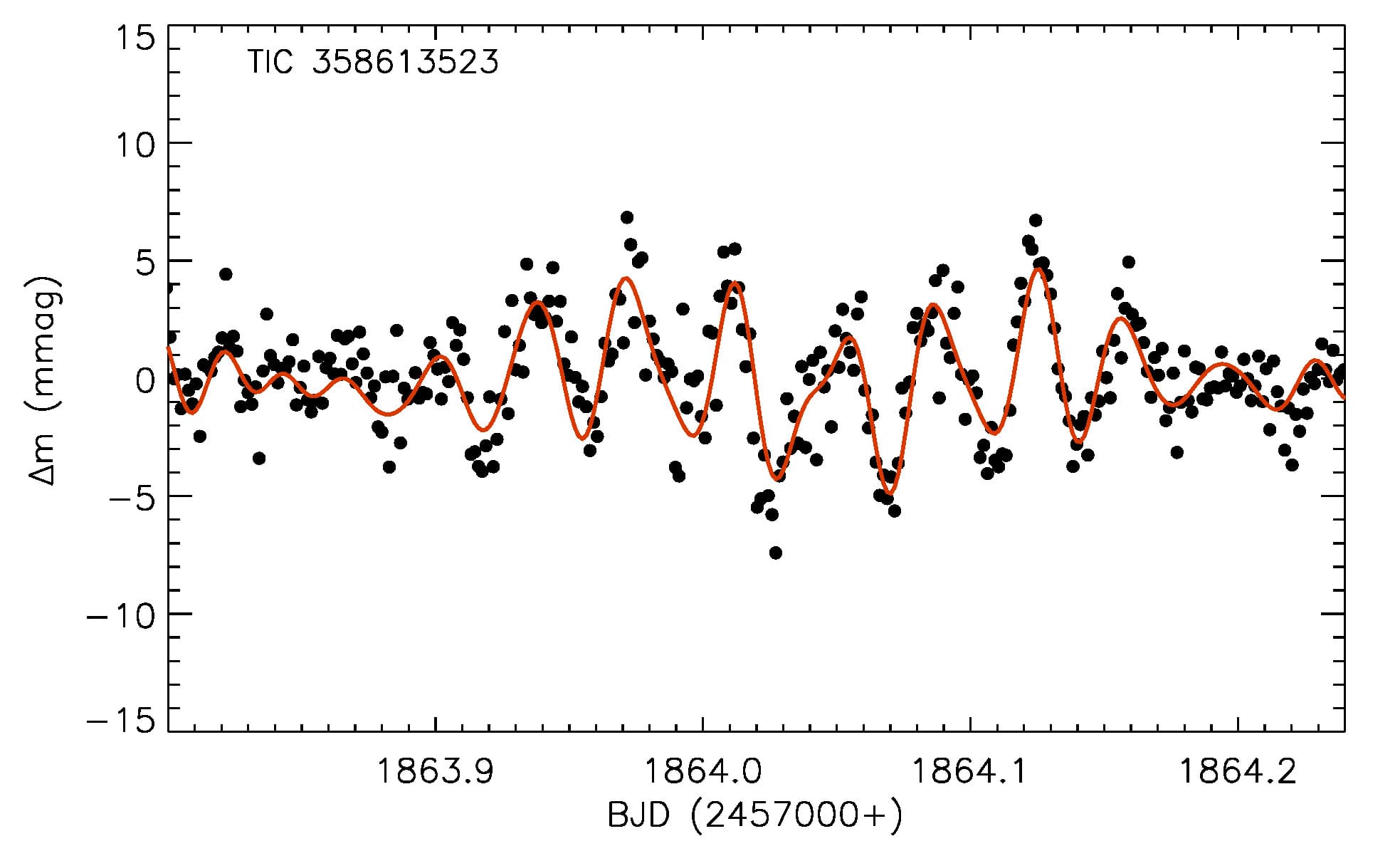}
 \end{minipage}
 \begin{minipage}[b]{0.24\textwidth}
  \includegraphics[height=3.5cm, width=1\textwidth]{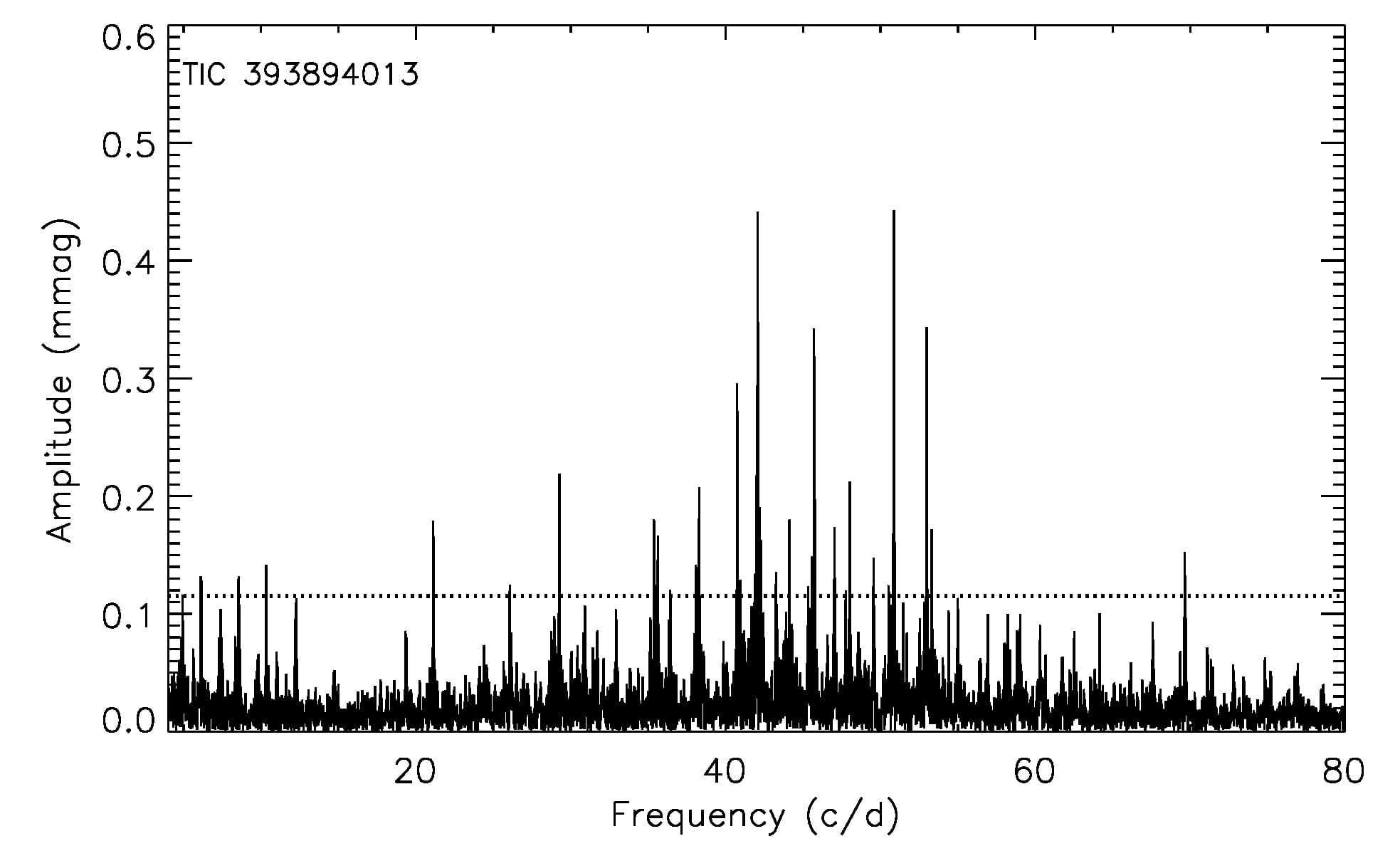}
 \end{minipage}
  \begin{minipage}[b]{0.24\textwidth}
  \includegraphics[height=3.5cm, width=1.0\textwidth]{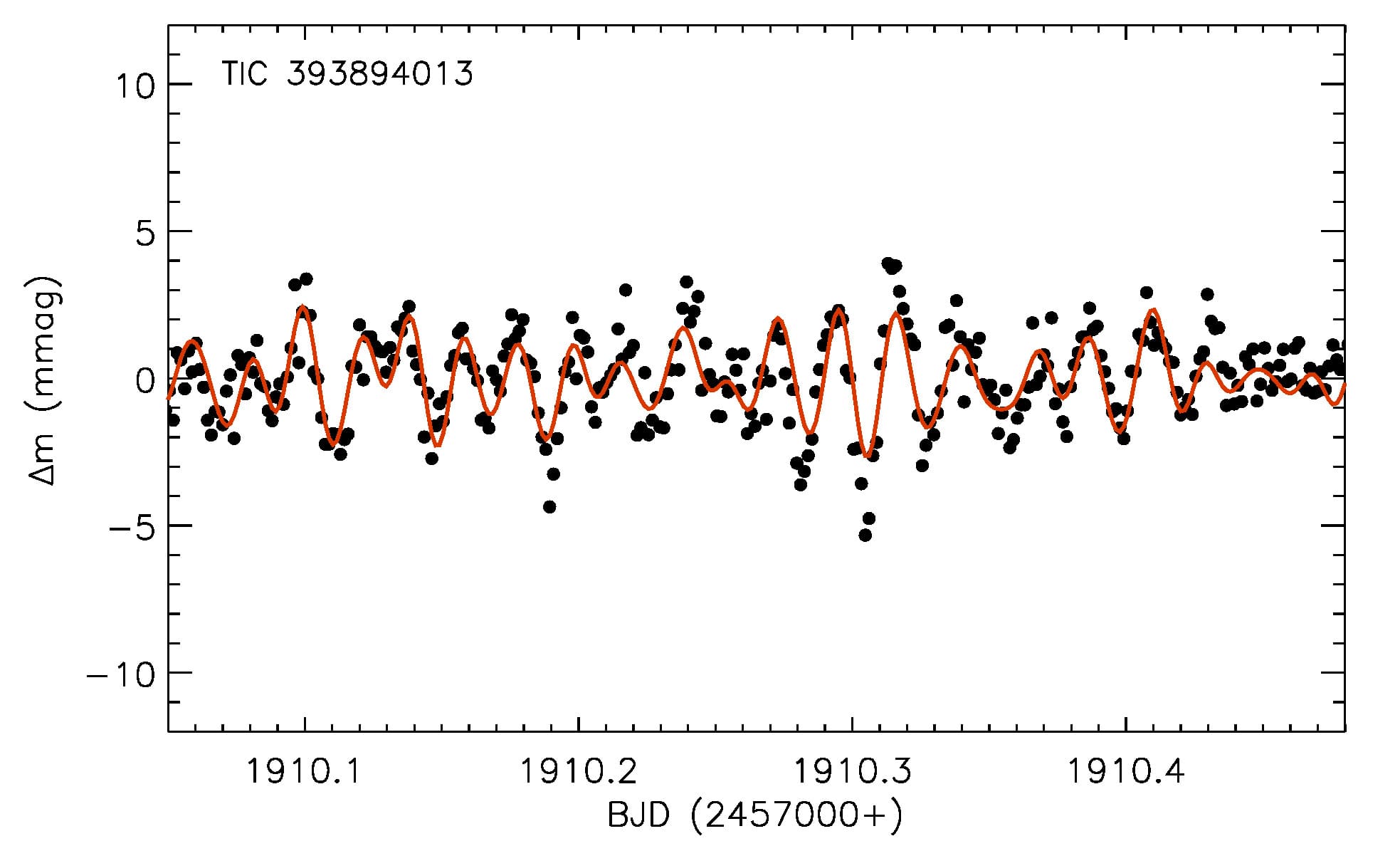}
  \end{minipage}
 \begin{minipage}[b]{0.24\textwidth}
 \includegraphics[height=3.5cm, width=1\textwidth]{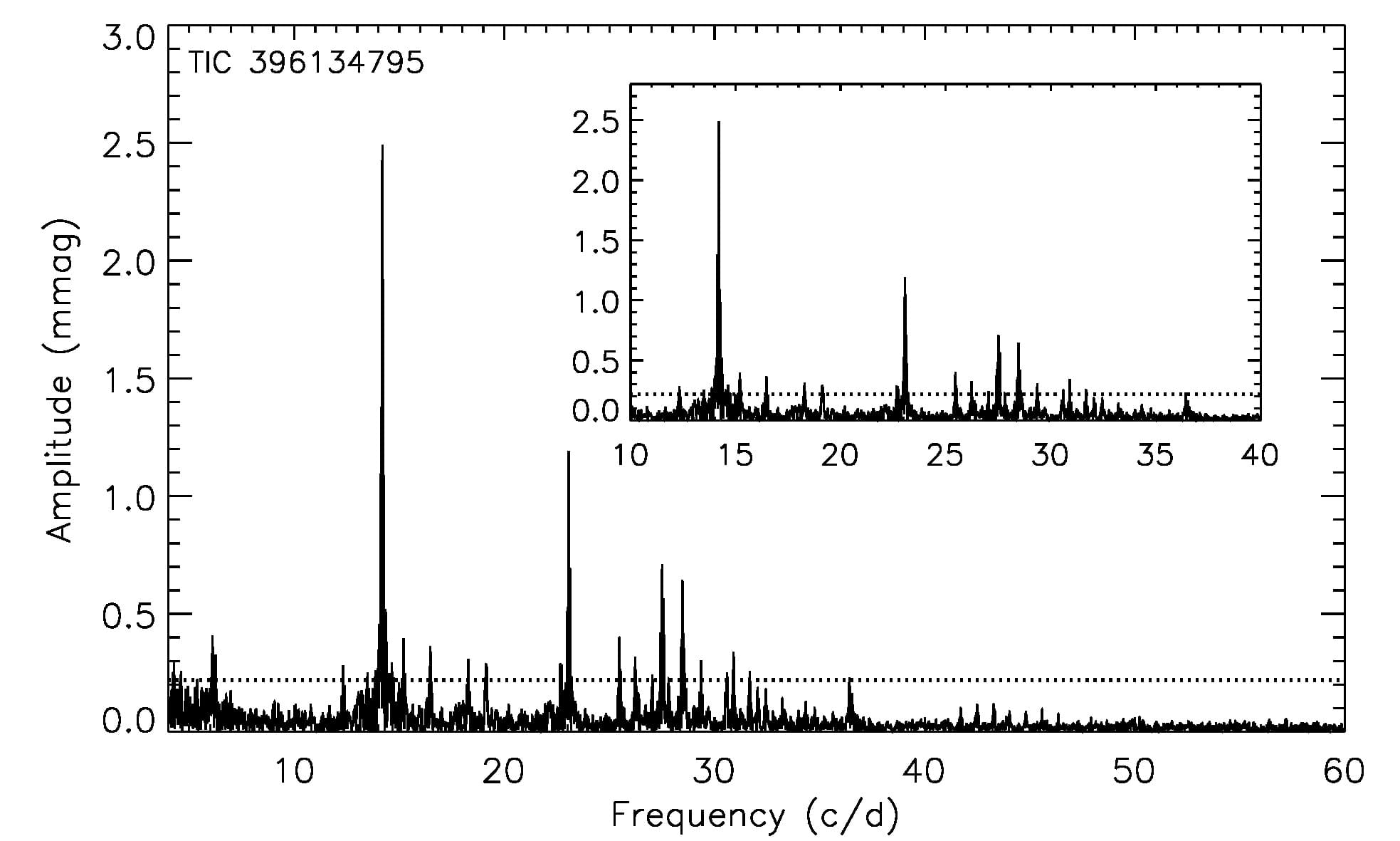}
 \end{minipage}
 \begin{minipage}[b]{0.24\textwidth}
 \includegraphics[height=3.5cm, width=1\textwidth]{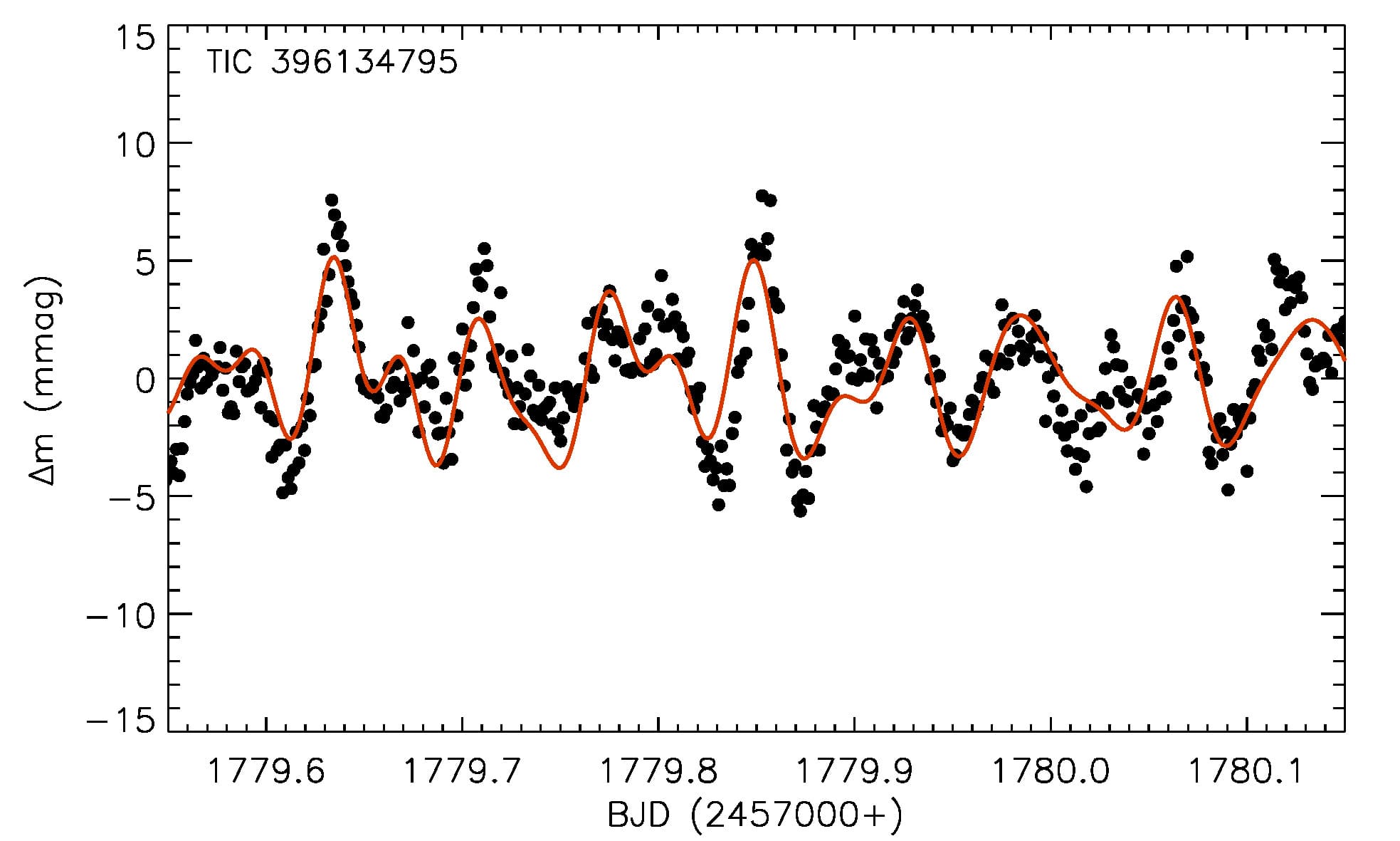}
 \end{minipage}
 \begin{minipage}[b]{0.24\textwidth}
 \includegraphics[height=3.5cm, width=1\textwidth]{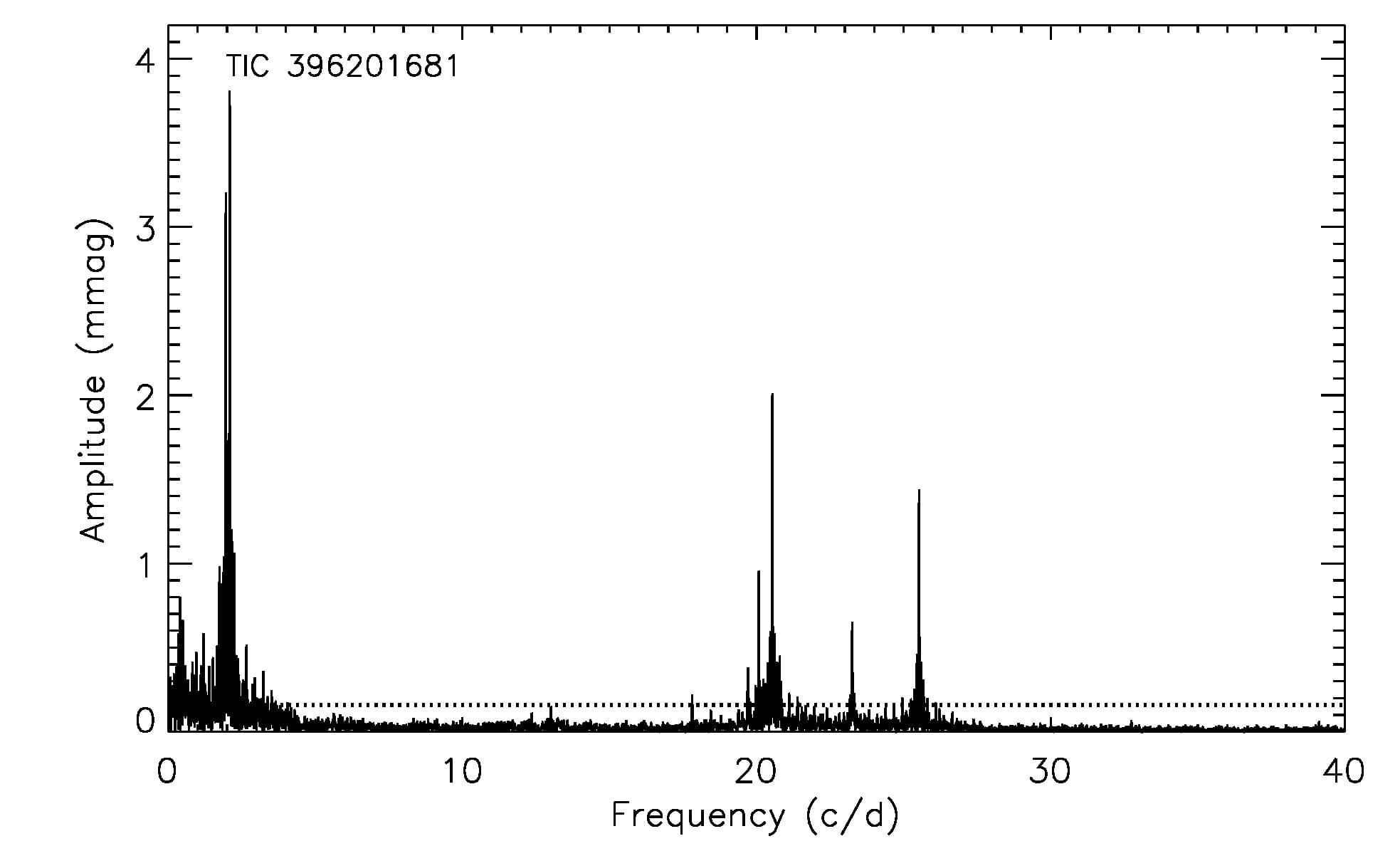}
 \end{minipage}
 \begin{minipage}[b]{0.24\textwidth}
 \includegraphics[height=3.5cm, width=1.0\textwidth]{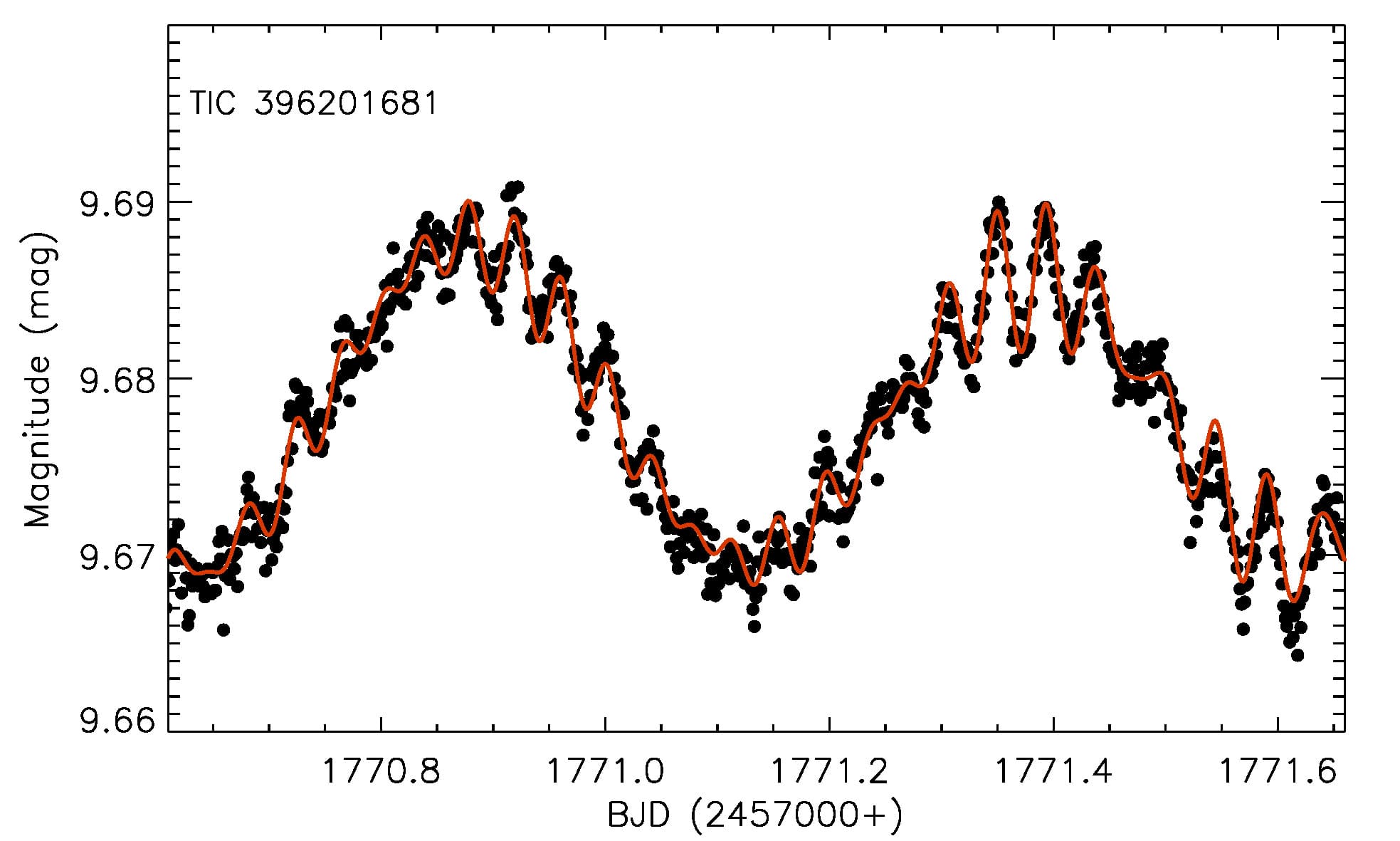}
 \end{minipage}
 \begin{minipage}[b]{0.24\textwidth}
 \includegraphics[height=3.5cm, width=1\textwidth]{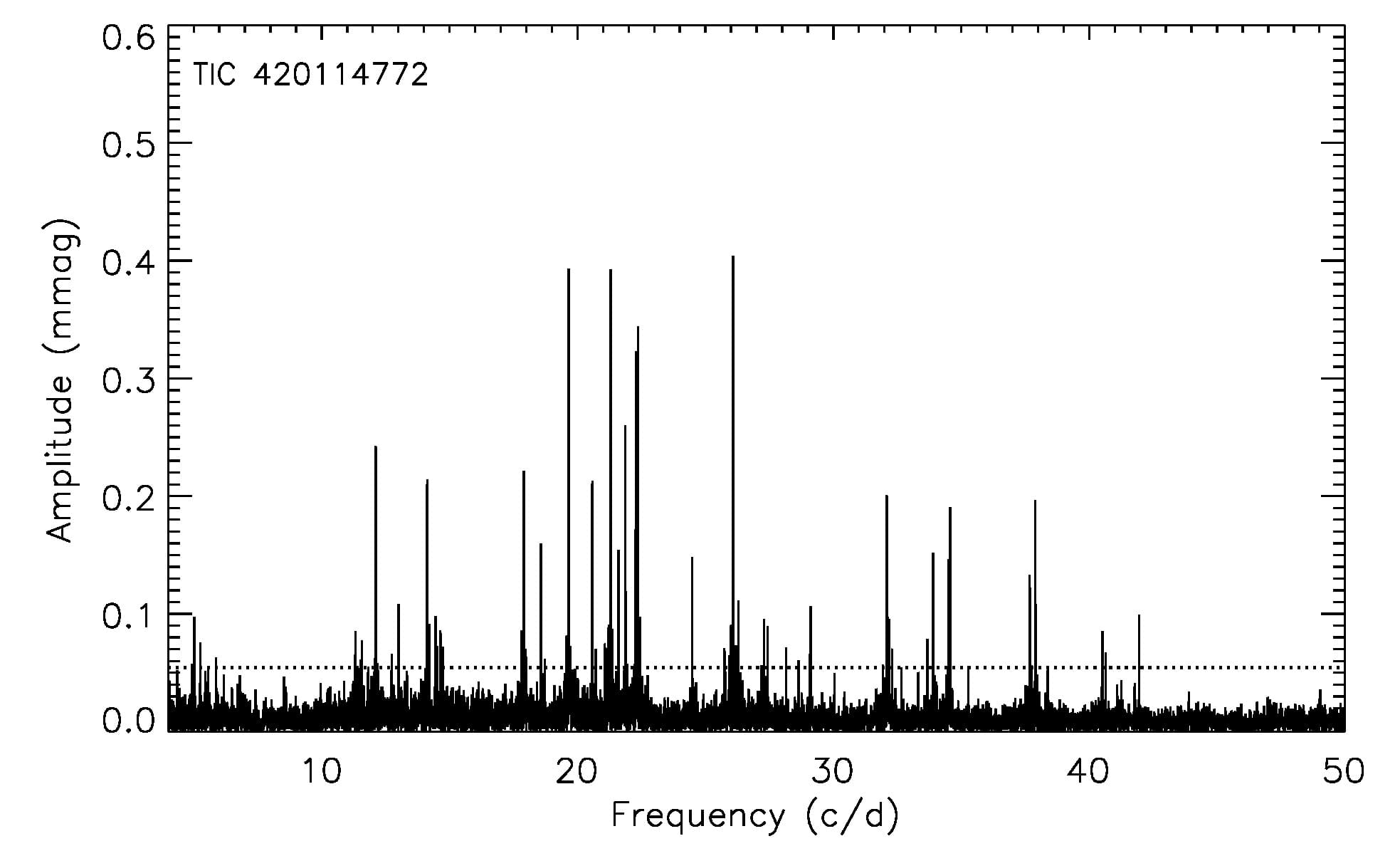}
 \end{minipage}
 \begin{minipage}[b]{0.24\textwidth}
 \includegraphics[height=3.5cm, width=1.0\textwidth]{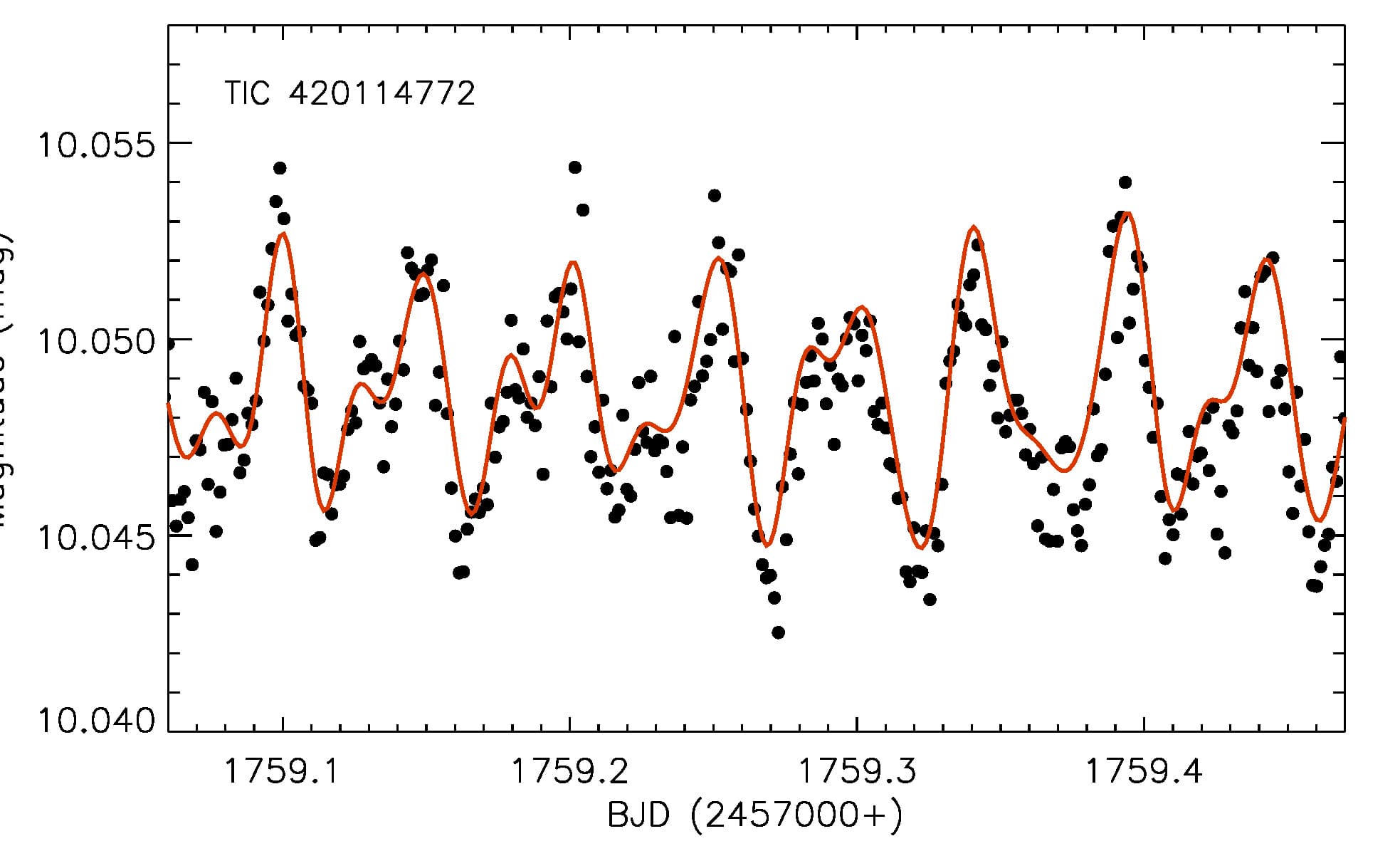}
 \end{minipage}
  \begin{minipage}[b]{0.24\textwidth}
  \includegraphics[height=3.5cm, width=1.0\textwidth]{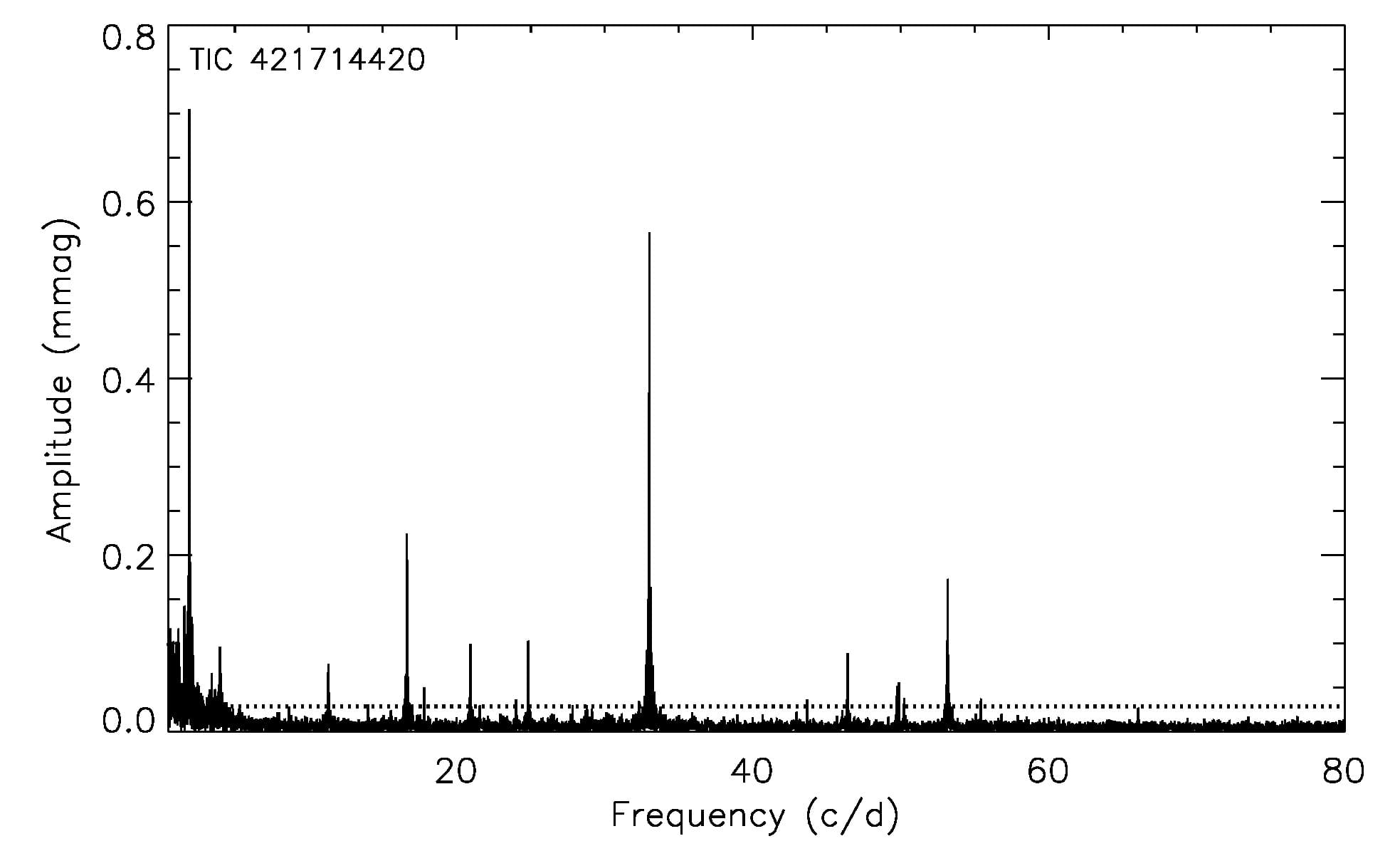}
  \end{minipage}
 \begin{minipage}[b]{0.24\textwidth}
  \includegraphics[height=3.5cm, width=1.0\textwidth]{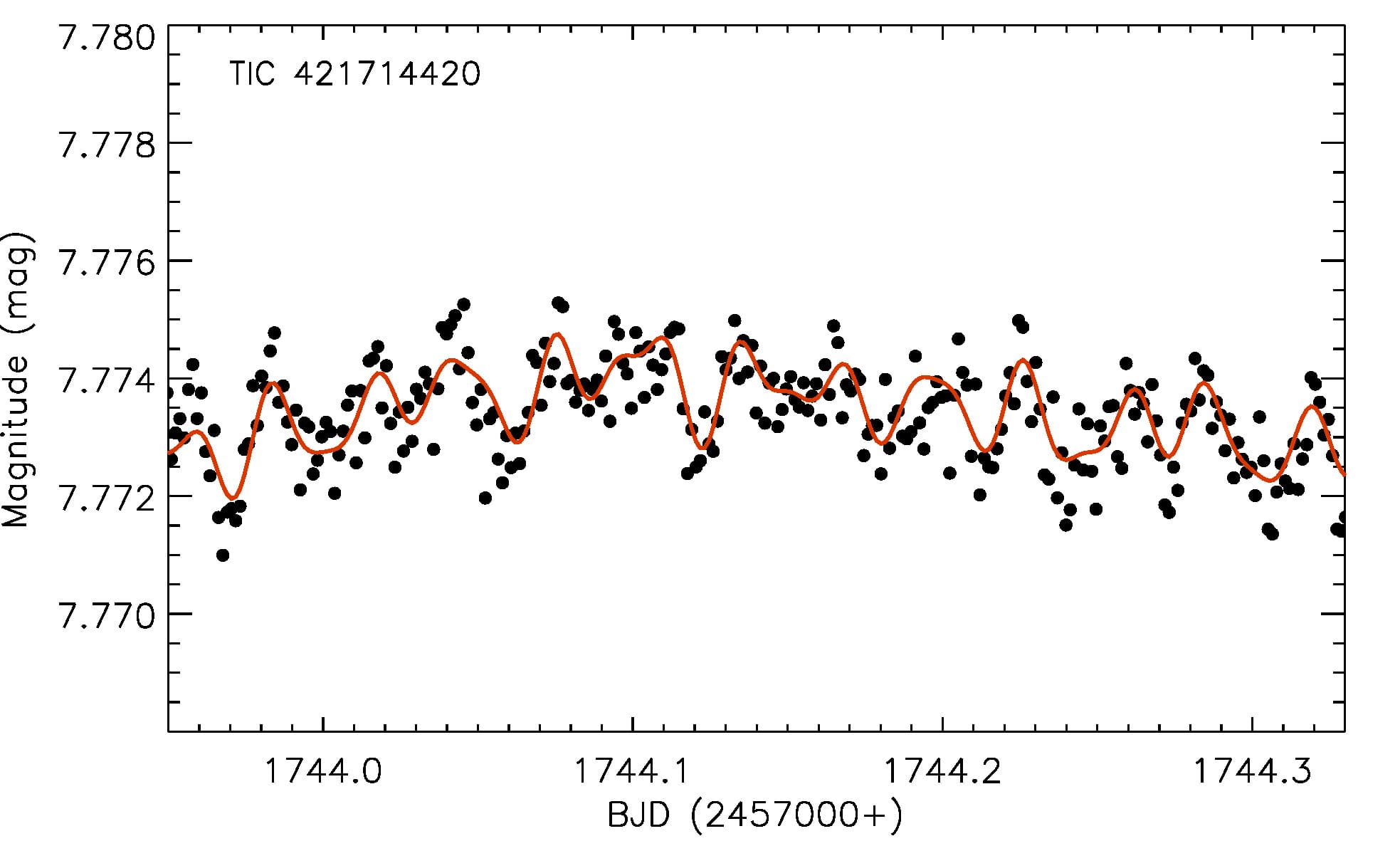}
 \end{minipage}
 \begin{minipage}[b]{0.24\textwidth}
  \includegraphics[height=3.5cm, width=1\textwidth]{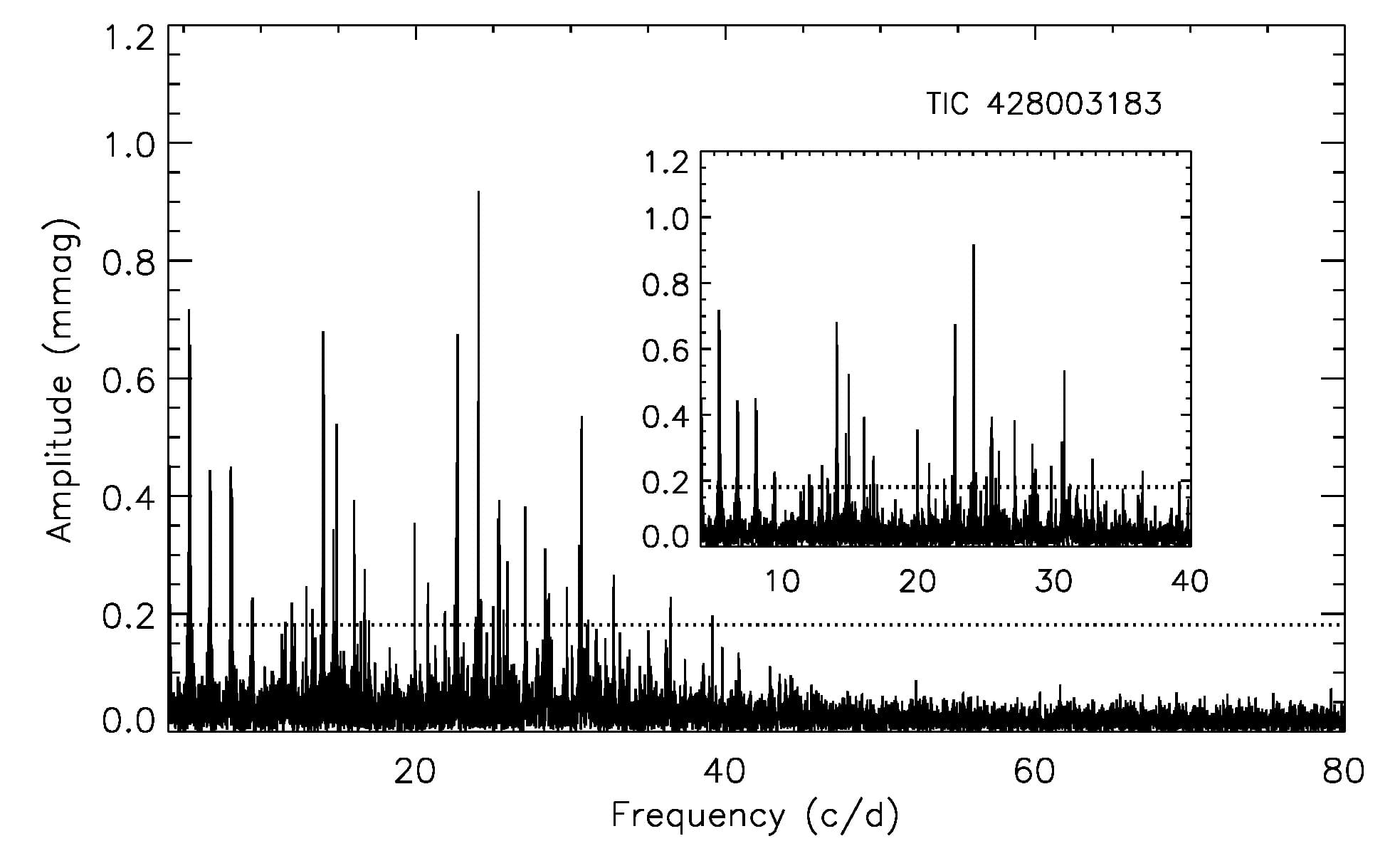}
 \end{minipage}
\begin{minipage}[b]{0.24\textwidth}
 \includegraphics[height=3.5cm, width=1\textwidth]{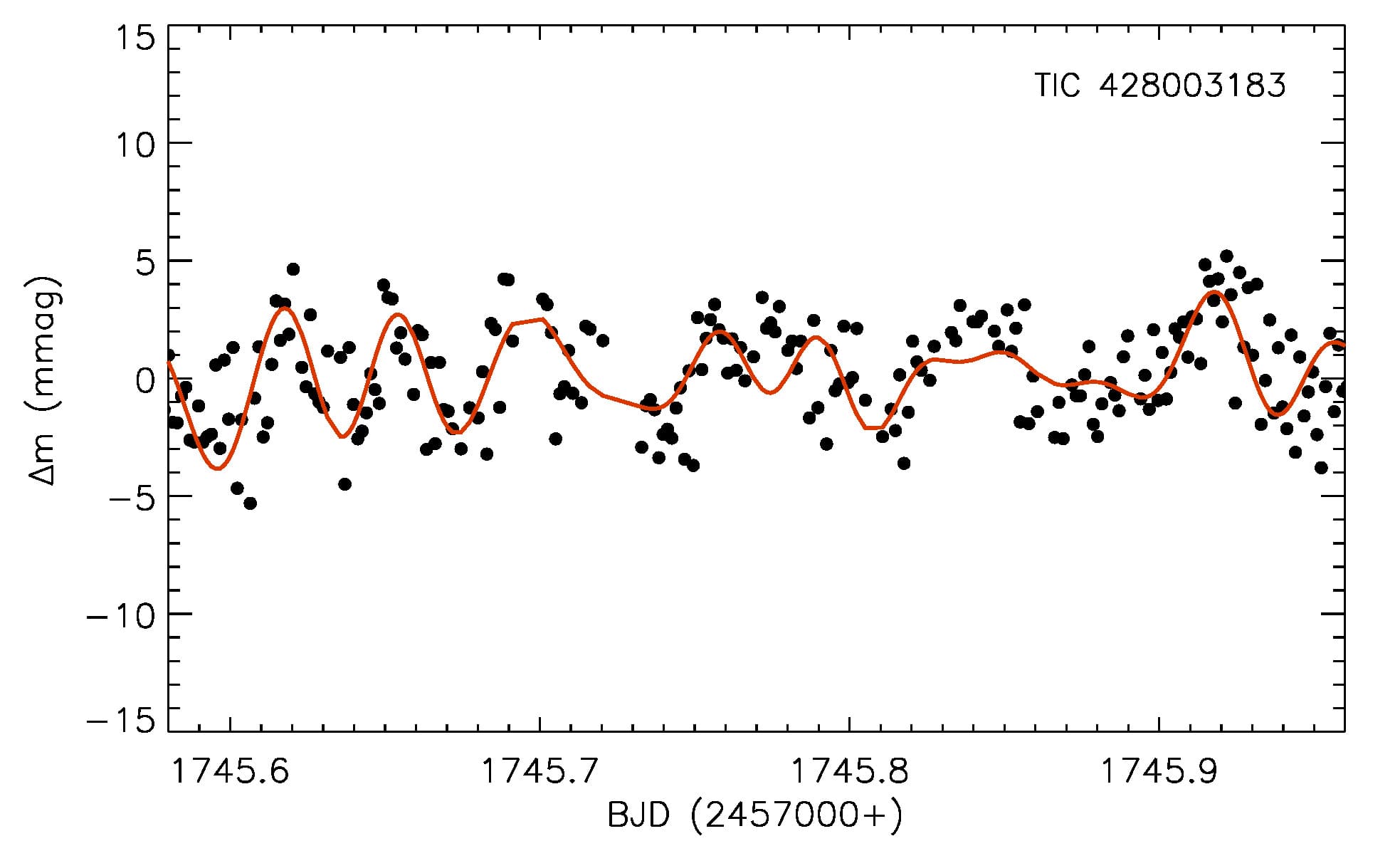}
 \end{minipage}
\begin{minipage}[b]{0.24\textwidth}
\includegraphics[height=3.5cm, width=1\textwidth]{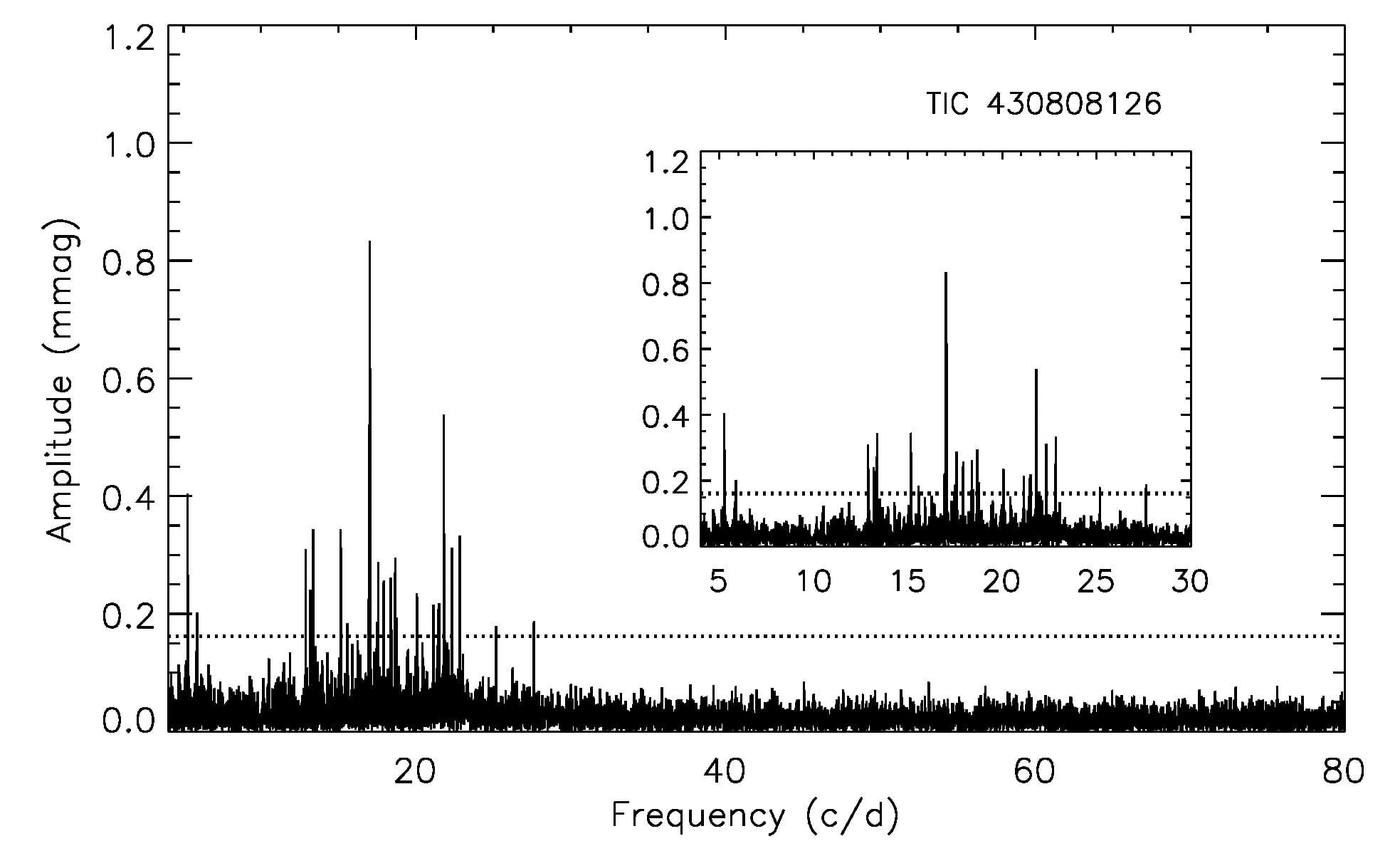}
\end{minipage}
 \begin{minipage}[b]{0.24\textwidth}
  \includegraphics[height=3.5cm, width=1\textwidth]{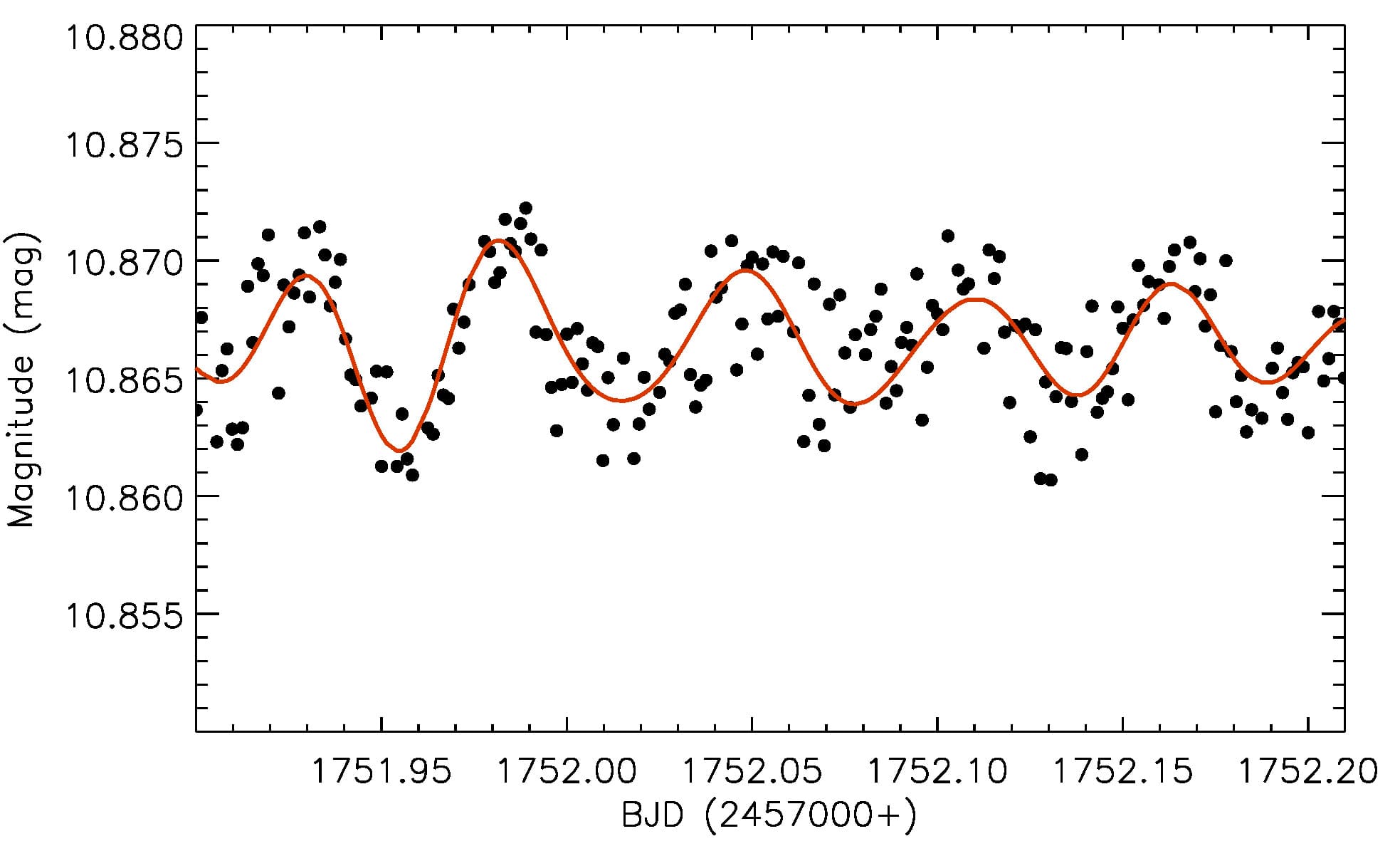}
 \end{minipage}
  \begin{minipage}[b]{0.24\textwidth}
  \includegraphics[height=3.5cm, width=1\textwidth]{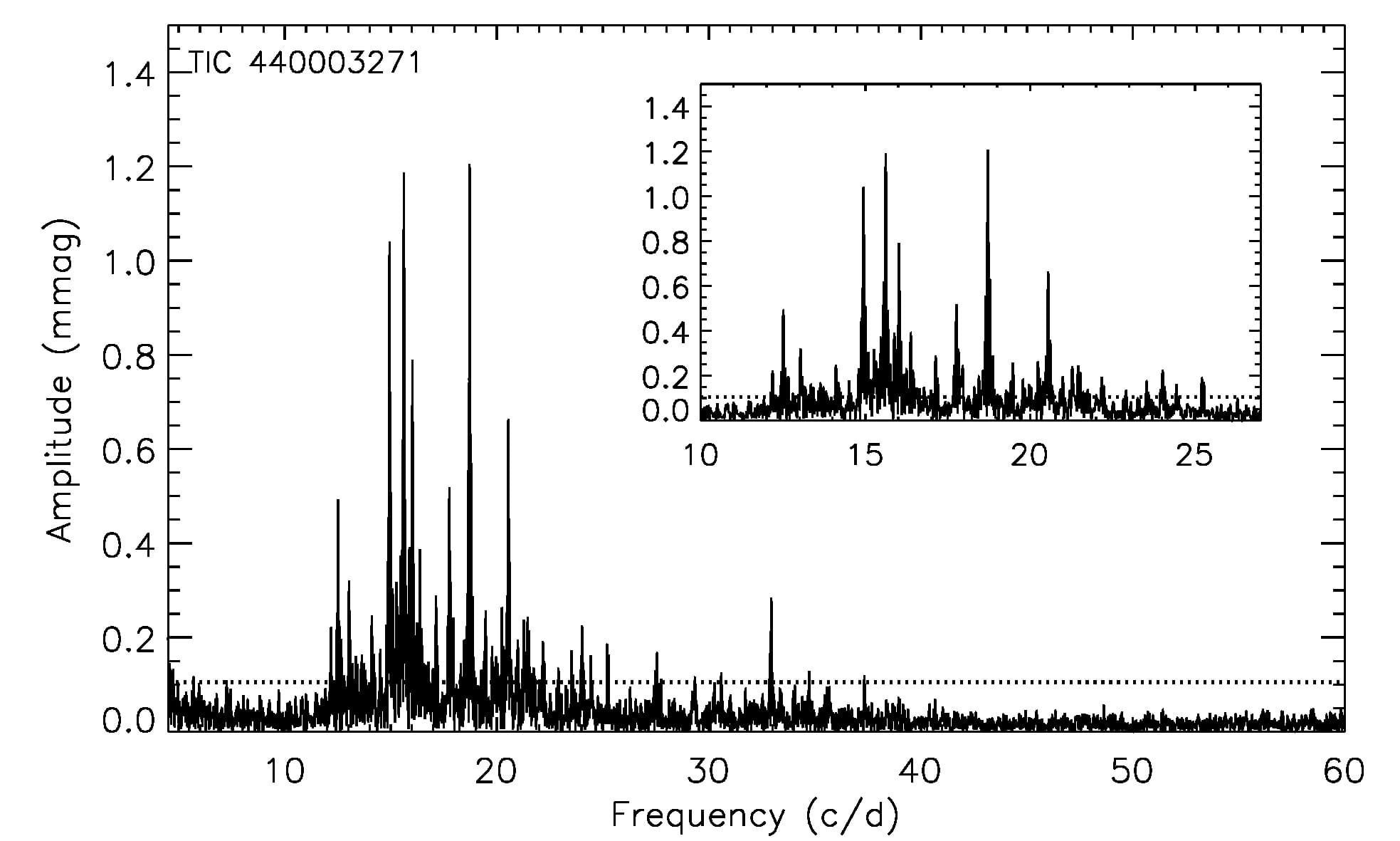}
%   \caption{1}
 \end{minipage}
 \begin{minipage}[b]{0.24\textwidth}
  \includegraphics[height=3.5cm, width=1\textwidth]{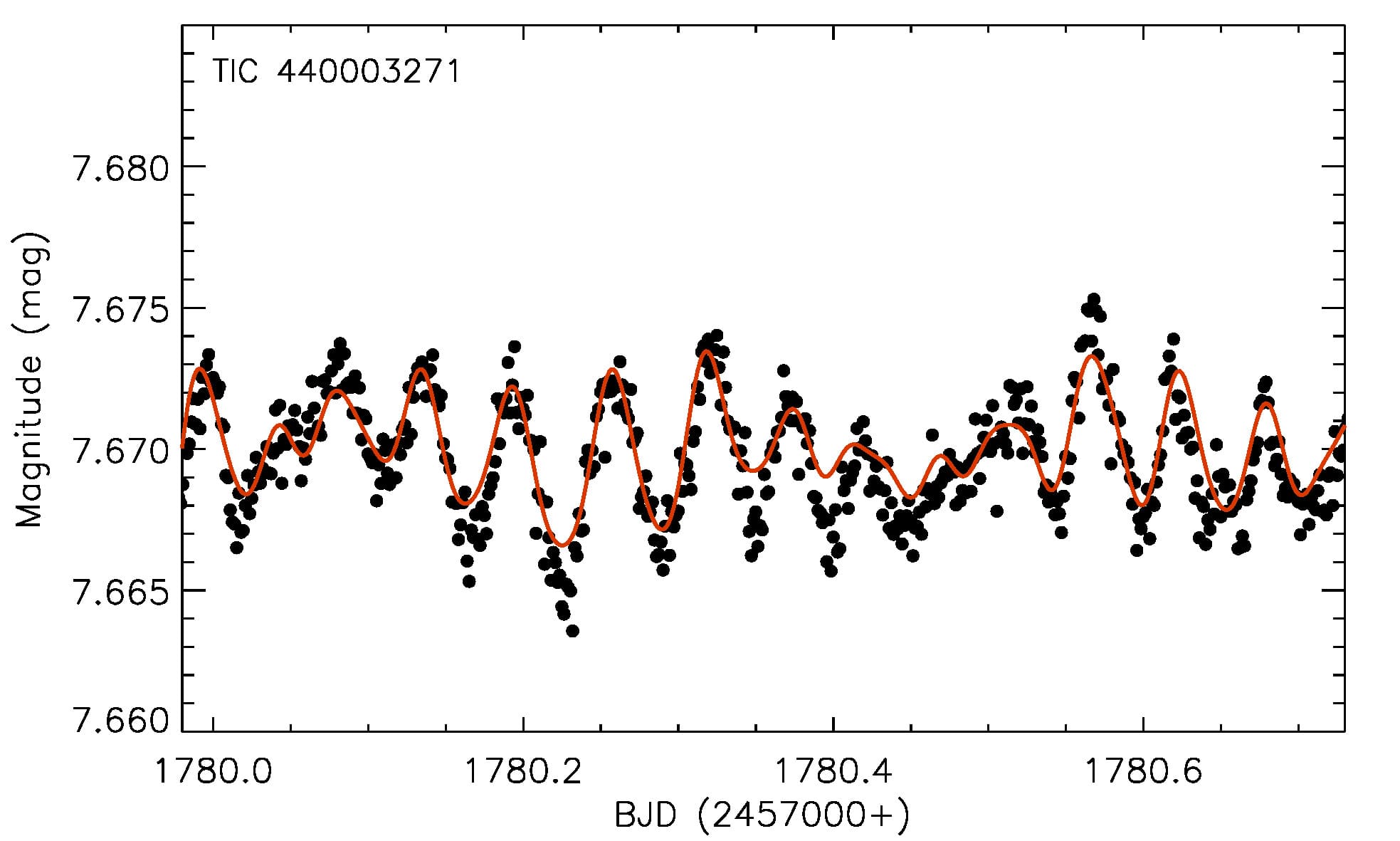}
\end{minipage}
 \begin{minipage}[b]{0.24\textwidth}
 \includegraphics[height=3.5cm, width=1\textwidth]{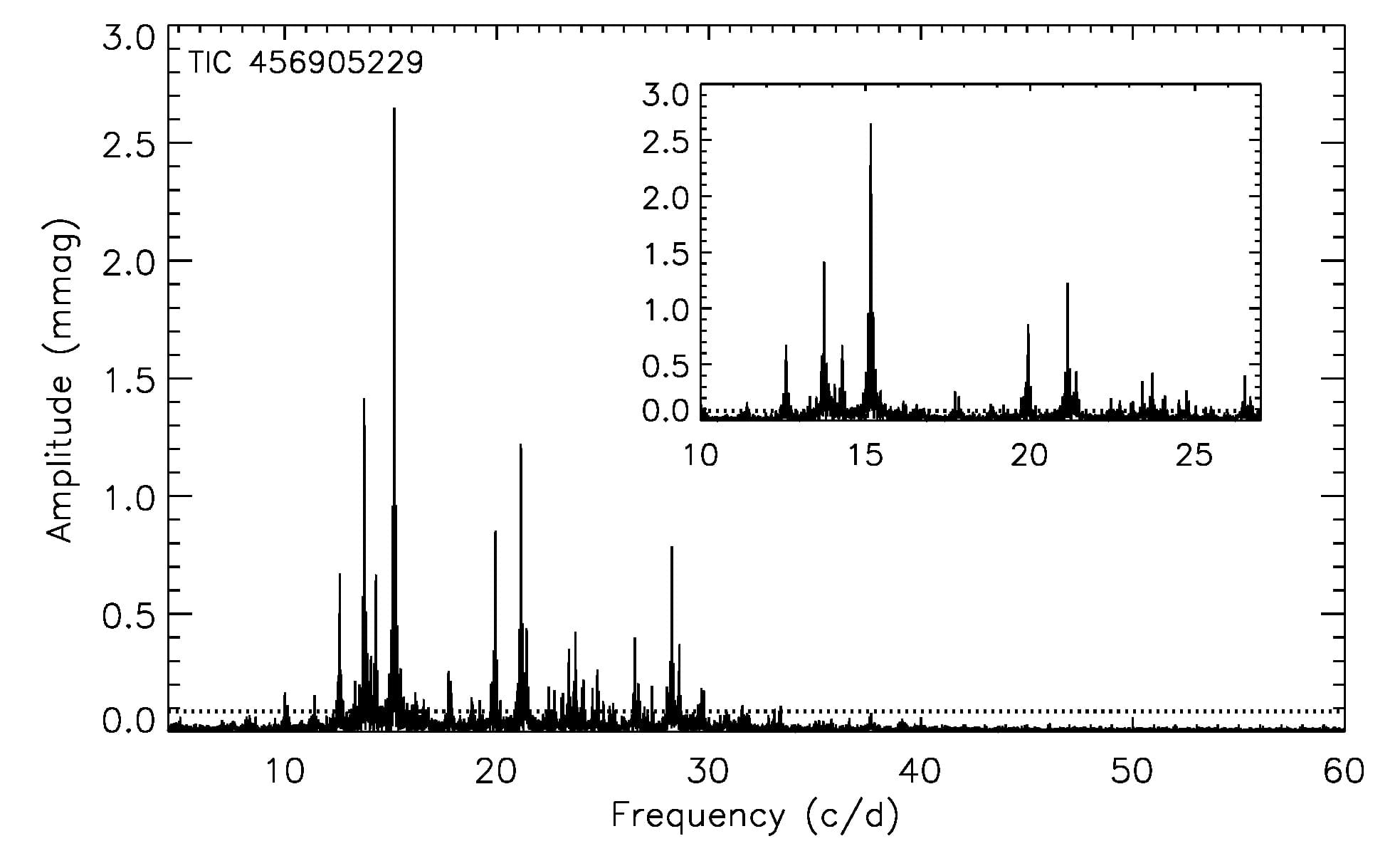}
\end{minipage}
 \begin{minipage}[b]{0.24\textwidth}
  \includegraphics[height=3.5cm, width=1.0\textwidth]{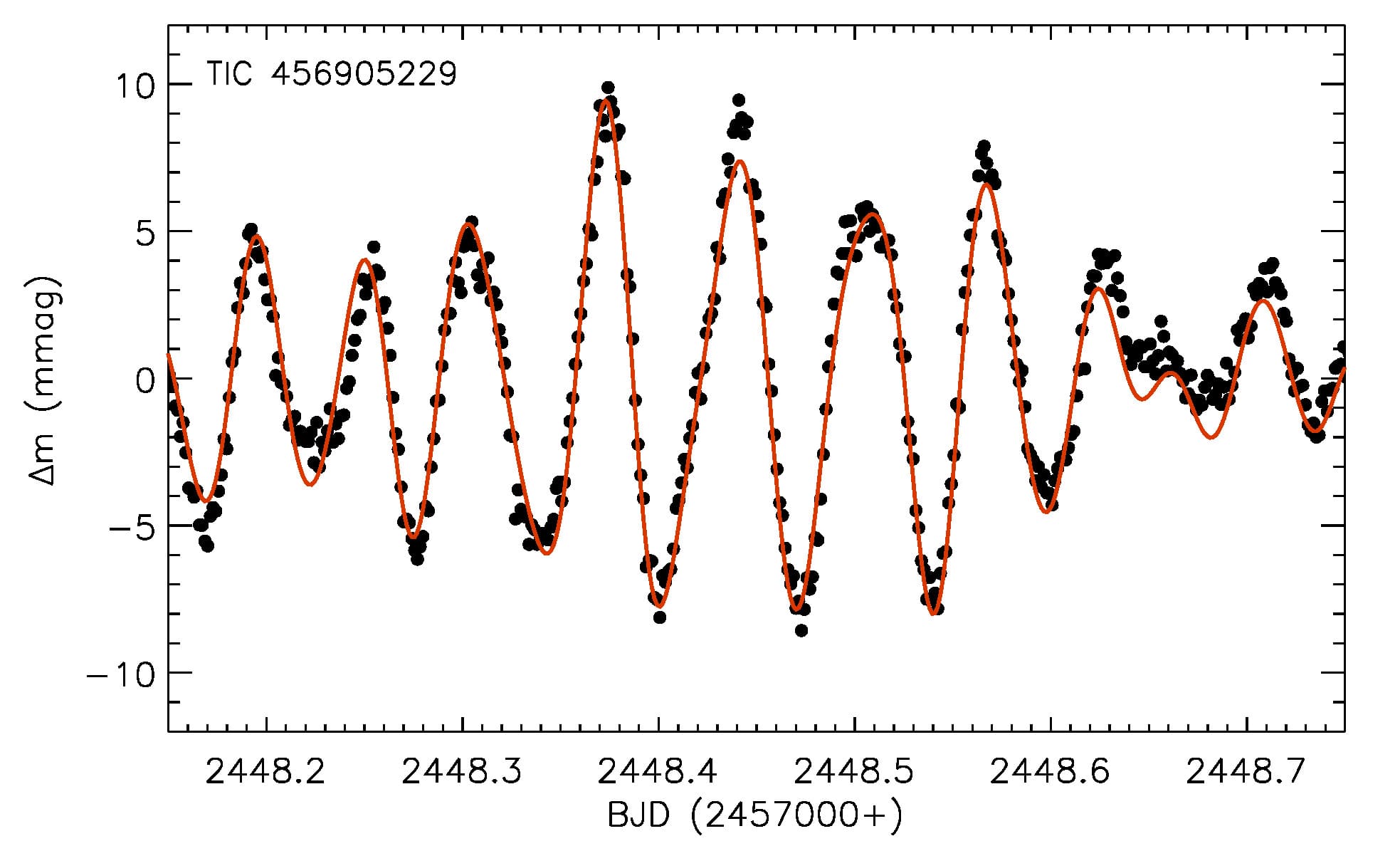}
  \end{minipage}
\begin{minipage}[b]{0.24\textwidth}
 \includegraphics[height=3.5cm, width=1\textwidth]{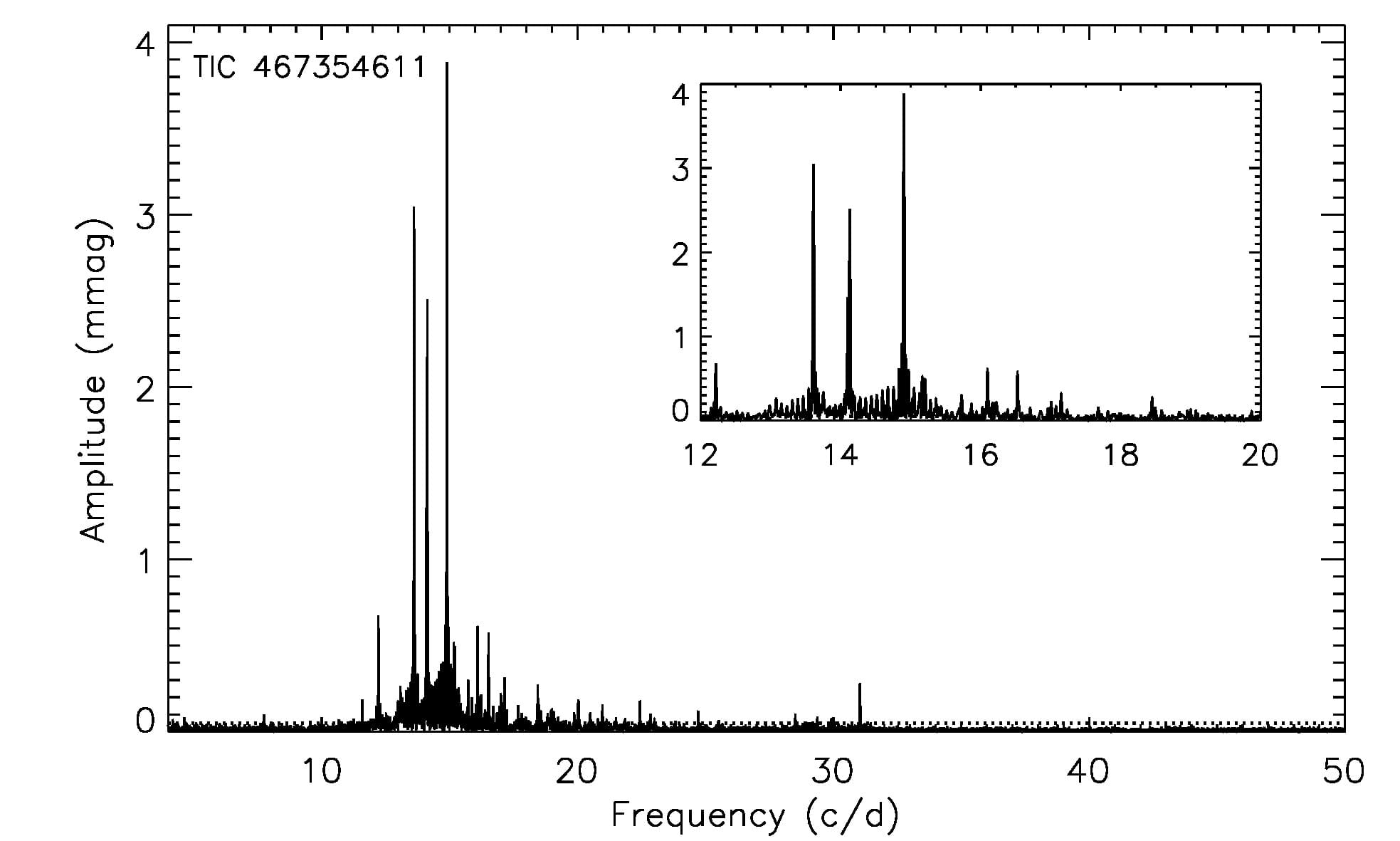}
  \end{minipage}
 \begin{minipage}[b]{0.24\textwidth}
  \includegraphics[height=3.5cm, width=1\textwidth]{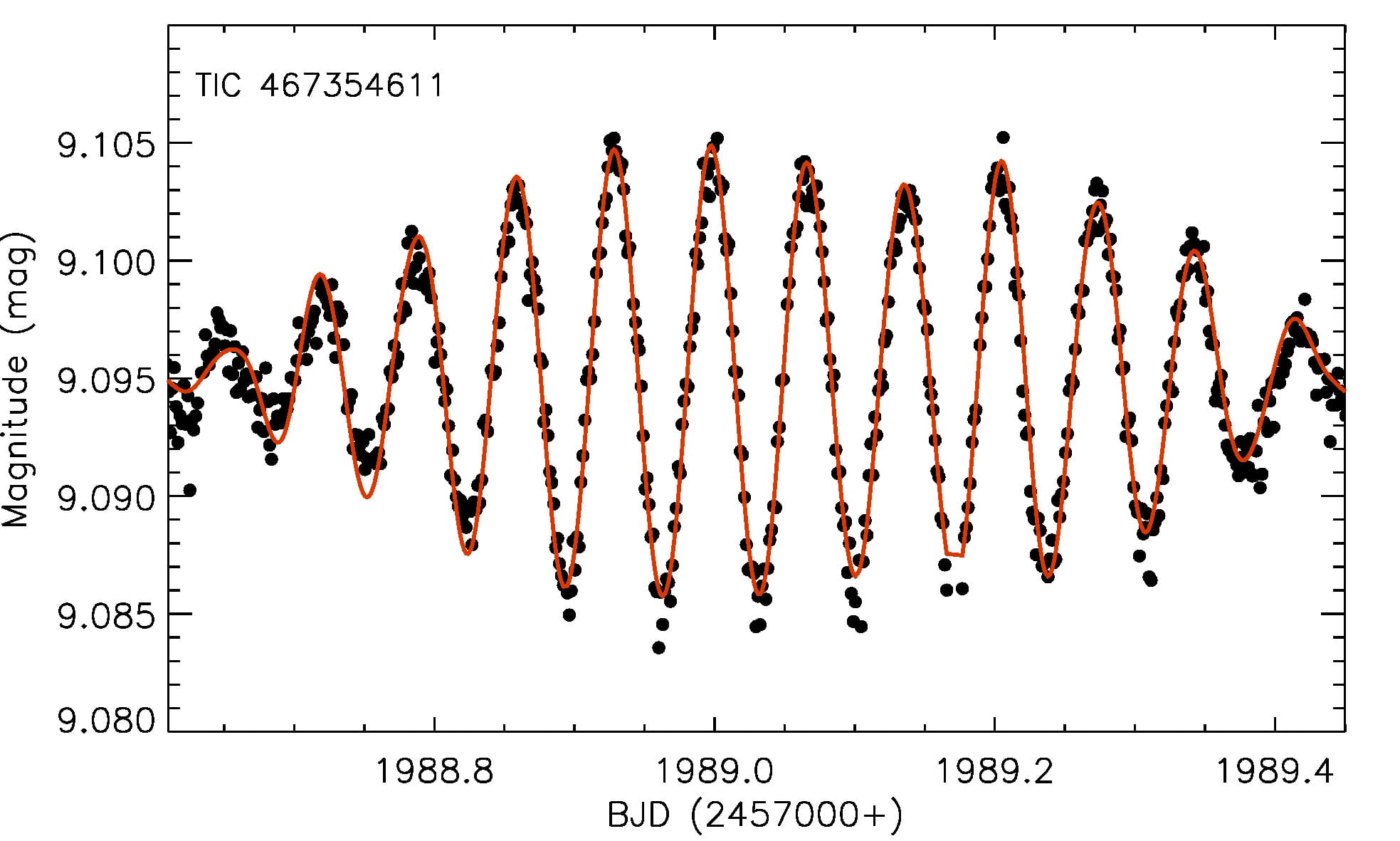}
  \end{minipage}
    \caption{Continuation.}\label{spec2}
\end{figure*}
\end{document}